\documentclass[a4paper,11pt]{article}
\usepackage{jheppub} 
\usepackage{lineno}
\usepackage[dvipsnames]{xcolor}

\usepackage{caption}
\usepackage{subcaption}
\usepackage{mathtools}

\usepackage{slashed}
\usepackage{cleveref}
\usepackage[normalem]{ulem}
\usepackage{multirow}
\usepackage{feynmp-auto}
\usepackage{amsmath}
\usepackage{csquotes}
\usepackage{comment}
                                                                  
\definecolor{all}{rgb}{1,0,1}

\def\be{\begin{equation}}
\def\ee{\end{equation}}

\title{
\boldmath 
Opening the Parameter Space of sub-GeV Inelastic Dark Matter through Parity Violation\\
}

\author[a]{\textbf{Giovani Dalla Valle Garcia,}}
\affiliation[a]{Institut für Astroteilchen Physik, Karlsruher Institut für Technologie (KIT), Hermann-von-\\Helmholtz-Platz 1, 76344 Eggenstein-Leopoldshafen, Germany}
\emailAdd{giovani.garcia@student.kit.edu}

\author[b,c]{\textbf{Juan Herrero-Garc\'ia,}}

\affiliation[b]{Departamento de Física Teòrica, Universitat de València, 46100 Burjassot, Spain}

\affiliation[c]{Instituto de Física Corpuscular (IFIC), CSIC‐Universitat de València\\
Parc Científic UV, c/ Catedrático José Beltrán, 2, E-46980 Paterna (València), Spain}

\emailAdd{juan.herrero@ific.uv.es}

\author[d]{\textbf{Joel Jones-P\'erez,}}
\affiliation[d]{Sección Física, Departamento de Ciencias, Pontificia Universidad Católica del Perú\\
Apartado 1761, Lima, Peru.}
\emailAdd{jones.j@pucp.edu.pe}

\author[d]{\textbf{Javier Silva-Malpartida}}
\emailAdd{javier.silvam@pucp.edu.pe}

\abstract{Sub-GeV dark matter (DM) has emerged as a particularly compelling target in light of the persistent null results from conventional DM searches. 
While $s$-wave annihilating DM candidates with masses below the GeV are strongly constrained by indirect-detection bounds, inelastic scenarios can naturally evade these limits. 
In this work, we show that parity violation can play an important role in inelastic DM models featuring long-lived excited states by inducing small diagonal couplings that significantly relax experimental constraints. A precise determination of the excited-state abundance is essential for assessing the phenomenology of such models.  To this end, we solve the integrated Boltzmann equation, fully accounting for up- and down-scattering with electrons and positrons as well as dark-sector conversion processes. Using the resulting abundance, we update the viable parameter space in light of the most recent experimental constraints and demonstrate that parity-violating interactions can reopen broad regions of parameter space that would otherwise be excluded. Moreover, the forthcoming LDMX experiment will probe a significant portion of the parameter space. The framework developed in this work can be readily applied to other exothermic sub-GeV DM scenarios.
}

\keywords{Sub-GeV Dark Matter, Dark Sectors, Inelastic, Exothermic, Parity Violation, Dark Sector Temperature}

\makeatother

\begin{document}

\maketitle

\section{Introduction} \label{sec:intro}

Dark matter (DM) constitutes a major component of the Universe’s energy budget, yet its fundamental nature remains unknown. Over recent decades, an extensive variety of theoretical models has been proposed, accompanied by a substantial experimental effort to test them. Nonetheless, no definitive signals have been detected so far.\footnote{A very recent study of Fermi-LAT data has identified a halo-like excess in the Galactic diffuse gamma ray emission at energies $\sim 20$--$30\,\mathrm{GeV}$, which could be interpreted in terms of DM annihilations. Nevertheless, the result is not statistically conclusive and requires independent confirmation \cite{Totani:2025fxx}.}   This situation has spurred growing interest in scenarios that can evade existing constraints. One such possibility is that DM consists of particles with masses below the GeV scale, which can naturally escape current direct-detection (DD) bounds. However, standard ($s$-wave) sub-GeV DM candidates are strongly constrained by indirect-detection (ID) searches. It is therefore crucial to identify and investigate sub-GeV DM models capable of satisfying all existing phenomenological constraints, thereby providing well-motivated targets for future searches, specially at collider and intensity-frontier experiments.

In this work we study one of such scenarios: inelastic DM (iDM)~\cite{TuckerSmith:2001hy}.\footnote{Inelastic (endothermic) DM was initially proposed to resolve the tension between the annual modulation signal observed by the DAMA Collaboration~\cite{Bernabei:2010mq} and other experimental DD upper limits~\cite{TuckerSmith:2001hy,TuckerSmith:2004jv,Bozorgnia:2013hsa}. Many models implementing iDM with a rich phenomenology have been suggested~\cite{Hall:1997ah,Cui:2009xq,Batell:2009vb,Chang:2010en,Aprile:2020tmw}, and some proposals have aimed to reconcile DD results via exothermic iDM scatterings~\cite{Graham:2010ca,Frandsen:2014ima,Chen:2014tka}. Inelastic DM has also been explored in a wide variety of other applications, for example ID signals~\cite{Arkani-Hamed:2008hhe, Chen:2009dm, Finkbeiner:2009mi}, capture in the Sun~\cite{Blennow:2015hzp, Blennow:2018xwu}, self-interactions~\cite{Blennow:2016gde} and collider searches~\cite{Izaguirre:2015zva}. Moreover, iDM was explored as a possible explanation for the now gone small excess measured in electron recoils in the XENON1T experiment~\cite{Harigaya:2020ckz,Baryakhtar:2020rwy,Bramante:2020zos,CarrilloGonzalez:2021lxm,Bloch:2020uzh,Lee_2021,Leerdam:2024jxn}.} We consider a dark sector (DS) that consists of a ground state $\chi$ with mass $m_\chi$, and a (long-lived) excited state $\chi^*$ with mass $m_{\chi^*}$, typically with a small mass splitting,
\begin{equation}
\delta = m_{\chi^*} - m_\chi \ll m_\chi \,.
\end{equation}
The scattering event involving the conversion of $\chi$ into $\chi^*$ (dubbed `up'-scattering) is endothermic, whereas   the conversion of $\chi^*$ into $\chi$ (dubbed `down'-scattering) is exothermic. 
The mass splitting (although small compared to the DM mass) makes the kinematics and phenomenology of iDM interactions significantly different from the purely-elastic case.

One of the most interesting production mechanisms is via thermal freeze-out. Crucial for the viability and phenomenology of these models is the current abundance of the excited states, leftover after all conversion processes have frozen out. Dark matter observables, such as DD, ID and self-interactions depend crucially on the fraction of excited states. Of course, the latter is not an independent parameter, but depends non-trivially on the particle physics model (and the parameter range) considered. Therefore, for a given model, a study of the cosmological evolution of the number density of excited states is necessary to correctly take into account all relevant constraints.

Typically, the total DM abundance is determined by annihilations which freeze out at early times and are well understood; however, the abundance of iDM states can be affected by scatterings (off electrons, protons, other DS states, etc.), which are active at much later times. At high temperatures, much larger than the mass splitting $\delta$, the up-scattering processes occur at the same rate as the corresponding down-scattering ones. Generally, the important processes are $ \chi\chi \leftrightarrow \chi^*\chi^* $ and $ \chi X \leftrightarrow \chi^* X $, where $ X $ represents any Standard Model (SM) particle. Both processes contribute to keep $ \chi $ and $ \chi^* $ in chemical equilibrium, and in addition the latter keep the DS in kinetic equilibrium with the SM thermal bath. Thus, the total DM abundance is essentially made up equally of $ \chi $ and $ \chi^* $ states, and the DM temperature $T_\chi$ tracks the temperature of the universe, $T_{\rm SM}\equiv T$. 

Up-scattering requires an energy input, implying that the cross section is suppressed compared to the down-scattering one. As a result, once the temperature has cooled enough that the mass splitting $ \delta $ is relevant $T_\chi\lesssim \delta$, the up-scattering rate is suppressed compared to the down-scattering rate. Generically, this results in a decrease in the fraction of $ \chi^* $ states. We can calculate the precise evolution of the fraction by considering the scattering rates compared to the Hubble rate, in addition to the evolution of the DM temperature. If the scattering processes freeze out much before $T_\chi \gg\delta$, we expect to obtain equal abundances of $ \chi $ and $ \chi^* $, but if they are still active at $ T\sim\delta $ or below, the $ \chi^* $ fraction will be exponentially reduced. We note that, in principle, decays of the excited state into massless or light SM particles also decrease the number of excited states. One can easily include them by decoupling freeze-out and the decays, since, in the scenario we consider, the decay time is much longer than the freeze-out time.

In this work, we focus on the commonly considered pseudo-Dirac inelastic DM model with a dark photon mediator~\cite{Izaguirre:2015zva,Beacham:2019nyx,DeSimone:2010tf}, but in a more generic framework which allows for parity violation~\cite{Darme:2018jmx,Garcia:2024uwf}. The standard parity-conserving scenario faces stringent constraints from DD and ID searches~\cite{Baryakhtar:2020rwy,CarrilloGonzalez:2021lxm}, forcing recent works to study either a cosmologically fast decaying regime, $\delta >2m_e$, or extremely small mass splittings, $\delta \lesssim100$~eV~\cite{Berlin:2023qco}.  We will show that relaxing such an assumption simply turns the pseudo-Dirac iDM model into a viable exothermic DM model for mass splittings $\delta\gtrsim$~keV. In this scenario, the exact fraction of excited states is crucial to determine DD and ID constraints. To compute it, we numerically solve the Boltzmann equation for the fraction of excited DM states. Our framework can be easily adapted to other iDM and exothermic DM models. 

Finally, let us mention that similar analysis have been performed in the literature. In Ref.~\cite{CarrilloGonzalez:2021lxm}, the authors perform the first comprehensive analysis of how pseudo-Dirac DM with small mass splittings (100 eV–MeV) evolves cosmologically, but without explicitly considering the effect of parity-violating interactions. In Ref.~\cite{Baryakhtar:2020rwy}, it is studied how solar and terrestrial up-scattering can efficiently populate excited states, enabling MeV-scale DM to be probed via Sun-induced excitation and GeV-scale DM via Earth-induced excitation---in the latter, they consider an additional decay channel. Finally, in Ref.~\cite{Berlin:2023qco}, the authors consider extremely small mass splittings, $\delta \lesssim100$~eV, and study how late-time re-population of the excited state in the Galaxy and subsequent coannihilation yield observable $1–100$ MeV gamma rays that may be detected.

The remainder of the paper is structured as follows. We present the inelastic DM model with parity violation in \cref{sec:model}. The computation of the DM relic abundance, results for the DS temperature evolution and the Boltzmann equation providing the evolution of the fraction of excited states are given in \cref{sec:relic}. In \cref{sec:pheno} we discuss the different experimental constraints. Our numerical results are shown in \cref{sec:numres}. Finally, we conclude in \cref{sec:conc}. There are also three appendices with further technical details: in \cref{app:int} we outline the method used for the numerical integration, in \cref{app:temperatureDS} we provide a discussion and derivation of the DS temperature, and in \cref{app:Zinv} we obtain constraints from invisible $Z$-boson decays.

\section{Pseudo-Dirac inelastic dark matter with parity violation} \label{sec:model}

We consider the standard pseudo-Dirac iDM model with a dark photon mediator~\cite{Izaguirre:2015zva} in a generic parity setup, following Ref.~\cite{Garcia:2024uwf}. For concreteness, we introduce a DS containing a new $U(1)'$ gauge symmetry with a gauge boson $A'$ and gauge coupling $e'$ (the dark fine-structure constant is defined as usual, $\alpha' \equiv e^{\prime\,2}/4\pi$). Before the breaking of $U(1)'$, the DS contains a Dirac fermion $\chi_D = \chi_L + \chi_R$, which is a singlet under the SM gauge group but carries a $U(1)'$ charge $q_{D}$. We take $q_{D} \equiv 1$ without loss of generality, since any choice of charge may be absorbed in the gauge coupling $e'$. 

We assume that the $U(1)'$ gauge symmetry is broken at low energies in a way that allows Majorana mass terms for the dark fermion and a gauge boson mass.\footnote{We remain agnostic about the origin of the breaking. For instance, one could consider a dark Higgs mechanism, see Ref.~\cite{Garcia:2024uwf}.} Thus, we add to the SM Lagrangian the terms
\begin{equation} \label{eq:NewPhysicsLagrangian}
    \mathcal{L}_{\text{NP}} =  \mathcal{L}_{\chi} + \mathcal{L}_{V}   \, ,
\end{equation} 
where
\begin{align} 
 \label{eq:DarkFermionLagrangian}
    \mathcal{L}_{\chi}  = & i \bar{\chi} \slashed{D} \chi- m_d\,\bar{\chi} \chi -  \left(\frac{1}{2} m_L \bar{\chi}^c_L \chi_L + \frac{1}{2}m_R  \bar{\chi}^c_R \chi_R + \text{h.c.}\right)\,, \\
 \label{eq:DarkPhotonLagrangian}
\mathcal{L}_V = & -\dfrac{1}{4} A'^{\mu \nu} A'_{\mu\nu} - \dfrac{1}{2} \dfrac{\epsilon}{\cos{\theta_w}} B^{\mu \nu} A'_{\mu\nu} -\dfrac{1}{2} m^2_{A'} A'^{\mu \nu} A'_{\mu\nu} \, . 
\end{align} 
Here, $A'_{\mu\nu}$ and $B_{\mu\nu}$ are the field strengths for $A'_\mu$ and the SM hypercharge field $B_\mu$, respectively. Moreover, we denote the weak mixing angle by $\theta_w$, the covariant derivative by $ i D_{\mu} \chi = i \partial_{\mu} \chi  -  e' A'_{\mu} \chi $ and the charge-conjugated field by $\chi_{L/R}^c = C \gamma_0^T \chi_{L/R}^*$, where $C$ is the charge conjugation matrix. In \cref{eq:DarkFermionLagrangian}, all $\mathcal{CP}$-violating phases can be rephased into the Dirac mass $m_d$, however, we will take this parameter to be real, $m_d \in \mathbb{R}$.  Note that in the limit $m_{L/R} \to0$, we restore a global $U(1)$ symmetry, and therefore the latter may be naturally small in the 't Hooft sense. In \cref{eq:DarkPhotonLagrangian}, the term proportional to $\epsilon$ is known as kinetic mixing and it provides the portal between the DS and the SM. For $m_{A'}\lesssim10$~GeV, $A'$ essentially only mixes with the SM photon, and the diagonalization of the kinetic term induces a millicharge $Q'=-\epsilon Q$, with $Q$ the respective electromagnetic charge, to all SM particles. Hence, we will refer to $A'$ as the dark photon. On the other hand, DS particles remain electromagnetically neutral.

After diagonalizing the fermion mass matrix we find two physical Majorana states, $\chi$ and $\chi^{\ast}$, with masses
\begin{align}
m_{\chi,\,\chi^{\ast}}=  \sqrt{m_d^2+  \dfrac{1}{4} (m_R-m_L)^2} \mp \dfrac{1}{2} (m_L+m_R) \, ,
    \label{eq:mass_eigenstates}
\end{align} where we made the assumption $m_d \geq \sqrt{m_R \,m_L}$, motivated by the aforementioned naturalness argument.
On the mass basis, the interaction matrix between $\chi$'s and $A'$ generically contains both elastic (diagonal) and inelastic (off-diagonal) terms. The respective strengths of these two type of interactions are defined as \begin{align}\label{eq:el&inelFINEstructure}
    \alpha_{\rm el}'\equiv \alpha'\cos^22\theta && \text{and} && \alpha_{\rm inel}'\equiv \alpha'\sin^22\theta \,,
\end{align}   where \begin{equation}\label{eq:cos2theta}
    \cos 2\theta=-\dfrac{\delta}{2m_\chi+\delta}\dfrac{\delta_y}{2+\delta_y}~,
\end{equation}
with the normalized Majorana mass difference $\delta_y$ defined as 
\begin{align}\label{eq:deltaYdefinition}
       \delta_y= \dfrac{m_R-m_L}{m_L} \, .
\end{align}
For $\delta_y=0$, parity is preserved (the Lagrangian is invariant under a transformation changing $\chi_L\leftrightarrow\chi_R$) and interactions are purely off-diagonal. For this reason, we may also refer to $\delta_y$ as the parity-violating parameter.  We take $\delta_y\geq0$ without loss of generality.\footnote{A scenario with $\delta_y<0$ is equivalent to a relabeling of our initial fermions, e.g., $\chi_{L/R}\to (\chi_{R/L})^c$ and $q_D \to -q_D$. }

\subsection{Parameter space}
\label{sec:paramspace}

We are interested in thermal DM produced via the predictive visible freeze-out mechanism---that is, through annihilations into SM particles. This requires $m_\chi \gtrsim 10$~MeV due to Big Bang Nucleosynthesis (BBN) constraints on new light particles in thermal equilibrium with the SM bath~\cite{Sabti:2019mhn}, and $m_{A'} \gtrsim 1.5\, m_\chi$  to suppress the secluded annihilation channel $\chi\chi \to A'A'$~\cite{Fitzpatrick:2021cij,Yang:2022zlh}
.\footnote{Furthermore, $\chi\chi \to A'A'$ annihilations proceed via $s$-wave and, thus, the secluded scenario faces stringent constraints from CMB in the sub-GeV regime~\cite{Planck:2018vyg,Duerr:2020muu}.}

This choice of mass hierarchy allows us to consider large values for the dark gauge coupling. We adopt benchmark values for the model parameters, following standard practices in the literature. For the DS fine-structure constant, we choose:
\begin{align}
    \alpha' = 0.1 && \text{and} && \alpha' = 0.5 \,,
\end{align}
representing relatively optimistic values. Scenarios with $\alpha' \ll 1$ generally face challenges in achieving the correct DM abundance via visible freeze-out~\cite{NA64:2023wbi,NA64:2025ddk}. Additionally, to better understand the $\alpha'$ dependence, we also explore varying $\alpha'$ in the range $\alpha' \in [10^{-3},1]$ for fixed values of mass splitting $\delta$.

In addition to these choices, we focus on the region of parameter space where the excited state $\chi^\ast$ has a sufficiently long lifetime to affect DD experiments and cosmological observables---particularly the CMB. In this regime, $\chi^\ast$ is effectively stable on collider timescales. Thus, we need to study and restrict possible fast decay modes of $\chi^*$. In particular, we impose the condition $\delta < 2\,m_e$, thereby kinematically forbidding the $\chi^\ast$ decay channels into charged SM particles~\cite{Batell:2009vb}. Under these conditions, the model we consider generally features small normalized mass splittings $\delta / m_{\chi} \ll 1$, for which decays of the excited state into dark photons ($\chi^* \to \chi A'$) become also kinematically forbidden.

As we will see in \cref{sec:relic}, parity-violating effects become negligible at very small values of $\delta$. Thus, in the following, we limit our analysis to mass  splittings $\delta \gtrsim \text{keV}$.\footnote{Note that for smaller $\delta$ a dedicated treatment of inelastic re-excitation is required~\cite{Berlin:2023qco}.}  In particular, when fixing the mass splitting we will adopt the benchmark choices: 
\begin{align}
    \delta = 100\text{ keV} && \text{and} && \delta = 500\text{ keV} \,.
\end{align} 
Moreover, we will focus on DM masses below $m_\chi \lesssim100$~MeV. As we will demonstrate, heavier DM masses (for typical benchmarks) lead to equal populations of ground and excited states. This is strongly disfavored by CMB observations, as inelastic DM $s$-wave annihilations into $e^+e^-$ pairs lead to unobserved distortions in the temperature anisotropies. In addition, for the dark photon mass we fix:
\begin{equation}
    m_{A'} = 3\,m_\chi\,,
\end{equation}
a choice that avoids strong propagator suppression and lies just above the $s$-channel resonance at $m_{A'} = 2m_\chi$, providing only a mild resonant enhancement. Heavier dark photons are typically disfavored in visible freeze-out scenarios~\cite{Duerr:2019dmv}.

\subsection{Decays}
\label{sec:decays}
\begin{figure}[t]
    \centering
    \includegraphics[width=0.5\linewidth]{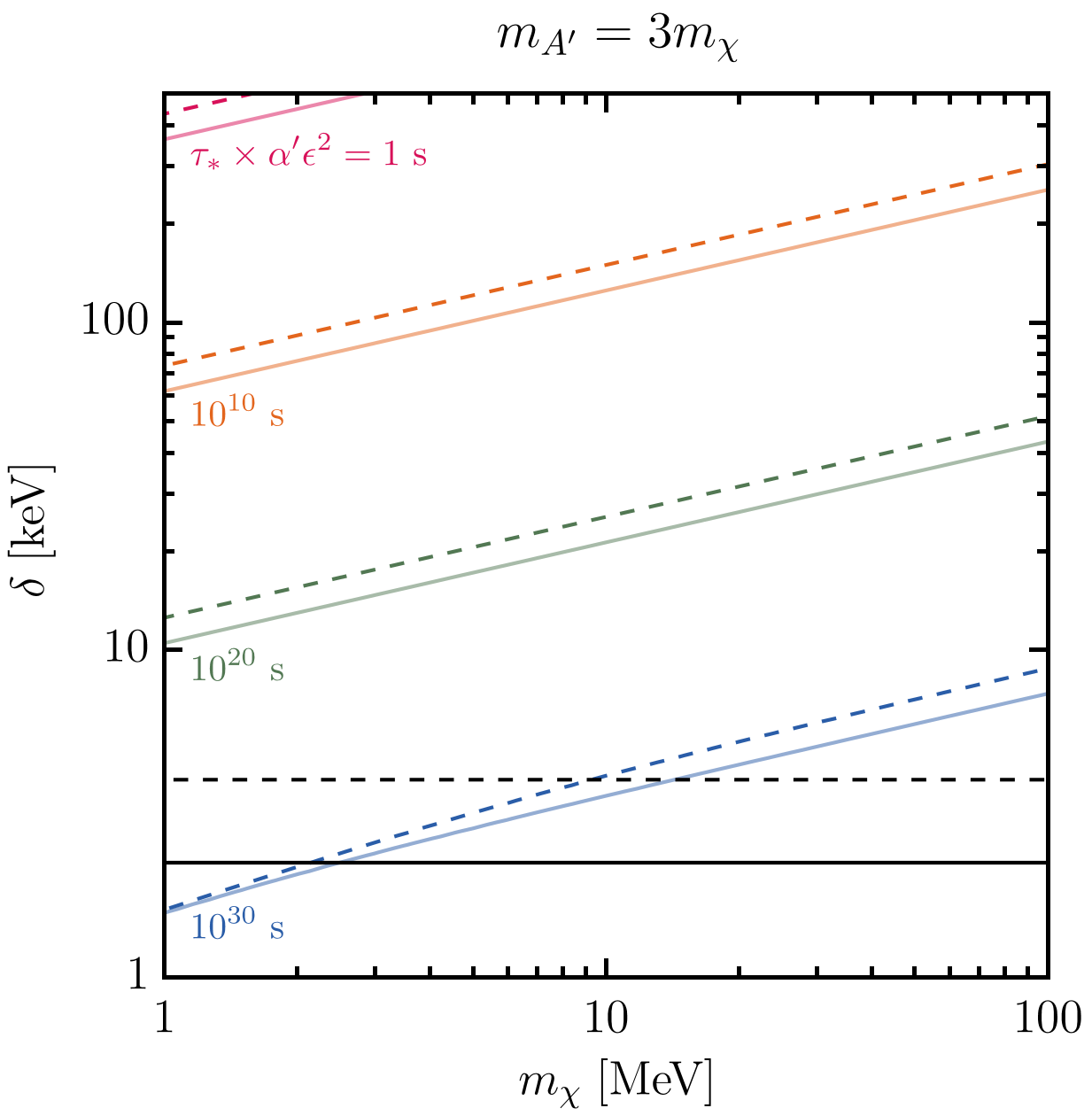}
    \caption{Contours of the excited-state lifetime $\tau_\ast \times(\alpha' \epsilon^2)$ are shown as colored lines, computed using the approaches of Ref.~\cite{Batell:2009vb} (solid) and Ref.~\cite{Bernal:2017mqb} (dashed). For the small mass splittings considered ($\delta / m_{\chi} \ll 1$), we have $\alpha' \simeq \alpha'_{\rm inel}$, and the results are effectively independent of $\delta_y$. The horizontal solid black line marks the value of $\delta$ below which the decay into neutrinos, $\chi^\ast \to \chi 2\nu$, dominates, while above the decay into photons $\chi^\ast \to \chi 3\gamma$ becomes dominant (and similarly for the horizontal dashed line). Note that kinetic mixing values in the range $\epsilon \sim 10^{-6}$--$10^{-4}$ are typically required to reproduce the observed DM relic abundance via standard thermal freeze-out, assuming benchmark choices of $\alpha' \sim 0.1- 0.5$ and $m_{A'} \simeq 3 \,m_\chi$, for the DM masses considered in this work---see \cref{fig:relicAbundance}.}
    \label{fig:lifetimes}
\end{figure}

Having defined the parameter space, we now discuss the remaining decay modes. These are: $\chi^* \to \chi + 3\gamma$, mediated by charged fermion loops, and $\chi^* \to \chi + 2\nu$, mediated by the suppressed $Z-$boson mixing. The corresponding decay widths, following the approach of Ref.~\cite{Batell:2009vb}, are given by~\cite{CarrilloGonzalez:2021lxm}:
\begin{align}
    \Gamma_\ast(\chi^* \to \chi + 3\gamma) &\sim \frac{17 \alpha^4\, \alpha'_{\rm inel}\, \epsilon^2 }{2^7 \cdot 3^6 \cdot 5^3 \,\pi^3 } \frac{\delta^9}{m_e^8} \frac{\delta^4}{m_{A'}^4} \,, \\
    \Gamma_\ast(\chi^* \to \chi + 2\nu) &\simeq \frac{4 \alpha\,  \alpha'_{\rm inel} \, \epsilon ^2}{315 \pi  \cos^4\theta_w } \frac{\delta ^9}{m_{A'}^4 m_Z^4} \,,
\end{align}
where $m_e$ and $m_Z$ denote the electron and $Z-$boson masses, respectively, and $\alpha$ is the electromagnetic fine-structure constant. Following Ref.~\cite{Bernal:2017mqb}, we can also estimate the three-photon decay width via:
\begin{equation}
     \Gamma_\ast(\chi^* \to \chi + 3\gamma) \sim \Gamma_\ast(\chi^* \to \chi + 2\nu) \left. \frac{\Gamma(A' \to 3\gamma)}{\Gamma(A' \to 2\nu)} \right\rvert_{m_{A'} \to \delta} \,,
\end{equation}
where $\Gamma(A' \to 3\gamma)$ is taken from Ref.~\cite{McDermott:2017qcg}. In \cref{fig:lifetimes} we plot contours of the excited-state lifetime in the plane of the DM mass $m_\chi$ and splitting $\delta$. We show both lifetime estimates in order to illustrate the theoretical uncertainty in $\Gamma_\ast(\chi^* \to \chi + 3\gamma)$. As can be seen, most of the parameter space presents cosmologically stable excited states, i.e. $\tau_\ast > t_U \equiv4.35\times 10^{17}$~s~\cite{Cirelli:2024ssz}. In the remainder, we will adopt the results from Ref.~\cite{Bernal:2017mqb} which lead to relatively stronger constraints given the larger lifetimes predicted.

Given that the decays into neutrinos are subdominant for $\delta\gtrsim2$~keV, we may neglect their contribution to the excited-state decay rate. The corresponding lifetime is then estimated as \begin{equation}
    \tau_\ast \sim 10^{22}\,\text{s} \left(\frac{100~\text{keV}}{\delta}\right)^{13} \left(\frac{m_{A'}}{40~\text{MeV}}\right)^4 \left(\frac{10^{-5}}{\epsilon}\right)^2 \left(\frac{0.5}{\alpha'_{\rm inel}}\right) \,,
\end{equation} however, we keep the  full decay width in our numerical analysis.

\section{Dark matter relic abundance and fraction of excited states} \label{sec:relic}

As mentioned in the previous section, we are interested in the case where the DM relic abundance is produced by visible freeze-out, meaning that we need to evaluate the chemical decoupling of DM annihilations into SM particles. However, as it is well known, this is not the end of the story. Given the long lifetime of $\chi^*$, one also needs to determine the relative abundance of excited states $f$, defined as
\begin{align}\label{eq:FratioDEF}
	f\equiv \dfrac{n_{\chi^\ast}}{n_{\chi^\ast}+n_\chi}=\dfrac{n_{\chi^\ast}}{n_{\chi,\rm tot}} \,,\qquad 0\leq f\leq 0.5\,,
\end{align}
where $n_i$ is the cosmological number density of the particle species $i$. This relative abundance, which is time-dependent, is crucial for the application of phenomenological constraints on the parameter space. To this end, it is essential to consider interactions that change the relative abundance of ground and excited DM states after freeze-out. This requires tracing the evolution of the temperature of the DS, $T_\chi$, which in turn needs the evaluation of the kinetic decoupling of DM from the SM plasma. In the following, we provide technical details of all of these steps.

\subsection{Chemical Freeze-Out}

\begin{figure}[t]
    \centering
    \begin{subfigure}[b]{0.5\textwidth}
        \centering
        \includegraphics[scale=0.35, trim={0 550 0 120}]{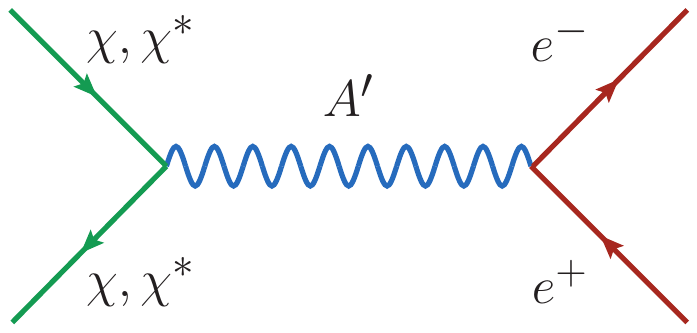}
        \caption{}
    \end{subfigure}%
    \hfill 
    \begin{subfigure}[b]{0.5\textwidth}
        \centering
        \includegraphics[scale=0.35, trim={0 480 0 120}]{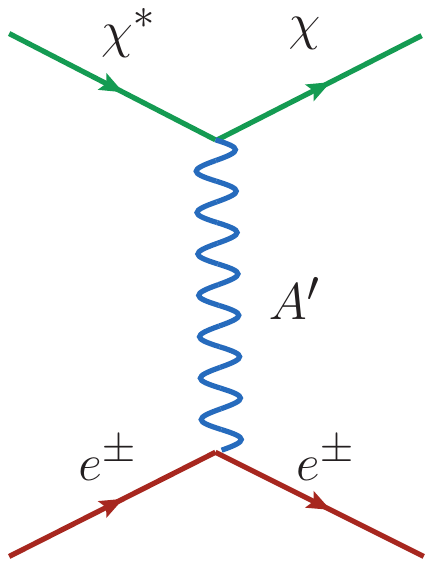}
        \caption{}
    \end{subfigure} %
    
    \vspace{1.5em}
    
    \begin{subfigure}[b]{0.5\textwidth}
        \centering
        \includegraphics[scale=0.35, trim={0 550 0 120}]{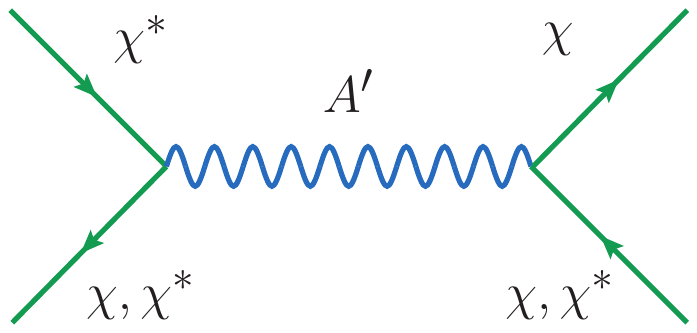}
        \caption{}
    \end{subfigure}%
    \hfill 
    \begin{subfigure}[b]{0.5\textwidth}
        \centering
        \includegraphics[scale=0.35, trim={0 480 0 60}]{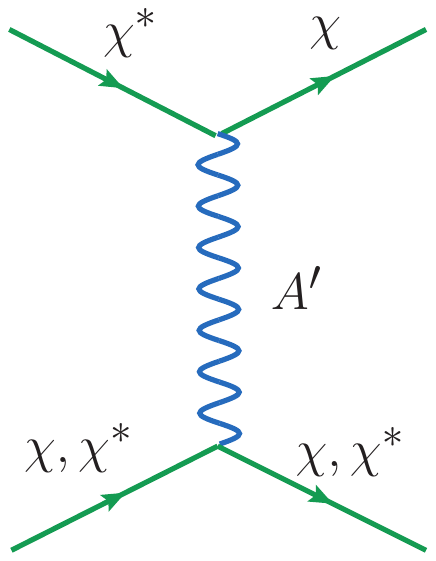}
        \caption{}
    \end{subfigure}%
    \caption{Key processes governing number evolution and state populations: (a) Annihilation and coannihilation into SM electrons and positrons; (b) DM–electron/positron inelastic scattering; Dark sector self-scatterings and conversions in $s$-channel (c) and $t$-channel (d).}
    \label{fig:diagrams}
\end{figure}


The possible total DM number-changing processes in our scenario include (co-)annihilations into SM fermions, $ \chi^{(\ast)} \chi^{(*)} \to \overline{\rm SM}\; {\rm SM}$, shown on the upper left panel of \cref{fig:diagrams} for the case of electrons. The relevance of elastic or inelastic interactions will depend on the presence of parity-violating interactions. In the regime $\delta_y = 0$, the total DM number density of the two fermionic states, $n_{\chi,\text{tot}} \equiv n_{\chi} + n_{\chi^*}$, is determined solely by the (inelastic) coannihilation process $\chi\, \chi^* \leftrightarrow e^+ e^-$~\cite{Foguel:2024lca, Berlin:2023qco}.\footnote{Interactions with electrons and positrons dominate due to their tiny masses, which lead to much larger abundances at low temperatures $T\lesssim 100$~MeV compared to other SM charged particles. } In contrast, once the elastic couplings are  available, i.e. $\delta_y \neq 0$, the (elastic) annihilation channels $\chi\, \chi \to e^+ e^-$ and $\chi^* \chi^* \to e^+ e^-$ must also be included~\cite{Garcia:2024uwf}. 

\begin{figure}
    \centering
    \includegraphics[width=0.46\linewidth]{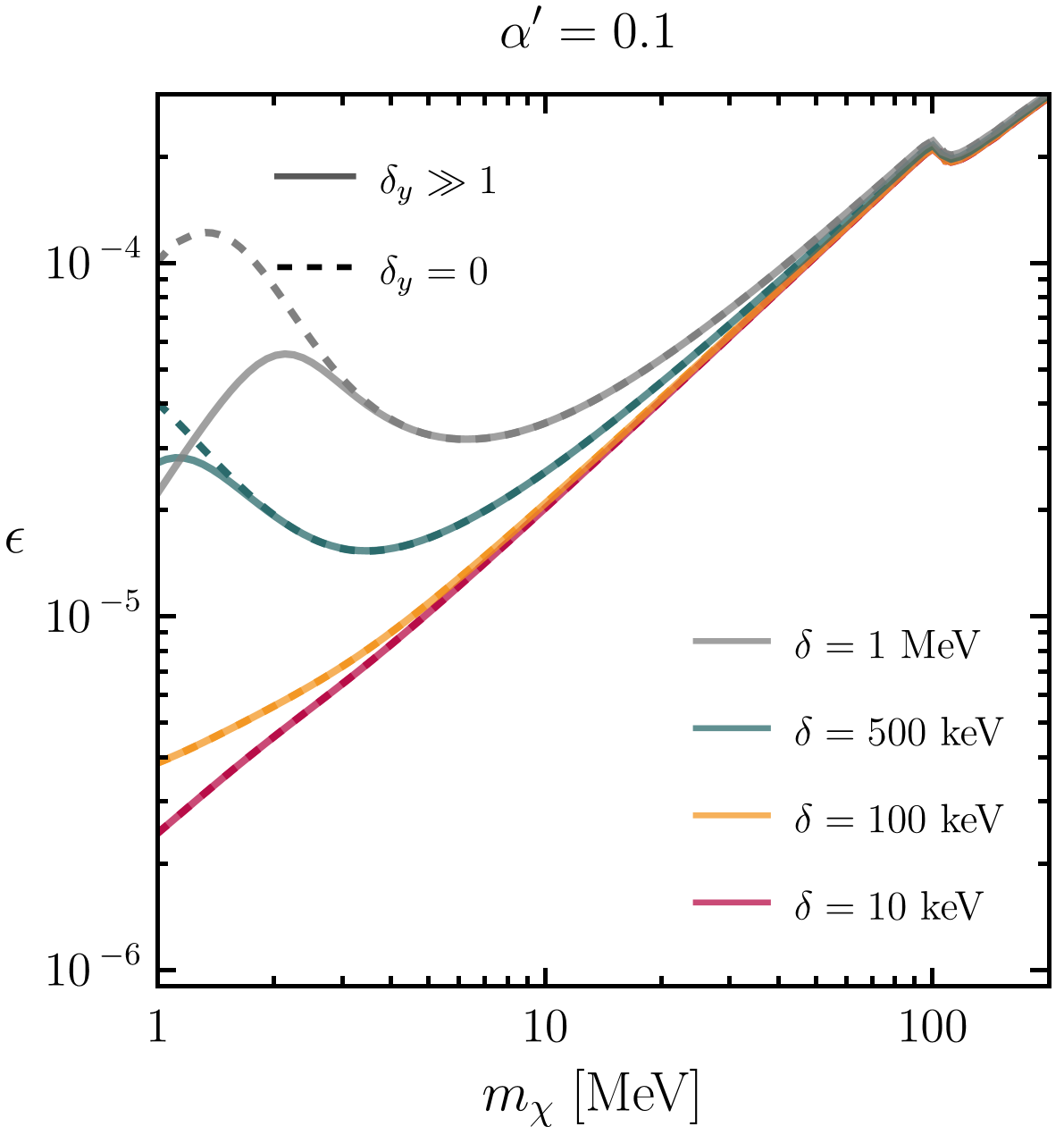}\hfill
    \includegraphics[width=0.46\linewidth]  {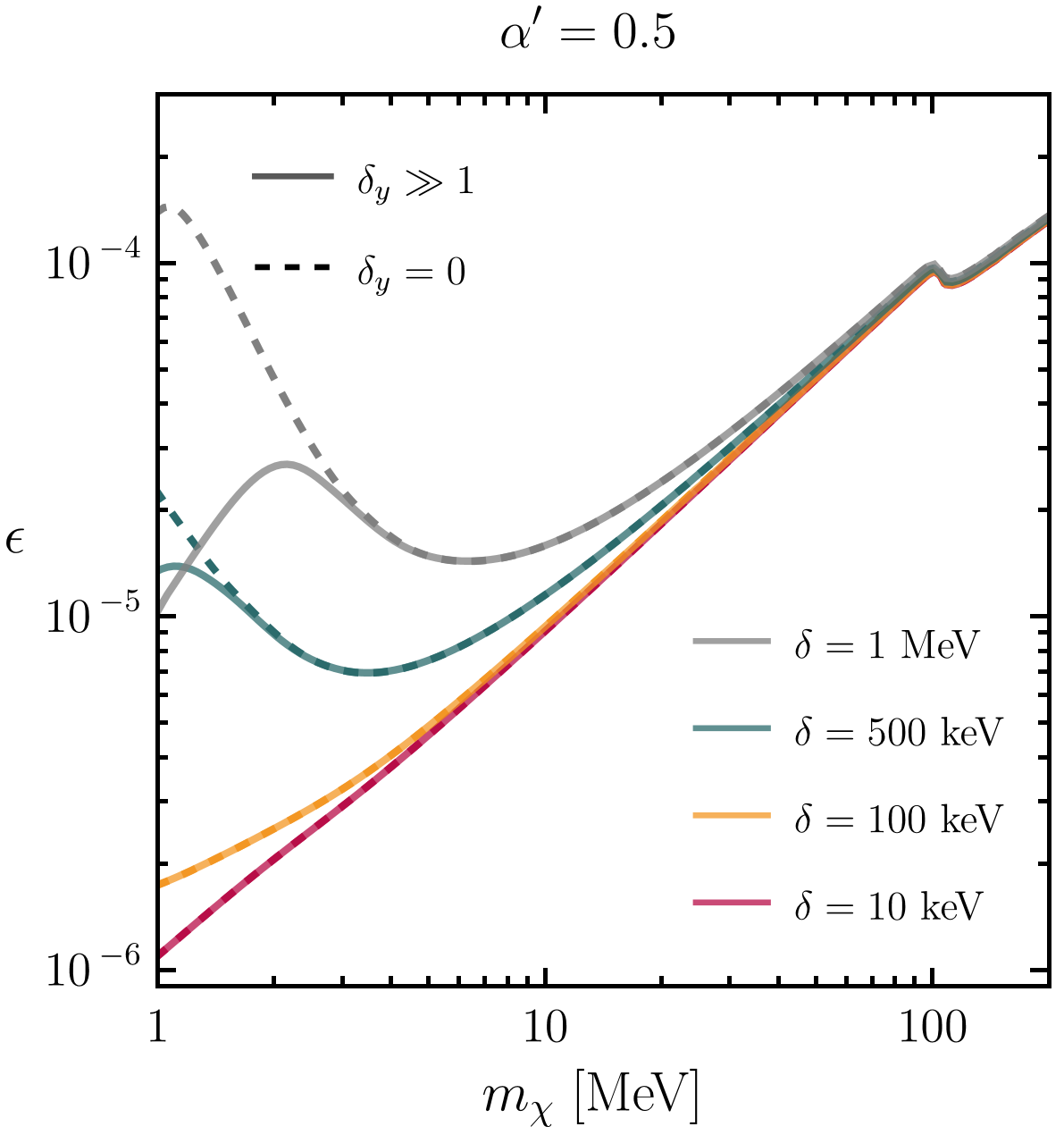}
    \caption{Thermal targets ($\Omega h^2 \simeq 0.12$~\cite{Planck:2018vyg}) for the benchmark choices $\alpha'=0.1$ (left) and $\alpha'=0.5$ (right) with $m_{A'}=3m_{\chi}$. We show results for both parity-conserving $\delta_y=0$ (dashed) and parity-breaking $\delta_y\gg1$ (solid) (not-so-)inelastic DM models for various mass splittings $\delta = \{10\text{ keV},\, 100\text{ keV},\, 500\text{ keV},\, 1\text{ MeV}\}$ in different colors.}
    \label{fig:relicAbundance}
\end{figure}

In \cref{fig:relicAbundance}, we show contours reproducing the observed relic abundance $\Omega_{\rm obs} \,h^2$, calculated using the \texttt{micrOMEGAs} package~\cite{Belanger:2020gnr}, in the plane of the DM mass $m_\chi$ and the kinetic mixing $\epsilon$, for $\alpha'=0.1$ (left panel) and $\alpha'=0.5$  (right panel). We plot both parity-conserving (dashed lines) and parity-violating (solid lines) interactions, for  several values of the mass splitting $\delta$. As can be observed, elastic contributions, present for $\delta_y\neq 0$, are negligible except for relatively large normalized mass splittings, $\delta/m_\chi \gtrsim 0.2$~\cite{Garcia:2024uwf}.

The annihilation cross section for $\chi\,\chi^*\to e^+e^-$ has been calculated in Refs.~\cite{Berlin:2014tja,Berlin:2018bsc} and, in the limit where the mass splitting $\delta$ can be neglected so that $\alpha_{\rm inel}' \simeq \alpha'$, it is estimated as:
\begin{equation}
\label{eq:freezeoutapprox}
\sigma v\sim \frac{\epsilon^2 \alpha \,\alpha' m_\chi^2}{(m_{A'}^2-4m_\chi^2)^2}~.
\end{equation}
 For our benchmark scenarios with $m_{A'}=3m_{\chi}$, we expect that a reduction in $m_{\chi}^2$, for example, can be compensated by a corresponding decrease in $\epsilon^2\,\alpha'$.  That is, for fixed $\alpha'$, we expect $\epsilon \propto m_\chi$. Moreover, in this regime there is no dependence on $\delta_y$. As expected, the predictions from \cref{eq:freezeoutapprox} match our numerical results, shown in \cref{fig:relicAbundance}, for $\delta\ll T_{\rm fo}\sim m_\chi/10$.\footnote{Note that \cref{fig:relicAbundance} shows a change in slope at $m_\chi \sim m_\mu$ due to the kinematic opening of the $\chi \chi^\ast \to \mu^+ \mu^-$ channel.}

For larger $\delta/m_\chi$, the aforementioned expectations do not hold. First,  we find that a reduction in the mass eventually requires an increase in $\epsilon$. This is attributed to the fact that, when $\delta\sim T_{\rm fo}$, there exists a strong Boltzmann suppression of the $\chi^*$ number density, such that a larger $\epsilon$ is necessary to have the same thermally-averaged cross section.\footnote{For large mass splittings, the center-of-mass energy of the coannihilation can approach the dark-photon resonance, 
$E_{\rm CM} \simeq 2m_\chi + \delta \approx m_{A'}$, 
which explains the decrease in the required kinetic mixing $\epsilon$ for $m_\chi \lesssim 2$~MeV and $\delta=1$~MeV observed in \cref{fig:relicAbundance}. 
In this resonant regime, early kinetic decoupling may become important~\cite{Brahma:2023psr}. 
We neglect this possibility here, since the resonance region is rather limited, 
and such decoupling effects would only be relevant for $m_\chi \lesssim 1.2$~MeV (
following $\epsilon_R \gtrsim0.1$ from Ref.~\cite{Brahma:2023psr}), which are typically already excluded by BBN constraints~\cite{Sabti:2019mhn}.} Such a suppression implies that, for parity-violating scenarios ($\delta_y\gg1$), standard annihilations $\chi\chi\to e^+e^-$ begin to dominate over coannihilations $\chi\chi^\ast\to e^+e^-$ and the approximate proportionality of $\epsilon$ and DM mass is recovered.
 
We can relate the relic density to the asymptotic DM yield after freeze-out, $Y$, whose present day value (denoted as usual by the $0$ subindex, i.e. $Y_0$) must satisfy:
\begin{equation}
 m_{\chi}Y_0=\frac{\Omega_{\rm obs} h^2\,\rho_c}{s_0\, h^2}\approx 4.3\times10^{-10}\,{\rm GeV}~,
\end{equation}
where we have used the values of the observed relic density $\Omega_{\rm obs} h^2$, the critical density $\rho_c$ and the present day entropy density $s_0$ found in Ref.~\cite{ParticleDataGroup:2024cfk}. Considering that $Y=n_{\chi,{\rm tot}}/s$, and assuming a calculated relic abundance $\Omega \,h^{2}$, we can estimate the total DM number density at the temperature of the SM plasma, $T_{\rm SM}$, using:
\begin{align} \label{eq:ntot}
    n_{\chi,\text{tot}}(T_{\rm SM}) \approx 1.3 \times 10^{-3}\,{\rm cm}^{-3}\,\left(\frac{s(T_{\rm SM})}{s_0}\right)\left ( \frac{1 \,\rm MeV}{m_{\chi}} \right ) \, \left ( \frac{\Omega\, h^{2}}{\Omega_{\rm obs} h^{2}} \right )\,,
\end{align}
In the equation above, the SM entropy density is given by:
\begin{align} \label{eq:sSM}
    s(T_{\rm SM}) = \frac{2\pi^{2}}{45} \, g_{\rm s}(T_{\rm SM}) \, T_{\rm SM}^{3}\,,
\end{align}
with $g_{\rm s}$ denoting the effective number of SM degrees of freedom contributing to the entropy density. 

\subsection{Temperature of the dark sector}

The DS typically remains in kinetic equilibrium with the SM bath, $T_\chi = T_{\rm SM}$, even after the DM chemical decoupling of $\chi^\ast \chi \to e^+ e^-$  at $T_{\rm fo}$. This is ensured by inelastic scatterings with electrons and positrons $\chi^\ast e^\pm \leftrightarrow \chi \,e^\pm$,\footnote{Elastic scattering is neglected since $\alpha'_{\rm el} \ll \alpha'_{\rm inel}$ and, for $T_{\rm kin} \gg \delta$, both states have similar abundances, $f \simeq 1/2$.} which dominate due to their extremely large abundance in comparison to the Boltzmann suppressed DM number densities,   $n_{e} \gg n_{\chi^{(\ast)}}$. 

Kinetic equilibrium is lost at a temperature $T_{\rm kin}$ generally below the electron mass $m_e$, once the density of charged particles becomes strongly Boltzmann-suppressed. After this point, the DS thermally decouples from the SM bath. Given the larger DM abundance compared to that of visible matter (and the near charge neutrality of the Universe, both implying $n_\chi \gtrsim n_{e}$ for sub-GeV DM), as well as the strong DS coupling $\alpha' \gg \alpha \epsilon^2$, scatterings among dark states efficiently maintain internal kinetic equilibrium, justifying the definition of a common DS temperature $T_\chi \neq T_{\rm SM}$.  

In the following, we adopt a freeze-out approximation for the temperature evolution. This is justified by considering that the chemical decoupling between $\chi^\ast$ and $\chi$ typically occurs at $T_{f^\ast} \ll T_{\rm kin}$ (at least for $\delta \ll m_e$), so the detailed shape of the temperature transition around $T_{\rm kin}$ is not expected to affect $T_{f^\ast}$. In this approximation, we compute the DM–SM momentum exchange rate $\gamma(T)$ and assume that $T_\chi = T_{\rm SM}$ until
\begin{equation}
    \gamma(T_{\rm kin}) = H(T_{\rm kin}) \, ,
\end{equation}
where $H$ is the Hubble rate, after which $T_\chi$ evolves independently according to entropy conservation (see \cref{app:temperatureDS} for details). Following Refs.~\cite{Gondolo:2012vh,Bertoni:2014mva,Berlin:2023qco}, the momentum exchange rate can be written as
\begin{equation}\label{eq:transf_momenta}
    \gamma(T) \simeq -\frac{1}{3 m_\chi T} 
    \int \frac{d^3 p_e}{(2\pi)^3} \,
    f_e (1 - f_e) \, v_e 
    \int_{-4p_e^2}^{0} dt \; t \, \frac{d\sigma_{\chi e}}{dt} \, ,
\end{equation}
where $f_e$ is the electron phase-space distribution, $v_e$ is the electron velocity, $p_e$ the electron momentum, and ${d\sigma_{\chi e}}/{dt}$ the differential cross section for $\chi^\ast e^\pm \leftrightarrow \chi \,e^\pm$ with respect to the Mandelstam variable $t$ (the squared four-momentum transfer).

After kinetic decoupling, we assume that the two DM states remain in mutual kinetic equilibrium, allowing both the definition of a common DS temperature and entropy, until the state-conversion processes $\chi^\ast \leftrightarrow \chi$ freeze out at $T_{f^\ast}$.  By applying entropy conservation, one can track the DS temperature evolution, properly accounting for partial reheating due to the conversion of excited states into ground states. As $n_{\chi^*}$ becomes suppressed with the temperature drop, the associated mass splitting $\delta$ is converted into kinetic energy, heating the remaining DM population. This effect is relevant for $T_\chi \ll \delta$ and ceases once conversions between states become irrelevant (or kinetic equilibrium is lost, as a common temperature no longer exists). Further details on the computation of the DS temperature are presented in \cref{app:temperatureDS}.

In the parameter space considered, we find a maximal possible reheating of approximately $60\%$, corresponding to
\begin{equation}
    T_\chi \lesssim 1.6 \, \frac{T_{\rm SM}^2}{T_{\rm kin}} \, ,
\end{equation}
where the scaling $T_\chi \propto T_{\rm SM}^2$ simply arises from the momentum redshift of non-relativistic single-state DM particles after kinetic decoupling. The reheating effect increases for larger mass splitting $\delta$, since more mass energy is converted into kinetic energy. However, this enhancement eventually saturates and diminishes once $\delta / m_\chi \gtrsim 0.1$, due to the strong Boltzmann suppression of the excited-state abundance, leaving too few $\chi^\ast$ particles to contribute appreciably to reheating. We have checked that such temperature corrections do not have significant effects on the computation of the excited-state fraction $f$. Thus, we do not further analyze the kinetic decoupling of the DS itself, simply reporting the results for the internal kinetic equilibrium scenario.  

\subsection{Fraction of excited states}
\label{sec:FractionStudy}

As previously mentioned, the relative abundance of excited states $f$, presented in \cref{eq:FratioDEF}, is needed to properly apply constraints on the model. Therefore, in the following we study the time evolution of the excited state $\chi^\ast$ after chemical decoupling, when conversion processes can still redistribute the $\chi$ and $\chi^\ast$ populations. The corresponding Boltzmann equation for $n_{\chi^\ast}$ reads
\begin{align}
	 \frac{{\rm d} n_{\chi^\ast} }{{\rm d} t} + 3 H\, n_{\chi^\ast} &=   - \langle \sigma_{\chi^\ast e^{\pm} \to \chi\, e^{\pm}} v\rangle n_{\chi^\ast}n_{e} + \langle \sigma_{\chi\, e^{\pm} \to \chi^\ast e^{\pm}} v\rangle n_{\chi} n_{e}  \nonumber\\
     &\quad -  \langle \sigma_{\chi^\ast \chi^\ast\to \chi\, \chi}  v\rangle n_{\chi^\ast}^2 + \langle \sigma_{\chi\, \chi\to \chi^\ast \chi^\ast}  v\rangle n_{\chi}^2 \nonumber \\
    &\quad - \langle \sigma_{\chi^\ast \chi \to \chi\, \chi}  v\rangle n_{\chi^\ast} n_{\chi} + \langle \sigma_{ \chi\, \chi \to \chi^\ast \chi}  v\rangle n_{\chi}^2 \nonumber\\
    &\quad - \langle \sigma_{ \chi^\ast \chi^\ast \to \chi^\ast \chi}  v\rangle n_{\chi^\ast}^2 + \langle \sigma_{\chi^\ast \chi \to \chi^\ast \chi^\ast}  v\rangle n_{\chi^\ast} n_{\chi} \,, \label{eq:be2}
\end{align}
where the right-hand side are the momentum-integrated collision operators, with $\langle \sigma_{i j \to k l} v\rangle$ the thermally-averaged cross sections of processes that reshuffle the populations of $\chi$ and $\chi^\ast$. The right-hand side of the first line accounts for inelastic scatterings with electrons and positrons, while the remaining lines correspond to conversions within the DS. It is also worth noting that the last two lines are absent in the case of purely inelastic interactions (i.e. parity-conserving), $\delta_y=0$. Representative Feynman diagrams for these processes are shown in \cref{fig:diagrams}. 

The Boltzmann equation in \cref{eq:be2} can be written in terms of the excited-state fraction $f$:
\begin{align}
\dot{f} = -&\Gamma_{\chi^\ast e^{\pm} \to \chi\, e^{\pm}}\, f + \Gamma_{\chi\, e^{\pm} \to \chi^\ast e^{\pm}}\, (1-f) \nonumber\\ 
- & \Gamma_{\chi^\ast \chi^\ast\to \chi\, \chi}\, f^2 + \Gamma_{\chi\, \chi\to \chi^\ast \chi^\ast}\, (1-f)^2 \nonumber\\
-&\Gamma_{\chi^\ast \chi \to \chi\, \chi}\, f(1-f) + \Gamma_{\chi\, \chi \to \chi^\ast \chi}\, (1-f)^2 \nonumber\\
-&\Gamma_{\chi^\ast \chi^\ast \to \chi^\ast \chi}\, f^2 + \Gamma_{\chi^\ast \chi \to \chi^\ast \chi^\ast}\, f(1-f) \,,
\label{f boltzmann eq}
\end{align}
with the effective rates reading:  
\begin{align}
\Gamma_{i\, e^{\pm} \to k \, e^\pm}\,&\equiv\, \langle \sigma_{i \,e^{\pm} \to k \,e^\pm} \,v \rangle \, n_{e}\,, \nonumber\\
\Gamma_{i \,k \to l \, m}\,&\equiv\, \langle \sigma_{i\,k \to l\,m} v \rangle \, n_{\chi,{\rm tot}}\,,
\label{eq:gamma2} 
\end{align} with $i,k,l,m=\chi, \chi^*$.
Here, $n_{\chi,{\rm tot}}(T_{\rm SM})$ is taken from \cref{eq:ntot,eq:sSM}. The density of $e^{\pm}$ entering the scattering rates is
\begin{equation}
    n_{e} \;=\; g_e\, \frac{ m_e^2\, T_{\rm SM}}{2\pi^2}\, K_2\!\left(\frac{m_e}{T_{\rm SM}}\right),\qquad g_e=4,
\label{ele_number_den}
\end{equation}
where $K_2$ is the modified Bessel function of the second kind.

\begin{figure}
    \centering
    \includegraphics[width=0.8\linewidth]{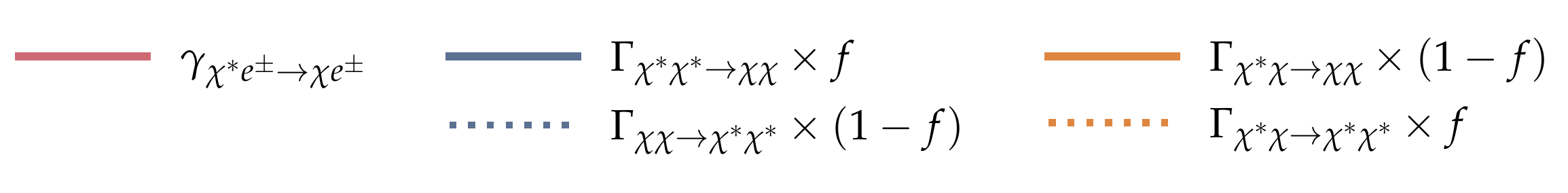}\vspace{0.2cm}
    \includegraphics[width=0.48\linewidth]{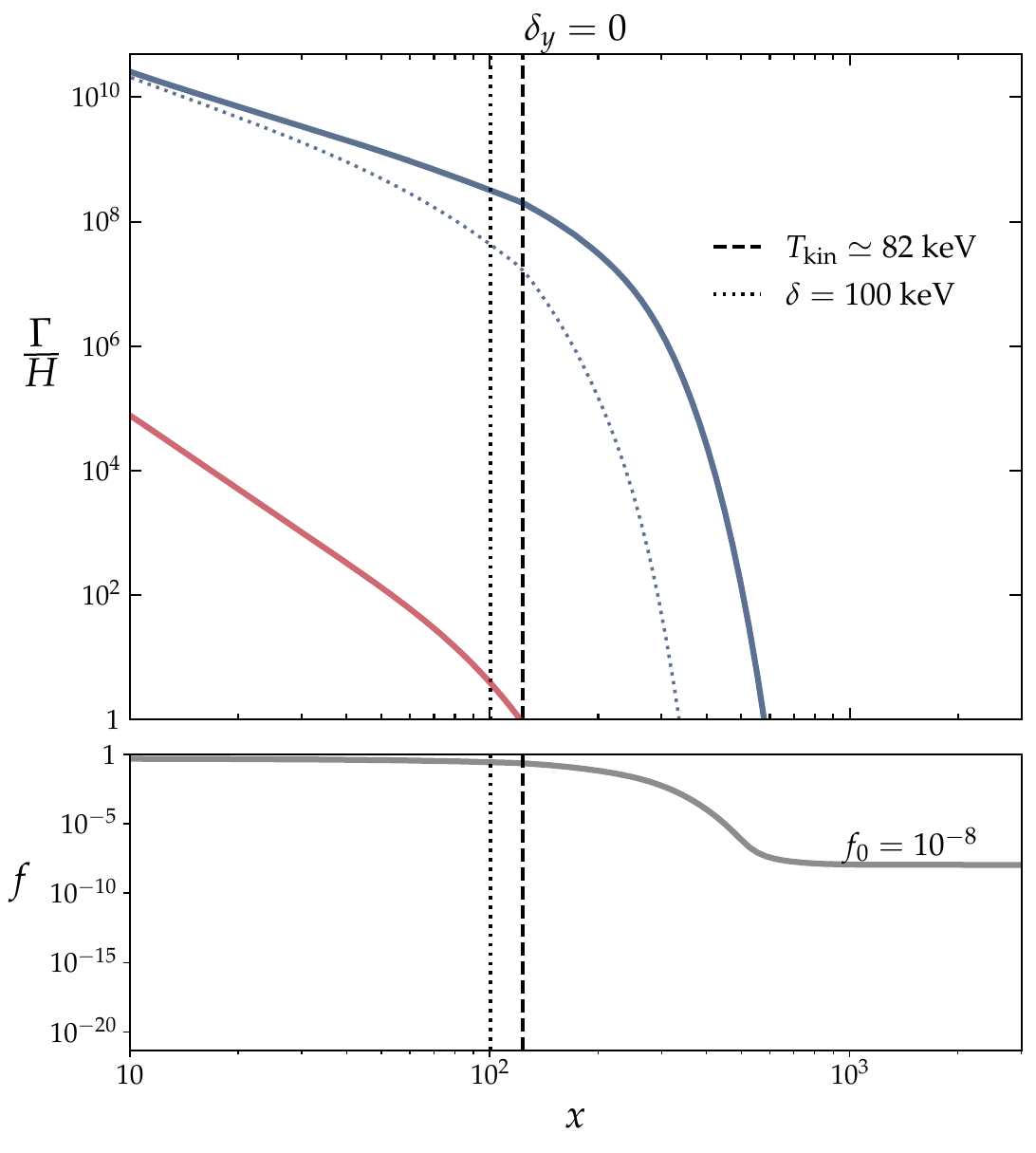}\hfill
    \includegraphics[width=0.48\linewidth]{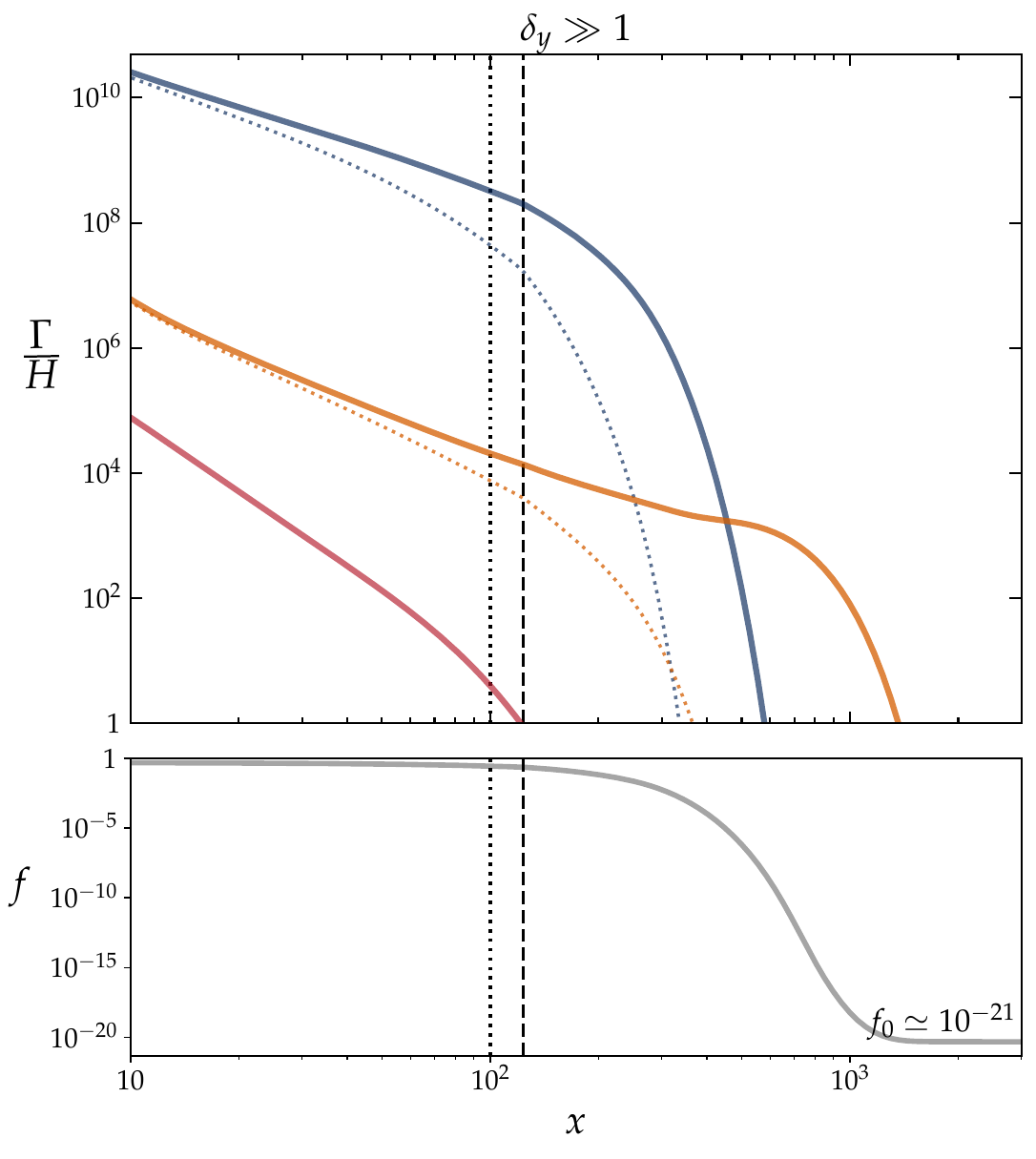}
    \caption{Evolution of the ratio between the relevant rates and the Hubble parameter, $\Gamma/H$ (upper panels), and of the excited-state fraction $f$ (lower panels) as a function of $x \equiv m_\chi/T_{\rm SM}$. Here, $\gamma_{\chi^\ast e^\pm \to \chi e^\pm}$ denotes the momentum exchange rate (see \cref{eq:transf_momenta}). The left panel corresponds to $\delta_y = 0$, where only off-diagonal (inelastic) interactions are present, while the right panel shows the maximal parity-violating case $\delta_y \gg 1$, in which diagonal processes are also active. The parameters are fixed as follows: $m_\chi = 10~\text{MeV}$, $\delta = 100~\text{keV}$, $\alpha' = 0.5$, $m_{A'}=3m_\chi$ and $\epsilon$ is chosen to reproduce $\Omega_{\rm obs} h^2$. Note that we do not include $\chi^\ast$ decay processes, as they are  suppressed.}
    \label{fig:rates_ds}
\end{figure}
 When parity is violated, $\delta_y \gg 1$, the additional diagonal interactions encoded in the last two lines of \cref{f boltzmann eq}, namely $\chi^\ast \chi \leftrightarrow \chi\chi$ and $\chi^\ast\chi^\ast \leftrightarrow \chi^\ast\chi$,  become efficient and modify the late-time evolution of $f$. This behaviour is illustrated in \cref{fig:rates_ds}, where we show the relevant conversion rates normalized to the Hubble parameter, $\Gamma/H$ (top panels), together with the evolution of the excited-state fraction $f$ (bottom panels) as a function of $x \equiv m_\chi/T_{\rm SM}$ for $\delta_y = 0$ (left plot) and $\delta_y \gg 1$ (right plot). In the parity-conserving case, $f$ effectively freezes out once the down scattering process $\chi^\ast\chi^\ast \to \chi\chi$ becomes slower than $H$. By contrast, for $\delta_y \gg 1$ the mixed channel $\chi^\ast\chi \to \chi\chi$, whose rate scales as $1-f$,   dominates the $\chi^\ast \leftrightarrow \chi$ conversions after $\chi^\ast\chi^\ast \to \chi\chi$ processes have frozen out, driving $f$ down to much smaller values.

For the lifetimes of interest there is a clear hierarchy between the epoch when $\chi^\ast\leftrightarrow\chi$ conversions freeze out ($t_{f^\ast}$) and the onset of $\chi^\ast$ decays ($\tau_\ast$), since $t_{f^\ast}\lesssim{\rm minutes}$ while $\tau_\ast\gg{\rm days}$. We therefore determine the freeze-out value,
\begin{equation}
    f_0 \equiv f|_{t=t_{f^\ast}}
\end{equation}
by solving \cref{f boltzmann eq} neglecting decays, and then obtain the fraction at later times taking into account the decays,
\begin{equation}\label{eq:decaysInfluenceInFranction}
    f(t) \;=\; f_0\, e^{-t/\tau_\ast}\,.
\end{equation}

As shown in \cref{fig:relicAbundance}, for $\delta/m_\chi \ll 1$, the diagonal (elastic) interactions present when $\delta_y \neq 0$ have a negligible impact on the DM–SM chemical freeze-out, since $\alpha'{\rm el} \ll \alpha'{\rm inel}$.
At lower temperatures, however, these same diagonal interactions remain relevant for depleting the population of excited states: despite their suppression, they can keep $\chi^\ast \leftrightarrow \chi$ conversions in equilibrium down to $T_\chi \lesssim \delta$, where the excited states become strongly Boltzmann suppressed.
Once $f$ becomes small, processes involving the ground state, such as $\chi^\ast \chi \to \chi \chi$, are effectively enhanced simply because $n_\chi \gg n_{\chi^\ast}$. 

\begin{figure}[p] 
    \centering
    \includegraphics[width=0.8\linewidth]{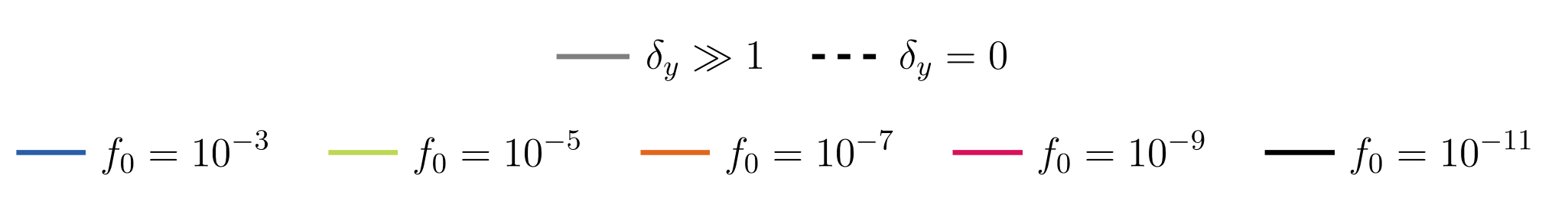}\vspace{0.2cm}
    \includegraphics[width=0.45\linewidth]{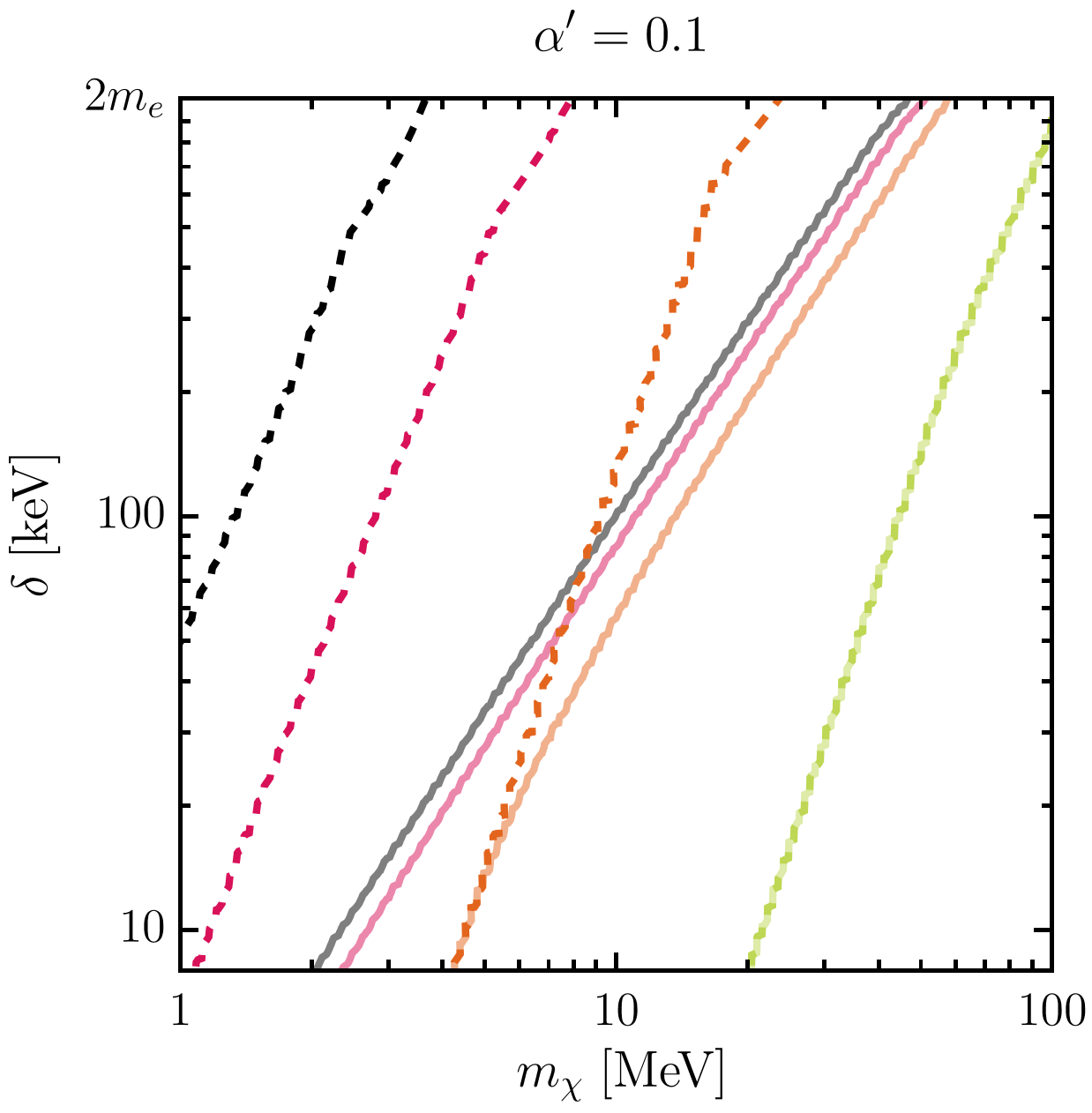}\vspace{0.25cm}\hspace{0.95cm}
    \includegraphics[width=0.45\linewidth]{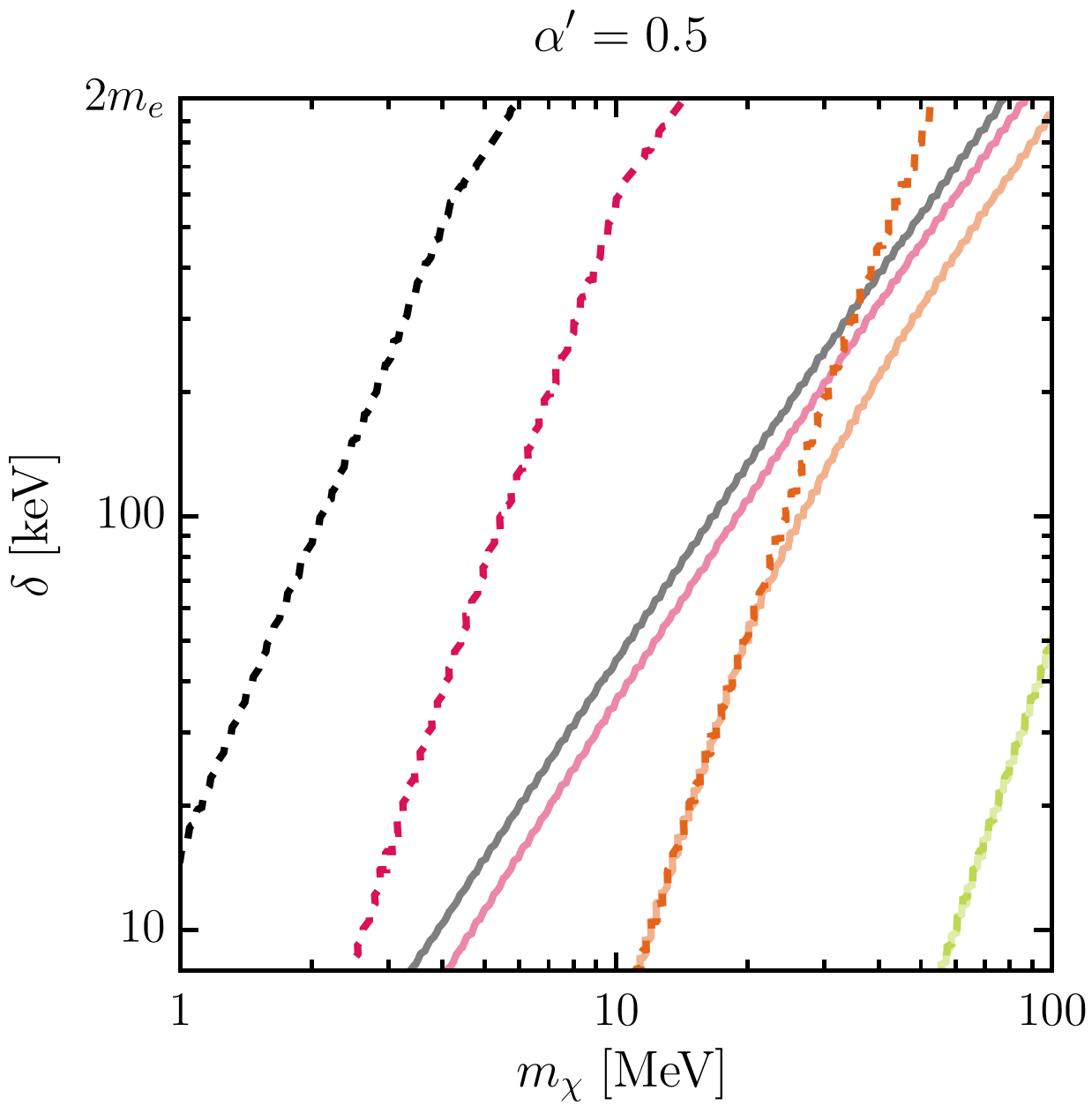}
    \includegraphics[width=0.465\linewidth]{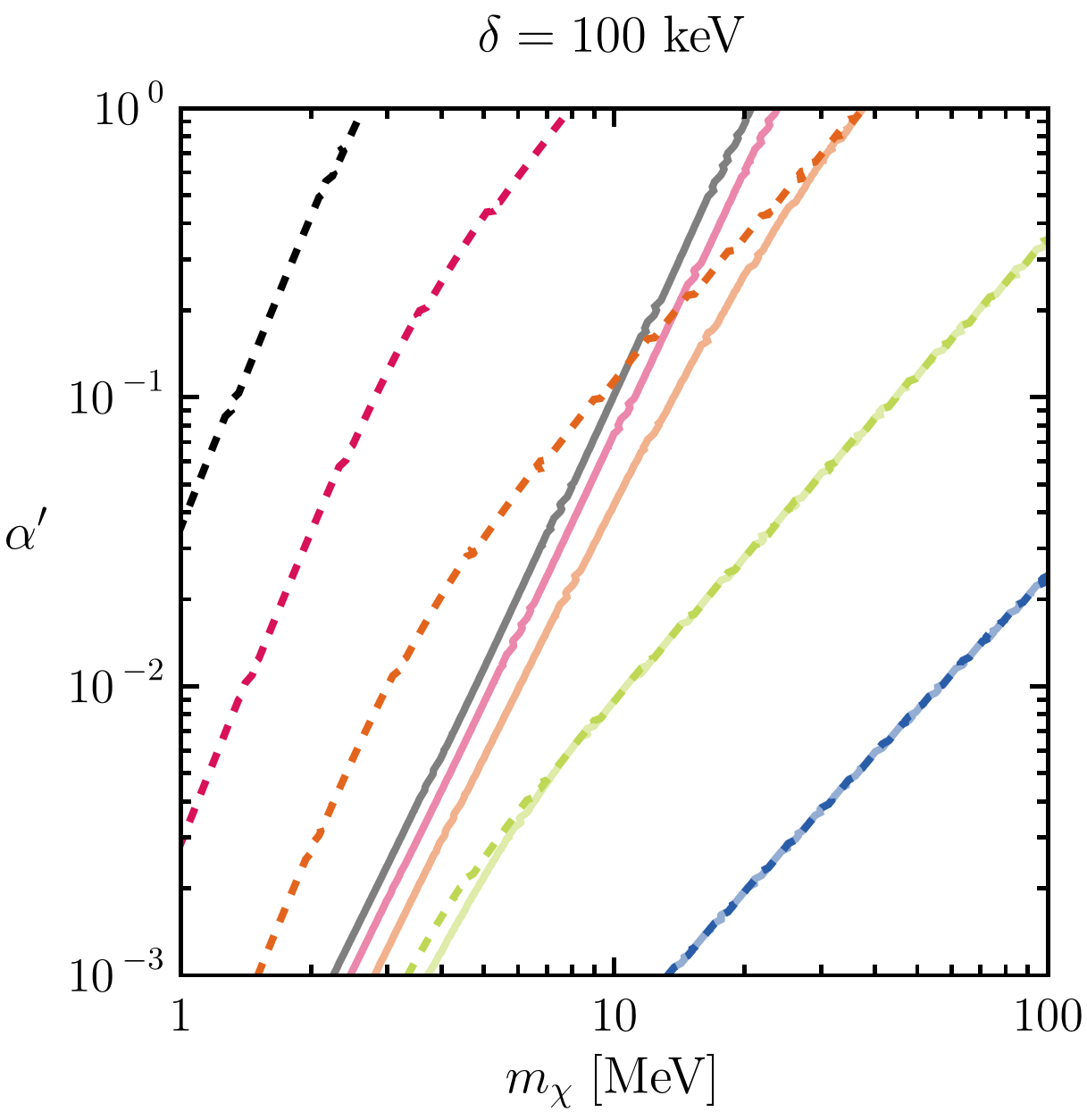}\hspace{0.7cm}
    \includegraphics[width=0.465\linewidth]{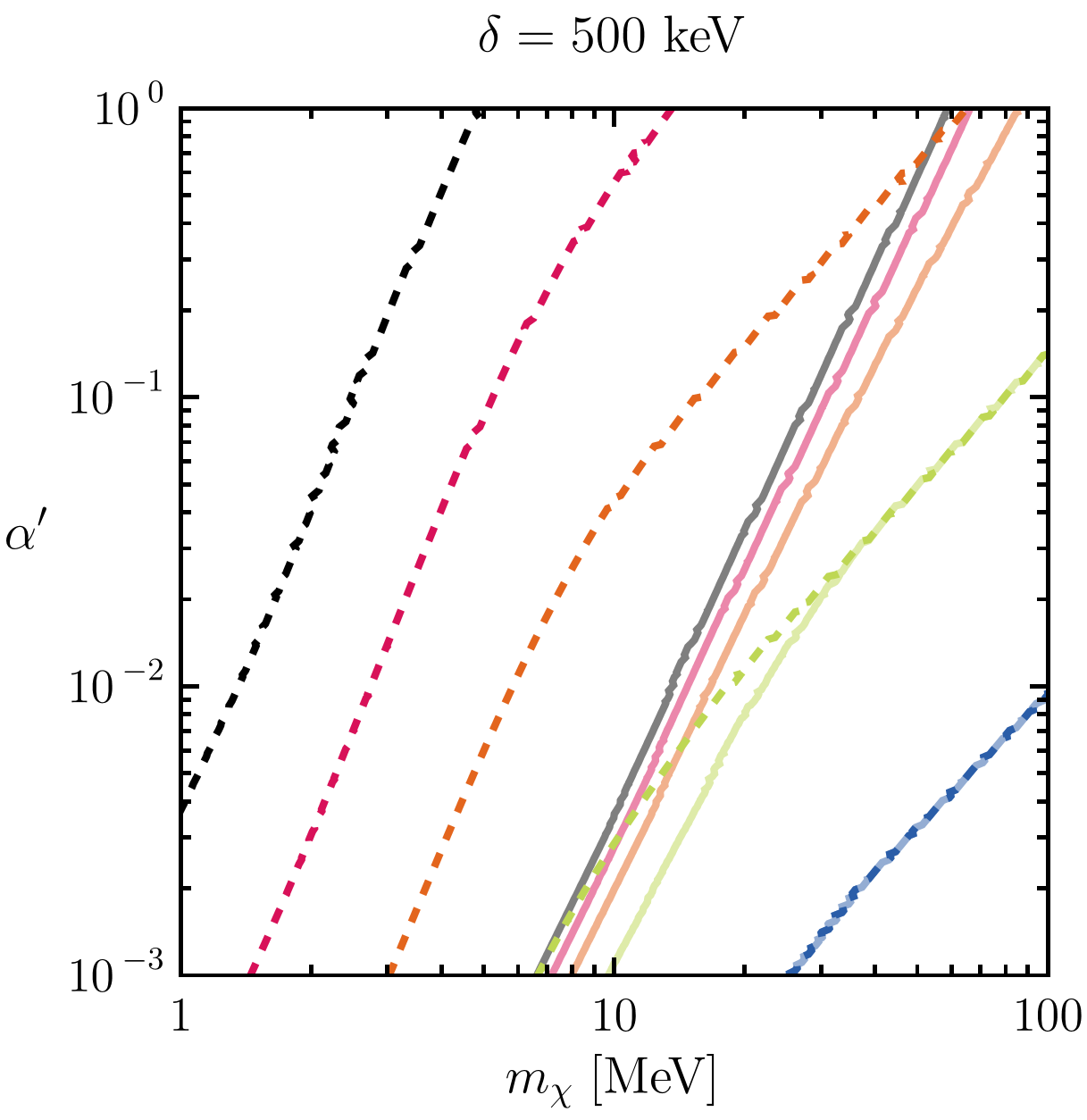}
    \caption{Contours  of the fraction of excited states at the freeze-out of $\chi^\ast\leftrightarrow\chi$ conversion processes ($f_0$) in the plane of mass splitting $\delta$  and DM mass $m_\chi$ (top) as well as in the plane of the dark fine-structure constant $\alpha'$ and DM mass $m_\chi$ (bottom). The solid translucent (dashed  opaque) lines correspond to parity-violating (-conserving) interactions.  The top left (top right) panel corresponds to $\alpha'=0.1$ ($\alpha'=0.5$), while the bottom left (bottom right) to $\delta=100$~keV ($\delta=500$~keV). At each point, the value of the kinetic mixing $\epsilon$ is such that the DM relic abundance is reproduced. Additionally, we take $m_{A'}=3m_\chi$.  } \label{fig:deltamass}
\end{figure}

In \cref{fig:deltamass} we shown the surviving excited-state fraction, $f_0$, in the plane of DM mass and mass splitting (dark fine structure constant) in the top (bottom) panels. We see that  $f_0 \ll 1$ throughout the parameter space considered, where the kinetic mixing parameter $\epsilon$ is fixed by the observed relic abundance. Moreover, in the parity-preserving limit (dashed curves), $f_0$ can remain large compared to the case where parity is broken (solid curves), since $\chi^\ast \leftrightarrow \chi$ conversions decouple earlier. Indeed, in the latter case, the additional diagonal interactions keep the two states in equilibrium longer, driving $f_0$ to significantly smaller values---effectively confirming the expectations of Ref.~\cite{CarrilloGonzalez:2021lxm}. Consequently, $f_0$ becomes negligible within the region where diagonal interactions are efficient.

\section{Experimental constraints} \label{sec:pheno}

In this section, we summarize the experimental constraints relevant for the pseudo-Dirac iDM model with a long-lived excited state. Readers interested only in the allowed parameter-space results may skip this section and go directly to \cref{sec:numres}. 

\subsection{Indirect detection}

\paragraph{DM annihilations:}

The annihilation of ground DM states is strongly suppressed due to both the small elastic coupling $\alpha'_{\rm el}$ and the $p$-wave nature of these processes in the non-relativistic limit~\cite{Duerr:2016tmh}. As a result, limits from annihilations do not lead to relevant constraints.\footnote{
Loop-induced annihilations $\chi\chi\to \bar f f$ are further suppressed either by velocity or helicity factors in addition to the loop suppression~\cite{Fitzpatrick:2021cij}.}

Coannihilations, by contrast, proceed via $s$-wave and are therefore potentially constrained. These bounds include searches for annihilations in celestial bodies~\cite{DelaTorreLuque:2023olp}, CMB limits on late-time energy injection~\cite{Slatyer:2015jla,Planck:2018vyg}, and other indirect searches~\cite{Cirelli:2024ssz}. The main modification in the iDM case is the reduced abundance of the excited state, studied in \cref{sec:FractionStudy}.

In celestial bodies the $J$-factor can be modified by the inelasticity of the DM~\cite{Berlin:2023qco}; however, in regions without significant excited-state repopulation—consistent with our assumptions—using the standard $J$-factor is expected to remain a good approximation. For this reason, the recasting of ID bounds from searches of DM annihilation products behaves similarly to that of the CMB limits. Namely, we follow Ref.~\cite{Berlin:2023qco} and map standard annihilation constraints to the coannihilation case via
\begin{equation}
   \langle \sigma v \rangle_{\rm ann} \rightarrow 2f\, \langle \sigma v\rangle_{\rm coa}\,,
\end{equation}
where the factor of 2 accounts for the distinct initial states and $\langle \sigma v\rangle_{\rm coa}$ is given in \cref{eq:freezeoutapprox} (generically with the replacement $\alpha'\to\alpha'_{\rm inel}$).  

Planck CMB data impose the bound~\cite{Berlin:2023qco,Planck:2018vyg}
\begin{equation}
    2f\,\langle\sigma v\rangle_{\rm coa} 
    \lesssim 2\times 10^{-26} \text{ cm}^3\text{/s}\,
    \left(\frac{m_\chi}{30~\mathrm{GeV}}\right),
\end{equation}
with $f$ evaluated at $z=600$, the epoch to which CMB energy-injection constraints are most sensitive~\cite{Finkbeiner:2011dx}. For the 1–100~MeV mass range, these are the dominant annihilation limits~\cite{Cirelli:2024ssz}.  Thus, we do not include the sub-dominant constraints from searches of DM annihilation products.

Off-diagonal interactions could in principle repopulate excited states via up-scattering. However, such repopulation is only relevant for splittings $\delta \lesssim 100$~eV~\cite{Berlin:2023qco}, well below our region of interest. Thus, throughout this work, $f$ is simply taken as the cosmological abundance $f_0$ modified only by decays---see \cref{eq:decaysInfluenceInFranction}.

\paragraph{Excited-state decays:}

Decays of the excited states inject energy into the SM at a rate~\cite{CarrilloGonzalez:2021lxm}
\begin{equation}
    R_{\rm dec} \equiv 
    \left(\frac{dE}{dt\, dV}\right)_{\rm iDM}= 
    F_{\rm 4-body}\, f\, 
    \frac{\delta}{m_\chi}\, 
    R_{\rm dec}^{\rm (s)}\,,
\end{equation}
where $R_{\rm dec}^{\rm (s)}$ is the standard decaying-DM injection rate~\cite{Mitridate:2018iag} and $F_{\rm 4-body}$ accounts for the visible fraction of the decay energy. This injection is highly suppressed by both $f\ll 1$ and $\delta/m_\chi \ll 1$. Decay constraints thus become relevant only for splittings $\delta\sim {\rm MeV}$, where the lifetime becomes sufficiently short. Since $f$ depends on  $\tau_\ast\propto \delta^{-13}$, these bounds rapidly weaken for smaller $\delta$.

The strongest decaying-DM limits usually come from diffuse X-ray measurements~\cite{Essig:2013goa,Linden:2024fby}, which require $\tau_\ast \gtrsim 10^{26}$--$10^{28}$~s. In the iDM setup the four-body decay softens the photon spectrum and reduces the visible energy per decay, weakening X-ray bounds. In contrast, constraints depending only on the total injected energy remain robust: CMB anisotropies~\cite{Slatyer:2016qyl}, CMB spectral distortions~\cite{Acharya:2019owx,Bolliet:2020ofj}, and dwarf-galaxy heating~\cite{Wadekar:2021qae}.

Robust bounds from energy injection, such as CMB limits (or BBN constraints for $\tau_\ast\lesssim 10^7$~\cite{Balazs:2022tjl}), can be found already for decaying particles with varying abundances. We  apply them by mapping standard limits on a decaying species $a$ with relative abundance $\zeta_a\equiv \Omega_a/\Omega_{\rm obs}$ to our scenario via
\begin{align}
    \zeta_a \to f_0 \,,\qquad 
    \tau_a \to \tau_\ast \,,\qquad
    m_a \to F_{\rm 4-body}\,\delta\,,
\end{align}
taking $F_{\rm 4-body}=3/4$ for simplicity. COBE-FIRAS constraints dominate for $\tau_\ast\lesssim 10^{12}$~s~\cite{Fixsen:1996nj,Acharya:2019owx}, while Planck limits dominate for longer lifetimes~\cite{Planck:2015fie,Slatyer:2016qyl,Balazs:2022tjl}. However, the latter constrains much smaller abundances reaching $\zeta_a \sim10^{-11}$ and, in contrast, the former only reaches down to $\zeta_a \sim10^{-4}$. Since these bounds depend only weakly on the injected energy in the keV–MeV range, the translation is straightforward.

For very long lifetimes $\tau_\ast\gg t_{\rm U}$, CMB limits become weaker, so we instead apply recasted limits from diffuse X-rays, neglecting detailed four-body kinematics. In this case, bounds on decaying-DM often assume $\tau_\ast\gg t_{\rm U}$, so that the DM density is unchanged over cosmic time. In our case, excited states may decay sufficiently fast such that this assumption fails. We therefore replace $f_0\to f(t_U)$ when applying such bounds (see \cref{eq:decaysInfluenceInFranction}). For $\tau_\ast\lesssim t_{\rm U}$ this becomes essential. Moreover, since these limits assume $f=1$ (a good approximation for $\tau_\ast\gtrsim t_{\rm U}$ in their analysis), we rescale them as
\begin{equation}
    \tau_\ast \gtrsim f(t_U)\, 
    \tau_{\rm limits}(F_{\rm 4-body}\,\delta)\,.
\end{equation}
This type of constraint is dominated by data from the INTEGRAL observatory~\cite{Bouchet:2008rp} across our parameter space~\cite{Linden:2024fby}.

\paragraph{Cooling/heating of astrophysical objects:}

Stellar bounds are irrelevant for the mass range considered here, $m_{\chi} \geq \mathcal{O}(\text{MeV})$~\cite{Li:2023vpv}. Similarly, bounds from supernova explosions are also irrelevant, for instance, the SN1987A observations exclude kinetic-mixing values in the range $\epsilon \sim 10^{-6}$--$10^{-10}$ for $m_{\chi,A'} \lesssim 100~\text{MeV}$~\cite{Chang:2018rso}, while we focus on $\epsilon \gtrsim 10^{-6}$. 

The high energies present in active galactic nuclei allow these environments to probe mass splittings up to the TeV scale via searches for anomalous cooling of cosmic rays emitted from these sources~\cite{Gustafson:2024aom}. However, such constraints depend sensitively on the assumed DM density profiles and carry substantial astrophysical uncertainties. Overall, they remain weaker than those derived from collider missing-energy searches, which we discuss below. For this reason, we do not include these bounds in our final analysis.

\paragraph{Big Bang nucleosynthesis:}New light particles that remain in thermal equilibrium with the SM bath during BBN can modify the predicted primordial light-element abundances. Ref.~\cite{Sabti:2019mhn} performed a detailed analysis of these effects and found that masses $m_\chi \lesssim 7$~MeV are excluded---where we adopt conservative results based solely on current measurements of the primordial helium and deuterium abundances, together with CMB determinations of $\Omega_b h^2$. The reported constraints are the results for a Dirac fermion, since both states of our pseudo-Dirac pair are expected to be present at BBN temperatures, $T_{\rm BBN}\sim{\rm MeV}$.\footnote{At CMB temperatures, $T_{\rm CMB}\sim{\rm eV}$, only the lighter Majorana state remains thermally populated. Including combined BBN+Planck limits for a Majorana particle would slightly strengthen the bound from $m_\chi > 7$~MeV to $m_\chi > 8$~MeV (or $m_\chi > 10.9$~MeV if DM is fully Dirac). As these differences are minor within the mass range of interest, we do not pursue a more detailed treatment of this constraint for the pseudo-Dirac case.}

\subsection{Direct detection}

\paragraph{Elastic scatterings:}

Due to parity violation the diagonal coupling $\alpha'_{\rm el}$ induces elastic scatterings, but these are highly suppressed both by the smallness of $\alpha'_{\rm el}$ and the velocity-suppression of the vector–Majorana interaction. In the sub-GeV mass range such signals are undetectable.

Loop-induced elastic scatterings can avoid the $\alpha'_{\rm el}$ suppression and generate spin-independent (SI) interactions~\cite{Bell:2018zra}. Applied to sub-GeV iDM, these remain weaker than generic LEP bounds, $\epsilon<3\times 10^{-2}$~\cite{Hook:2010tw}, and thus are irrelevant here---see also Ref.~\cite{Duerr:2019dmv}.

\paragraph{Up-scatterings:}

Off-diagonal interactions could induce $\chi\to\chi^\ast$ up-scattering in the detector, the Earth, or upstream environments.  
Within Earth, up-scattering is kinematically suppressed when
\begin{equation}
    \frac{\delta}{\mu_r} > \frac{v_{\rm esc}^2}{2} \approx 3\times 10^{-6},
\end{equation}
with $\mu_r$ the reduced mass of the scattering system~\cite{Emken:2021vmf}.  
Accounting for possible high-velocity components from the Large Magellanic Cloud or the Local Group does not significantly change this threshold~\cite{Smith-Orlik:2023kyl,Herrera:2023fpq}. 
Thus, we find we always work in a region where up-scatterings on Earth are negligible.  

Up-scatterings in the Sun can be relevant for $\delta\lesssim 10$~keV and $m_\chi\lesssim 10$~MeV due to hot electrons in the solar core~\cite{Baryakhtar:2020rwy}. Since we focus on $\delta\gtrsim 10$~keV, we do not include these signals.

Cosmic rays can boost DM to relativistic energies~\cite{Bell:2021xff}, yielding constraints comparable to those of elastic scatterings. These are weaker than missing-energy collider bounds  for $m_\chi\lesssim 100$~MeV~\cite{Guha:2024mjr}, so we omit them.  
Other astrophysical acceleration mechanisms (blazars~\cite{Bhowmick:2022zkj,CDEX:2024qzq,Jeesun:2025gzt}, AGN neutrino emitters~\cite{Mishra:2025juk,Gustafson:2025dff}) are highly uncertain and yield limits comparable to current collider reaches, so we do not include them either.

\paragraph{Down-scatterings (exothermic):}

Down-scatterings of the primordial excited-state population are the most distinctive DD signature. The mass splitting is converted into recoil energy, giving a mono-energetic line at \begin{equation}
    \bar{E}_{R,i} = \frac{\mu_{i}}{m_i}\,\delta
\end{equation} with $i \in \{e,N\}$. For sub-GeV DM, electron recoils dominate thanks to the higher values of $\bar{E}_{R,e}>\bar{E}_{R,N}$.

We adopt the rate of Ref.~\cite{Baryakhtar:2020rwy}, replacing $\alpha'\to \alpha'_{\rm inel}$ (numerically negligible here) and neglecting Earth-shielding effects on the DM flux~\cite{CarrilloGonzalez:2021lxm}.\footnote{
Ref.~\cite{CarrilloGonzalez:2021lxm} finds that shielding is negligible for $m_\chi\gtrsim 1$~GeV and never exceeds $\sim 50\%$ even for $m_\chi\sim\mathrm{MeV}$.}  
The resulting event rate is  
\begin{equation}
    R_{\rm exo} \approx 8\times 10^5\,\text{(t\,y)}^{-1}\,f \,\alpha'_{\rm inel}
    \left( \frac{\delta}{\mathrm{keV}} \right)^{1/2}
    \left( \frac{\epsilon}{10^{-5}} \right)^{2}
    \left( \frac{m_\chi}{m_{A'}} \right)^{4}
    \left( \frac{100~\mathrm{MeV}}{m_\chi} \right)^{5}\,, 
\end{equation} where we assume that local DM energy densities scale as the cosmological ones. XENON1T~\cite{XENON:2020rca} and XENONnT~\cite{XENON:2022ltv}  provide leading limits on line-like electron recoils~\cite{PandaX:2024cic}.  
We recast their dark photon DM absorption bounds using~\cite{An:2014twa}
\begin{equation}
    r_{A'} \simeq 
    \frac{\rho_{\rm DM}}{m_{A'}}\, \epsilon^{2}\,
    \sigma_{\gamma}(\omega=m_{A'})\,,
\end{equation}
where $\rho_{\rm DM}$ is the local DM energy density and   $\sigma_\gamma(\omega)$ is  the  photoelectric cross section of xenon evaluated at energy $\omega$; taken from Ref.~\cite{NIST_XCOM_2010}.  
Converting $r_{A'}$ into detector rates, $R_{A'}$ (measured in events per tonne per year), down-scattering is excluded whenever
\begin{equation}
    R_{\rm exo} < R_{A'}(\epsilon_{\rm exp}) \,,
\end{equation} with $\epsilon_{\rm exp}$ denoting the experimental upper limit on the kinetic mixing.

\subsection{Self-interactions}

Cluster mergers, in particular the Bullet Cluster, constrain~\cite{Robertson:2016xjh,Wittman:2017gxn}
\begin{equation}\label{eq:SIDM_bound}
    \sigma/m \lesssim 2~\mathrm{cm}^2/\mathrm{g}\,,
\end{equation}
with earlier results quoting $0.7~\mathrm{cm}^2/\mathrm{g}$~\cite{Randall:2008ppe}. Both are shown below in our plots. 

Elastic scatterings induced by diagonal interactions are negligible for $\delta\ll m_\chi$ since $\alpha'_{\rm el} \propto \delta^2/m_\chi^2$, see \cref{eq:el&inelFINEstructure}. Thus, their contribution remains sub-dominant to scatterings induced by the off-diagonal coupling, namely:
\smallskip

\emph{(i) Loop-induced elastic scatterings of identical states.}
Ref.~\cite{Fitzpatrick:2021cij} computed the one-loop amplitude at zero momentum transfer (neglecting the mass splitting for internal $\chi^\ast$-lines since $\delta\ll m_\chi$). For $m_{A'}=3m_\chi$, the authors find for $\chi \chi$ scatterings
\begin{equation}\label{eq:SIDM_cx}
    \sigma_{\rm SIDM,1}=
\frac{\overline{|\mathcal{M}|^2}}{128\pi m_\chi^2}
    \simeq
    0.06\frac{\alpha_{\rm inel}'^4}{m_\chi^2}\,,
\end{equation}
where we used $\overline{|\mathcal{M}|^2}\simeq 24\,\alpha_{\rm inel}'^4$ taken from appendix C of Ref.~\cite{Fitzpatrick:2021cij}.\footnote{
Earlier loop estimates can be obtained by translating the DM–quark calculation of~\cite{Bell:2018zra} via $Q_q^2\epsilon^2\alpha\to \alpha'$ in the operator $\bar\chi\chi\,\bar q q\to \bar\chi\chi\,\bar\chi\chi$.  
This yields cross section scaling as $\alpha_{\rm inel}'^4 (m_\chi/m_{A'})^8 F_3^2$, with the loop function $F_3$ given in Ref.~\cite{Bell:2018zra}.  
These estimates are smaller and align with earlier results such as Ref.~\cite{Baryakhtar:2020rwy}.} Note that $\chi^*\chi^*$ scatterings are negligible since $f\ll 1$.

\smallskip
\emph{(ii) Elastic scatterings of different states.}
The off-diagonal coupling also induces  $\chi\chi^\ast\to\chi\chi^\ast$, giving
\begin{equation}
    \sigma_{\rm SIDM,2}\sim 
    f\,\frac{\alpha_{\rm inel}'^{2} m_\chi^2}{m_{A'}^4}\,,
\end{equation}
which, however, is negligible since $f\ll 1$. Therefore, we simply apply the self-interaction constraint, \cref{eq:SIDM_bound}, via $\sigma_{\rm SIDM,1}$ given in \cref{eq:SIDM_cx}.

\subsection{Collider searches}

The excited state is cosmologically long-lived and, therefore, invisible in detectors, while the dark photon $A'$ decays promptly. Since visible decays are suppressed by the hierarchy $\alpha'\gg \epsilon^2\alpha$, collider constraints are generally dominated by missing-energy searches.

For $1~\mathrm{MeV}\lesssim m_{A'}\lesssim 300~\mathrm{MeV}$, the strongest limits come from the electron beam dump experiment NA64~\cite{NA64:2023wbi,NA64:2025ddk}. As both DM states are stable on detector scales, the published bounds apply directly. Similarly, via missing-momentum searches, the future LDMX experiment~\cite{LDMX:2018cma} will improve the sensitivity of NA64 by several orders of magnitude; we include its projections in our analyses.

Dark photons produced at high-intensity facilities can also yield boosted DM that scatters in downstream detectors. Experiments such as LSND~\cite{LSND:2001akn,deNiverville:2011it}, E137~\cite{Bjorken:1988as,Batell:2014mga}, and MiniBooNE~\cite{MiniBooNE:2017nqe} probe such scattering signals.  
At the relevant energies, inelastic kinematic suppressions are negligible, and we follow Ref.~\cite{Berlin:2018pwi} in rescaling their bounds by the appropriate choice of $\alpha'$. 

\section{Numerical Results} \label{sec:numres}

\begin{figure}[p] 
    \centering \includegraphics[width=0.775\linewidth]{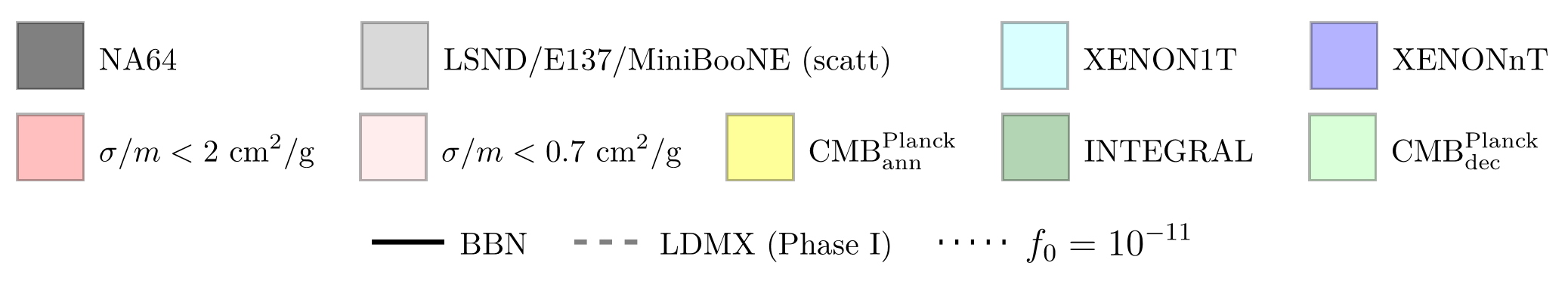}\vspace{0.1cm}\\
    \includegraphics[width=0.45\linewidth]{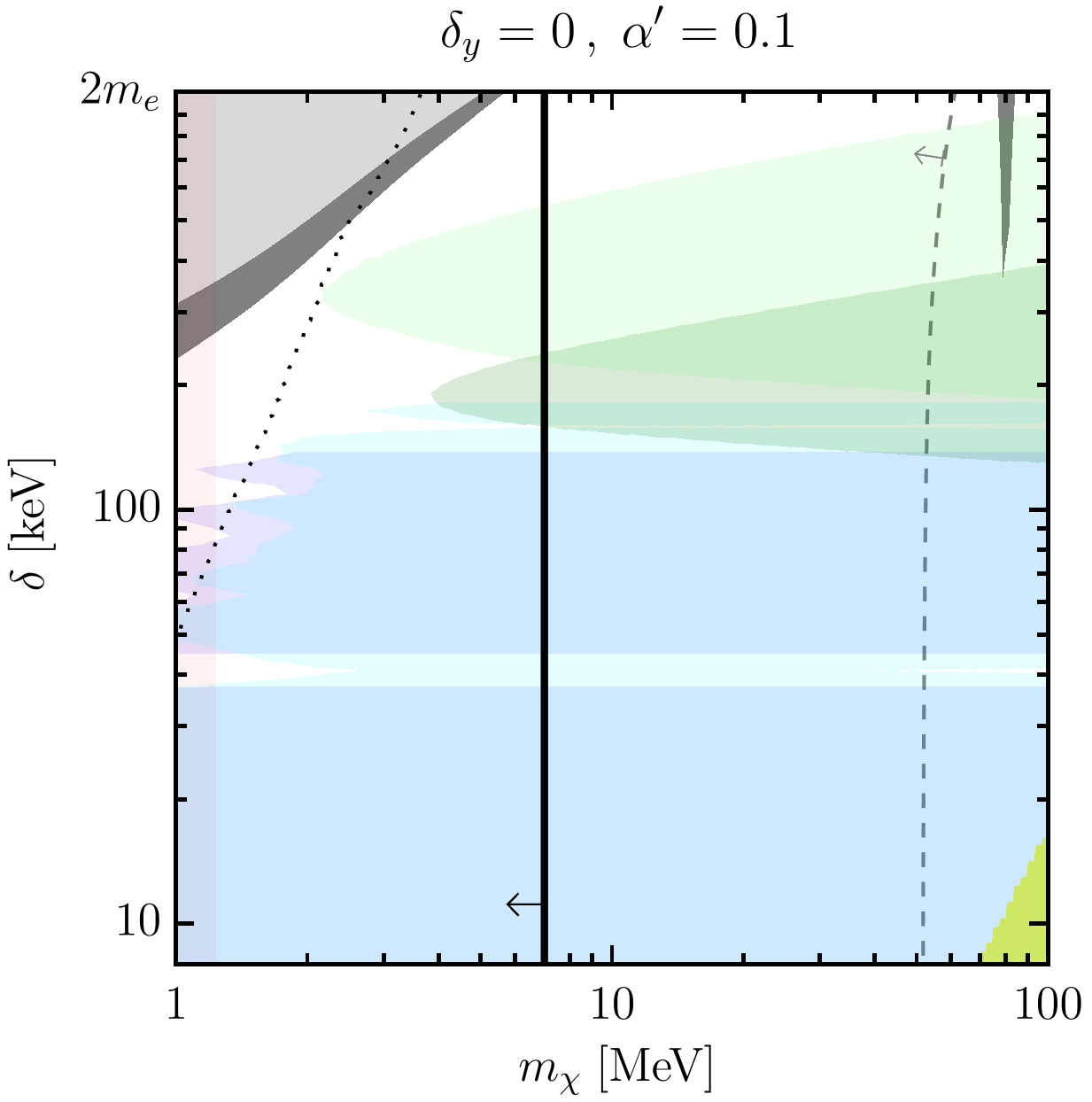}\vspace{0.25cm}\hspace{0.95cm}
    \includegraphics[width=0.45\linewidth]{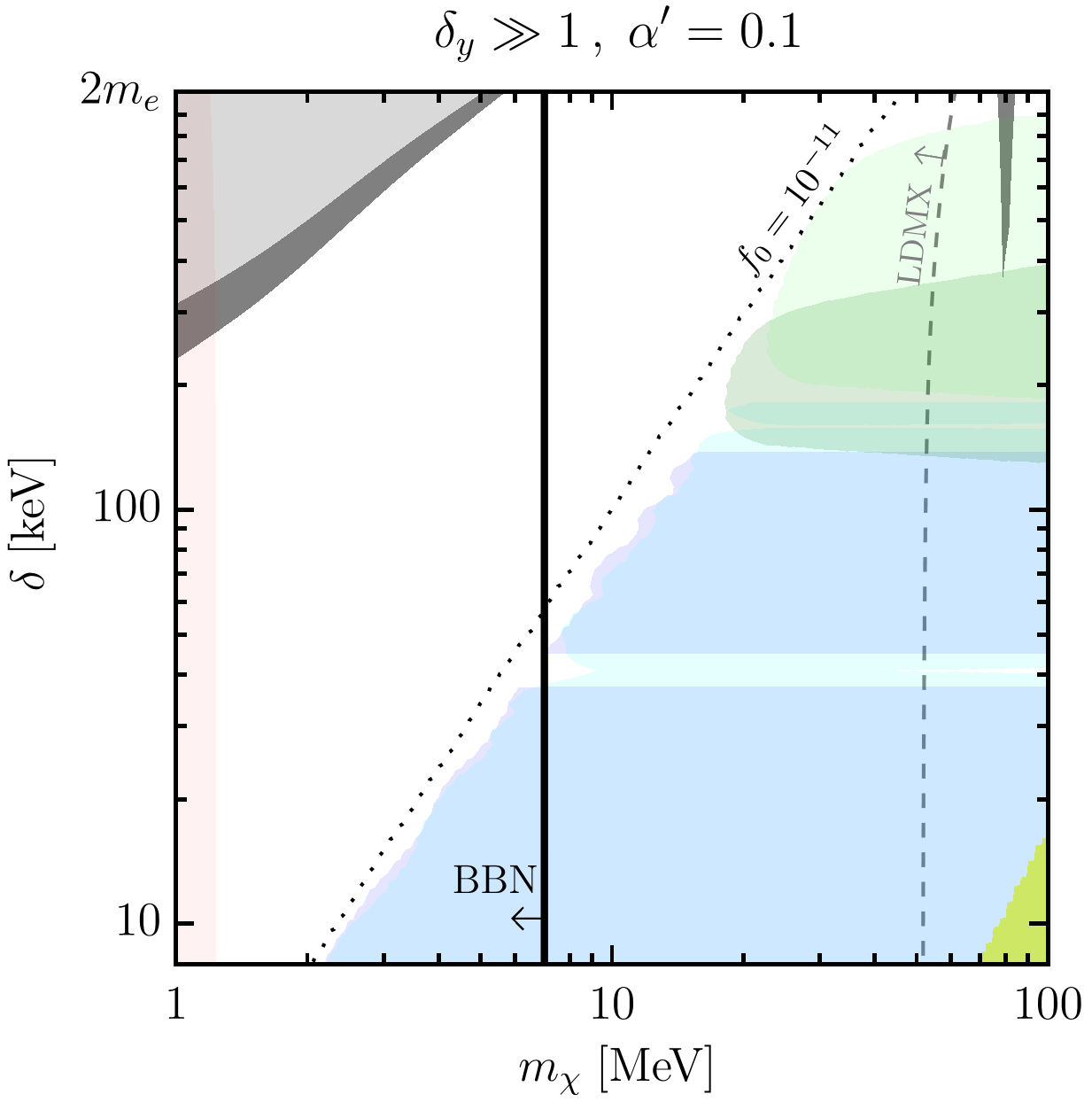}
    \includegraphics[width=0.45\linewidth]{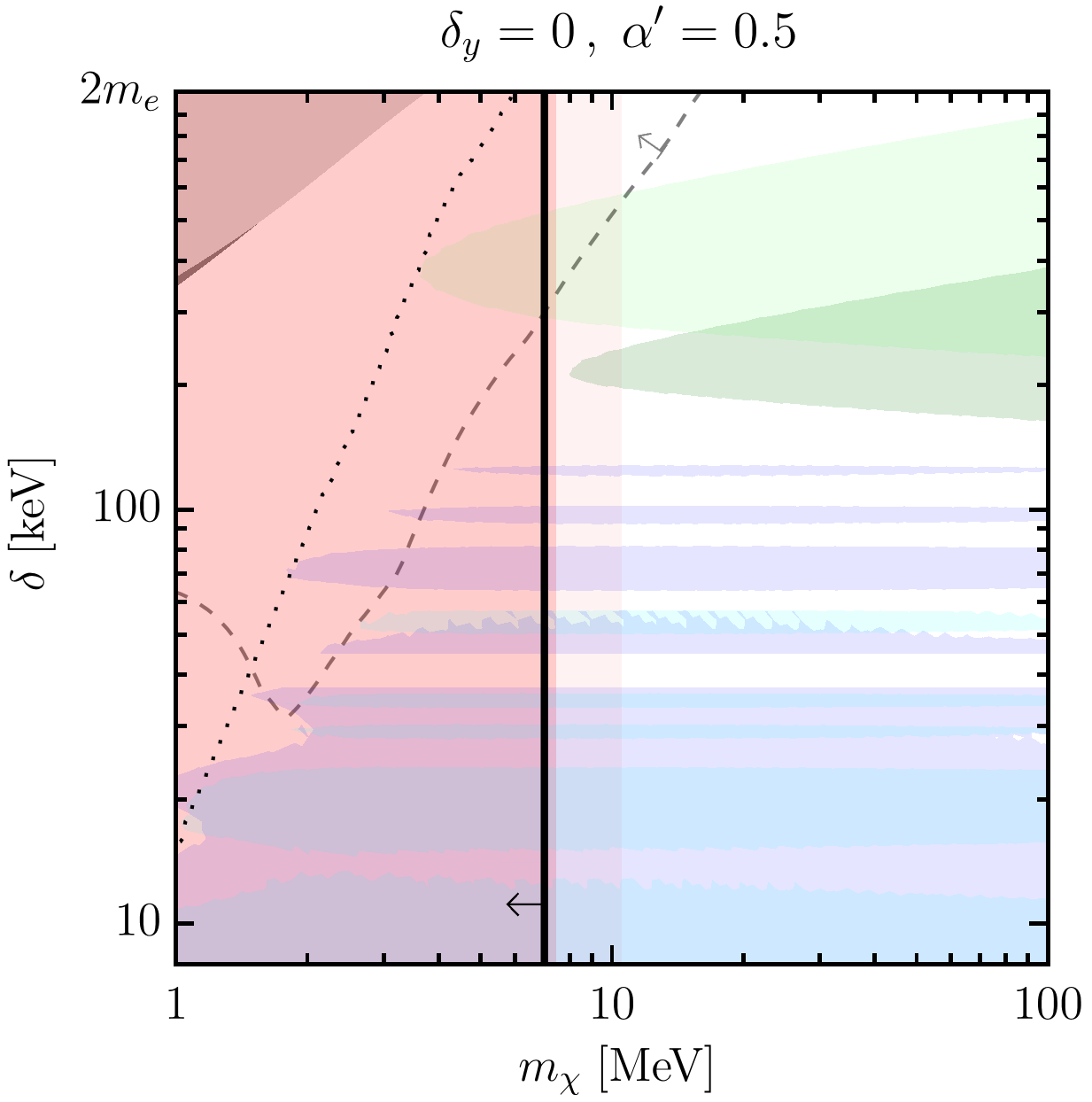}\hspace{0.95cm}
    \includegraphics[width=0.45\linewidth]{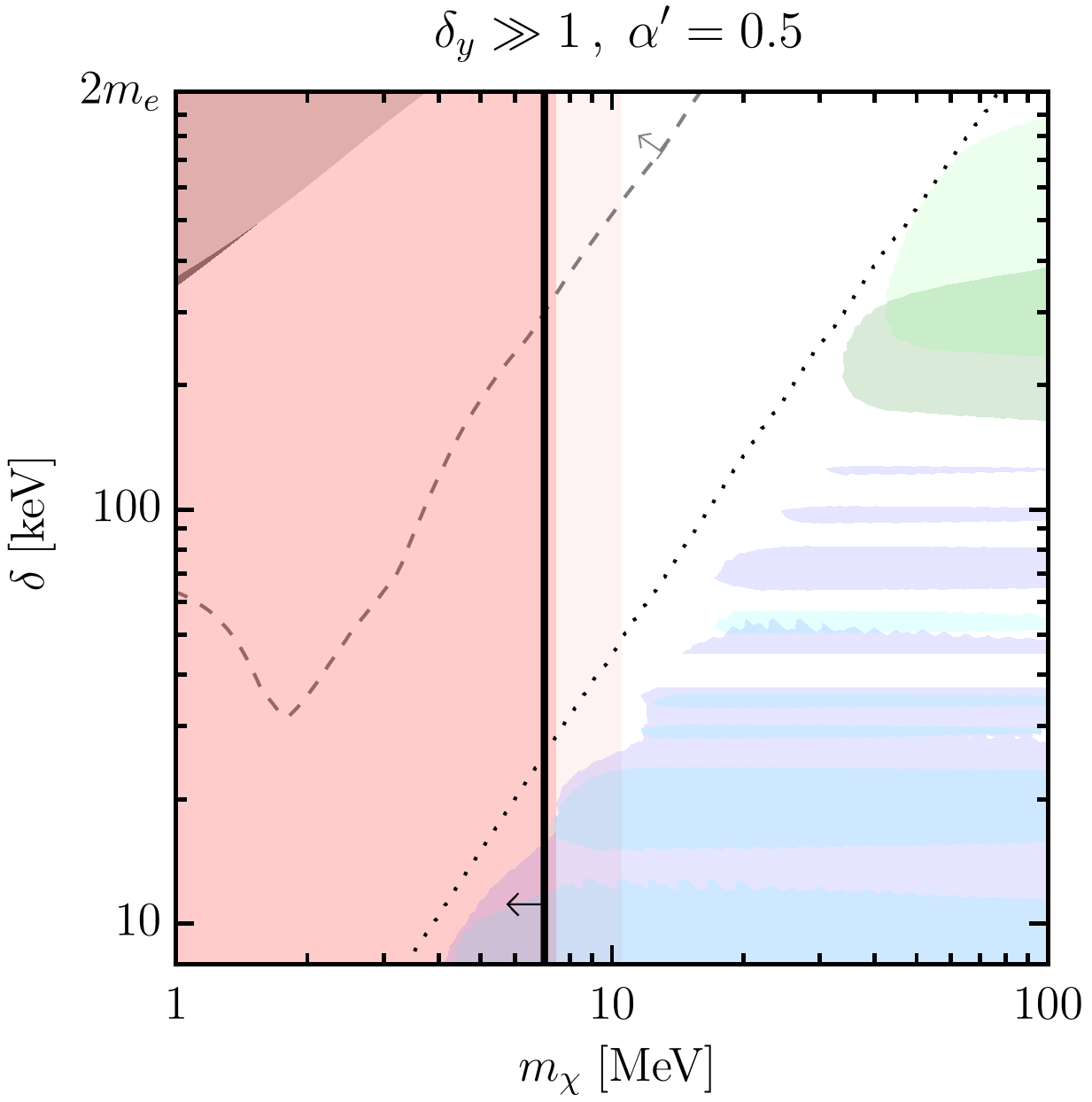}
    \caption{Experimental and cosmological constraints on the (not-so-)inelastic DM model in the plane DM mass~$m_\chi$ and of mass splitting~$\delta$. 
        At each point, the kinetic mixing~$\epsilon$ is fixed by the relic abundance, $\Omega h^2 = 0.12$. 
        Left panels correspond to the parity-conserving ($\delta_y=0$) case, while right panels correspond to maximal parity violation ($\delta_y\gg1$). 
        The dark fine-structure constant is fixed to $\alpha' = 0.1$ (top) or $\alpha' = 0.5$ (bottom). 
        The vertical solid line indicates the expected DM mass region ($m_\chi \lesssim 7$~MeV) excluded by BBN~\cite{Sabti:2019mhn}. 
        The projected reach of the ``Phase I'' LDMX run~\cite{LDMX:2018cma} is shown by the dashed gray lines. 
        Dotted contours correspond to $f_0 = 10^{-11}$. 
        }\label{fig:exclusionIDMvsNIDMmassxdelta}
\end{figure}

\begin{figure}[t] 
    \centering \includegraphics[width=0.775\linewidth]{plots/toplegend_fig1.png}\vspace{0.1cm}\\
    \centering 
    \includegraphics[width=0.45\linewidth]{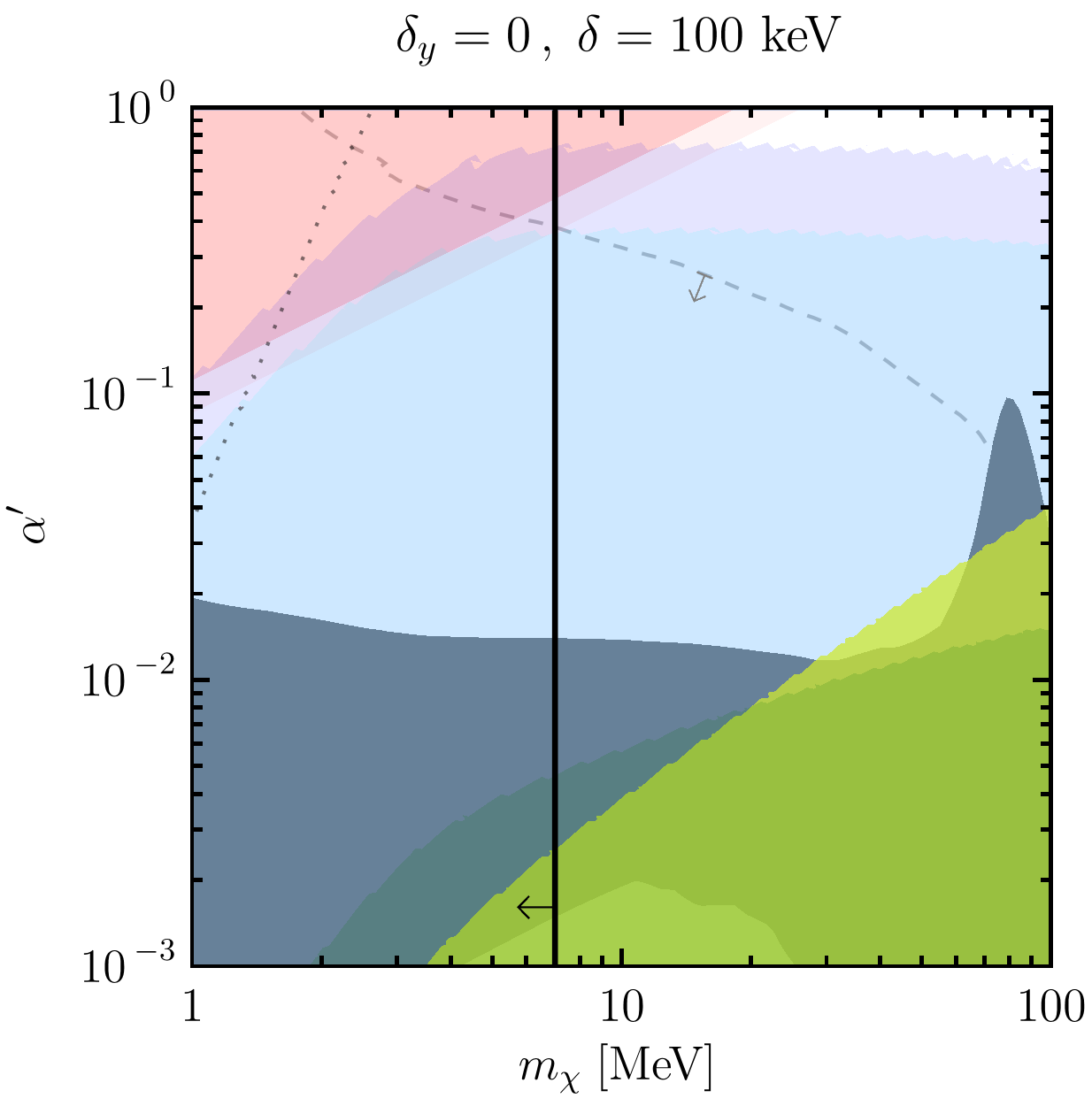}\vspace{0.25cm}\hspace{0.95cm}
    \includegraphics[width=0.45\linewidth]{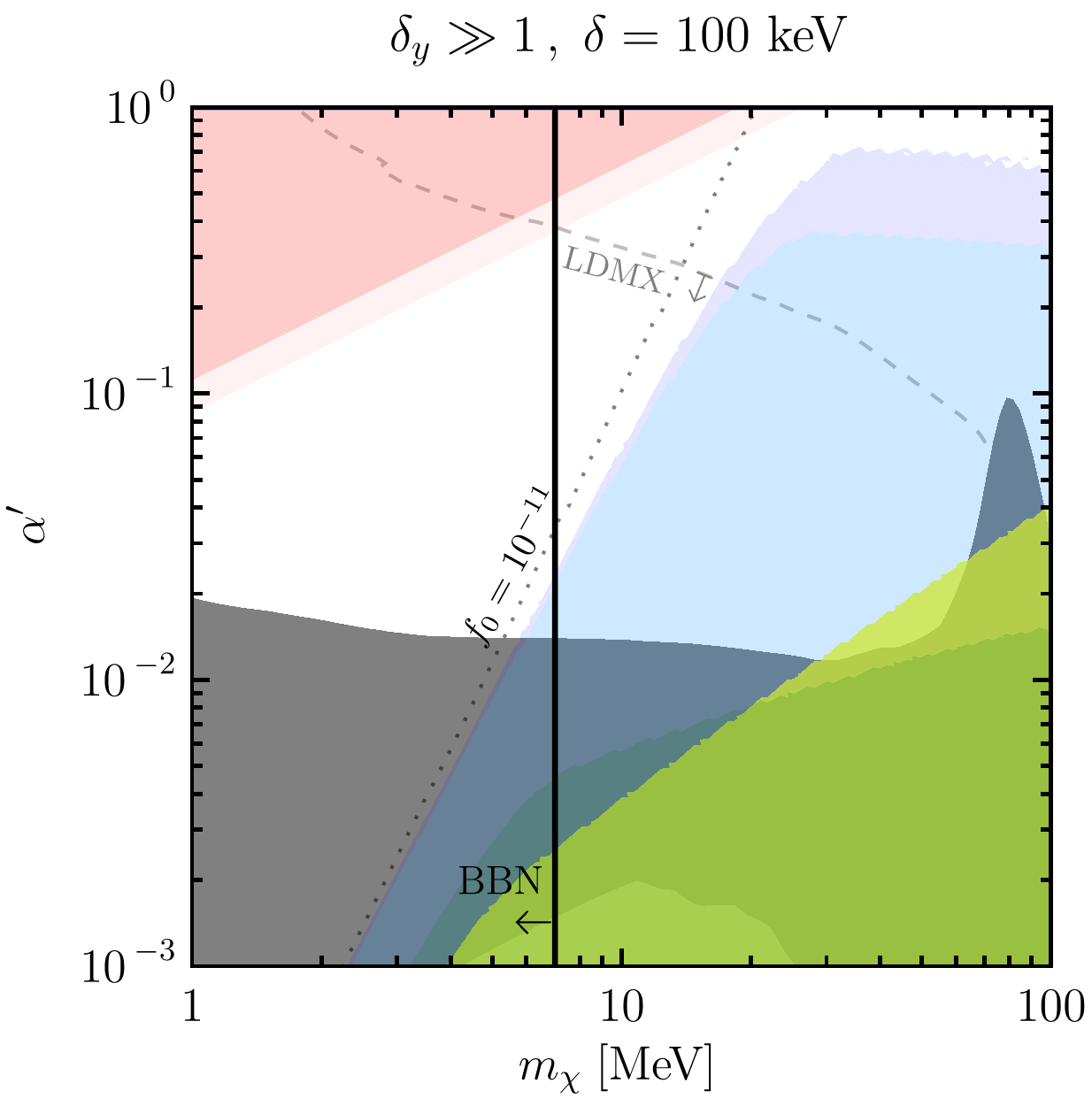}
    \includegraphics[width=0.45\linewidth]{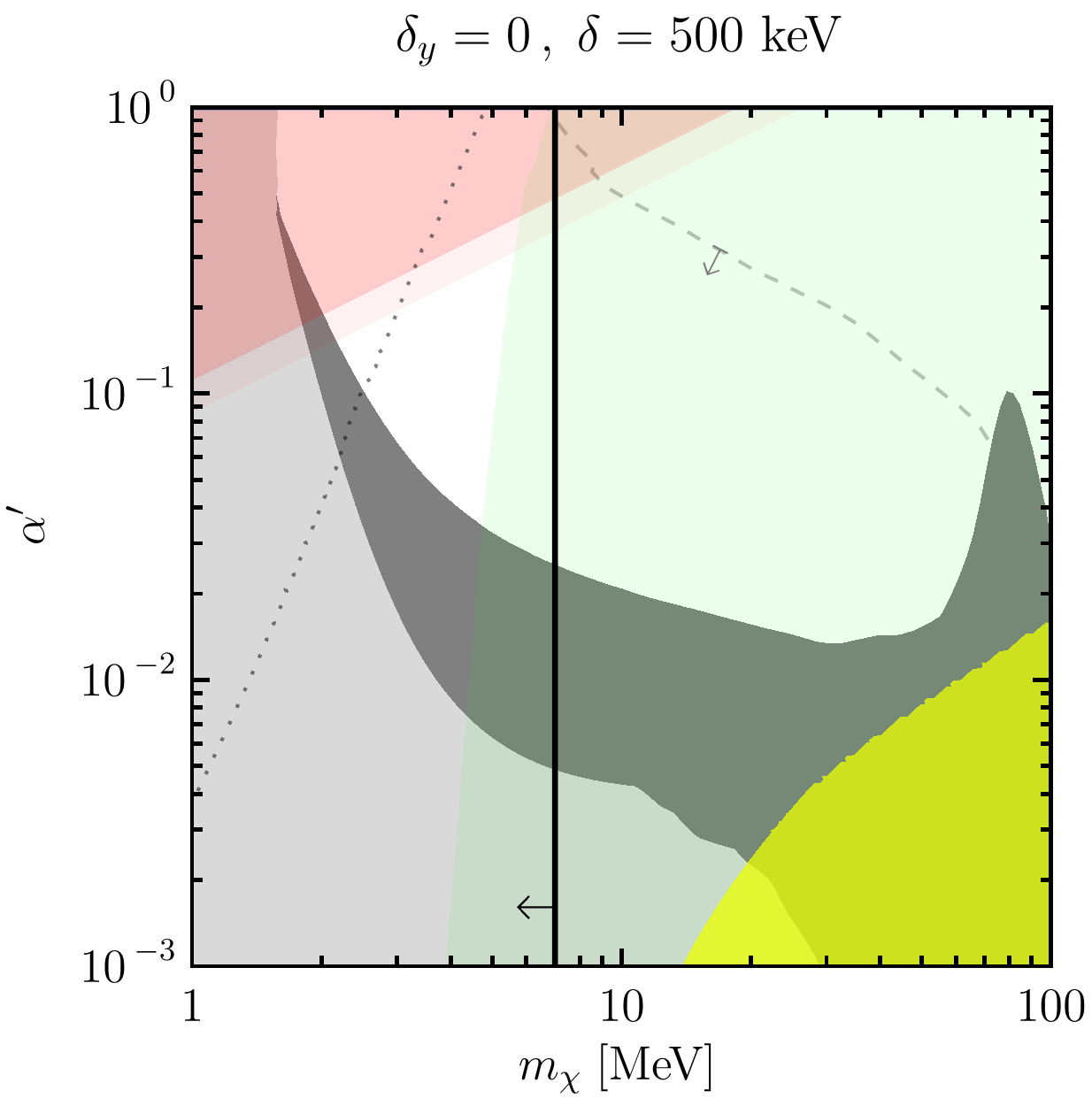}\hspace{0.95cm}
    \includegraphics[width=0.45\linewidth]{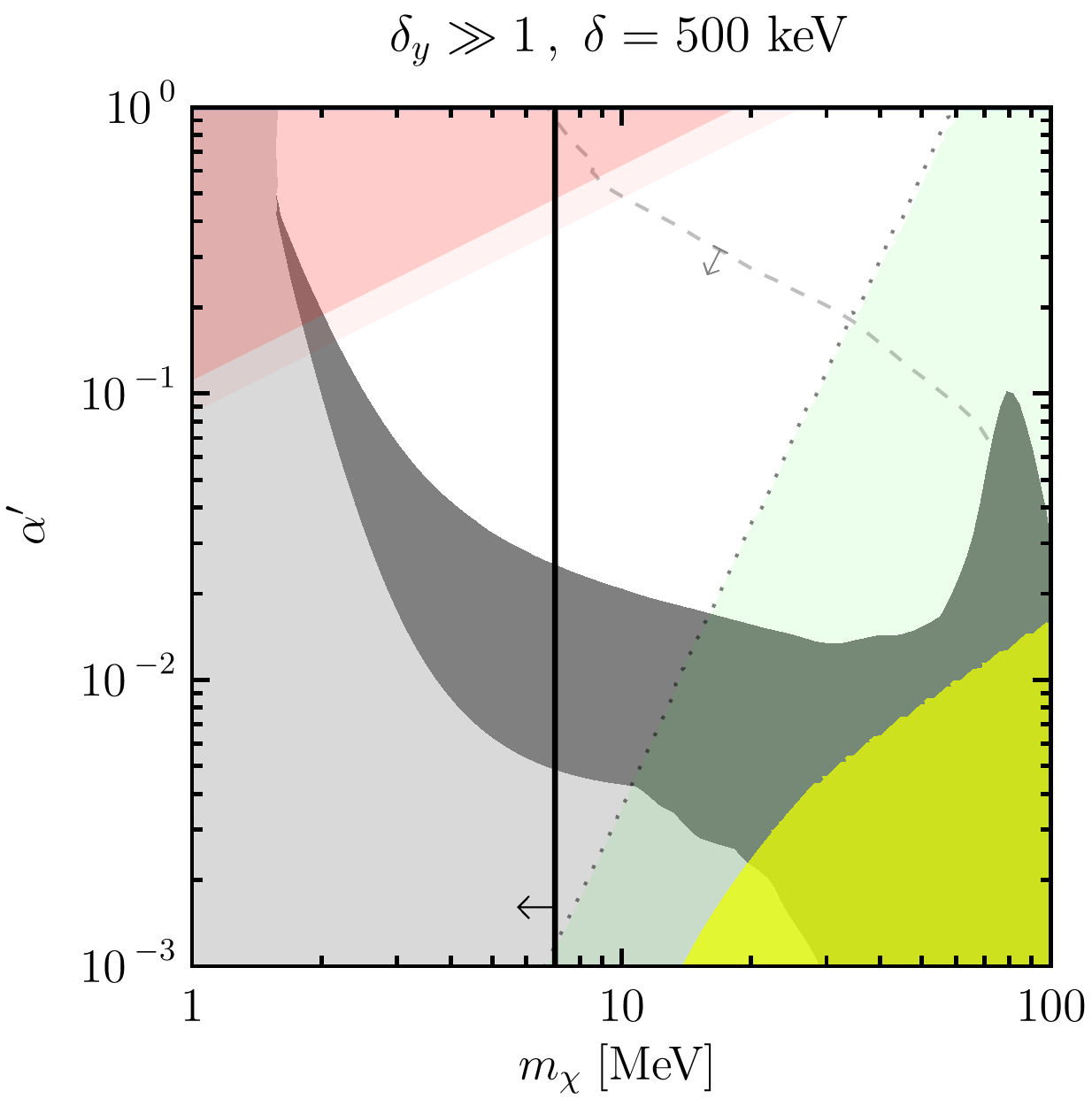}
    \caption{Same constraints as in \cref{fig:exclusionIDMvsNIDMmassxdelta}, shown in the $(m_\chi, \alpha')$ plane for fixed mass splittings 
        $\delta = 100~\text{keV}$ (top) and $\delta = 500~\text{keV}$ (bottom). 
        Left (right) panels correspond to the parity-conserving (maximally parity-violating) scenarios.   }\label{fig:exclusionIDMvsNIDMmassxalpha}
\end{figure}

In \cref{fig:exclusionIDMvsNIDMmassxdelta}, we summarize the current constraints on (not-so-)inelastic DM in the plane $\delta$ versus $m_\chi$ for different values of $\alpha'$  within the parity-conserving scenario $\delta_y=0$ (left panels)  and the maximally parity broken scenario $\delta_y \gg1$ (right panels). Similarly, in \cref{fig:exclusionIDMvsNIDMmassxalpha}, we summarize the constraints in the plane $\alpha'$ versus $m_\chi$ for different values of $\delta$. In each point, the value of the kinetic mixing $\epsilon$ is such that the DM relic abundance is reproduced.

In both \cref{fig:exclusionIDMvsNIDMmassxdelta,fig:exclusionIDMvsNIDMmassxalpha}, we find that DD (XENON1T, XENONnT)\footnote{Note that the oscillation of the DD bound is due to the statistical fluctuations of the background data and the photoelectric cross section.} and ID (INTEGRAL, CMB from decays) constraints are significantly weakened in the parity-violating scenario. This reduction originates from the strong suppression of $f$ in that case (see \cref{fig:deltamass}). To illustrate this effect, we show contours of the excited-state fraction corresponding to $f_0 = 10^{-11}$, which mark the approximate boundary where DD and ID constraints become negligible in the parity-violating regime. Regions with smaller excited-state fractions are therefore also allowed. CMB limits from annihilations, on the other hand, are only relevant for $f\gtrsim10^{-5}$ where parity effects are negligible, so their constraints are the same in both scenarios. Moreover, according to expectations, constraints from self-interactions ($\sigma/m$) and colliders (NA64, LSND) are  insensitive to parity violation within our evaluated parameter space.

In both figures, we also include the projected sensitivity of missing momentum searches at the future electron beam dump experiment LDMX~\cite{LDMX:2018cma} for its “Phase I” LDMX run with $4\times 10^{14}$ electrons on target (EOT).  On \cref{fig:exclusionIDMvsNIDMmassxdelta}, we find that for $\alpha'=0.1$ this experiment can exclude most of our parameter space, while for $\alpha'=0.5$ there will still exist regions free from bounds. Similarly, in \cref{fig:exclusionIDMvsNIDMmassxalpha} we find that LDMX has the potential to exclude most regions of the parameter space, excepting those with very large $\alpha'$ and $m_\chi$.  It is worth mentioning that the sensitivity reported for the extended LDMX run (also called ``Phase II" LDMX run) with $1.6\times 10^{15}$ EOT is expected to probe the entire parameter space explored in our analysis.

Regarding other future searches, NA64 is currently running and its sensitivity will increase in the following years, probing some of the free regions of parameter space  at large $\alpha'$ and low masses~\cite{Na64SPSC}. Furthermore, note that bounds on decaying particles on cosmological scales become very weak below $\tau_{\ast} \lesssim10^{12}$, which nearly corresponds to the upper boundary of the decay constraint from Planck on energy injection during CMB. In this case, for lower lifetimes, only mild improvements are expected from   future PIXIE~\cite{Kogut:2011xw} measurements~\cite{Balazs:2022tjl} and we do not include its projections in our figures.

\begin{figure}[t]
    \def\sepf{0.45}\centering \includegraphics[width=0.63\linewidth]{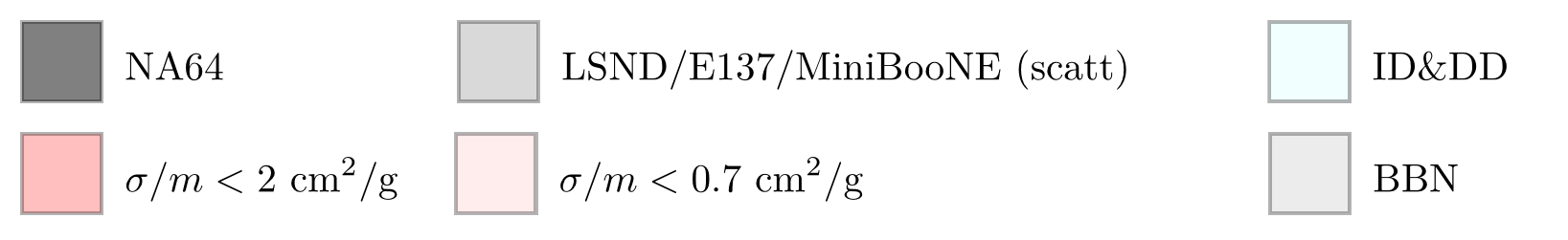}\vspace{0.1cm}\\
    \centering
    \includegraphics[width=\sepf\columnwidth]{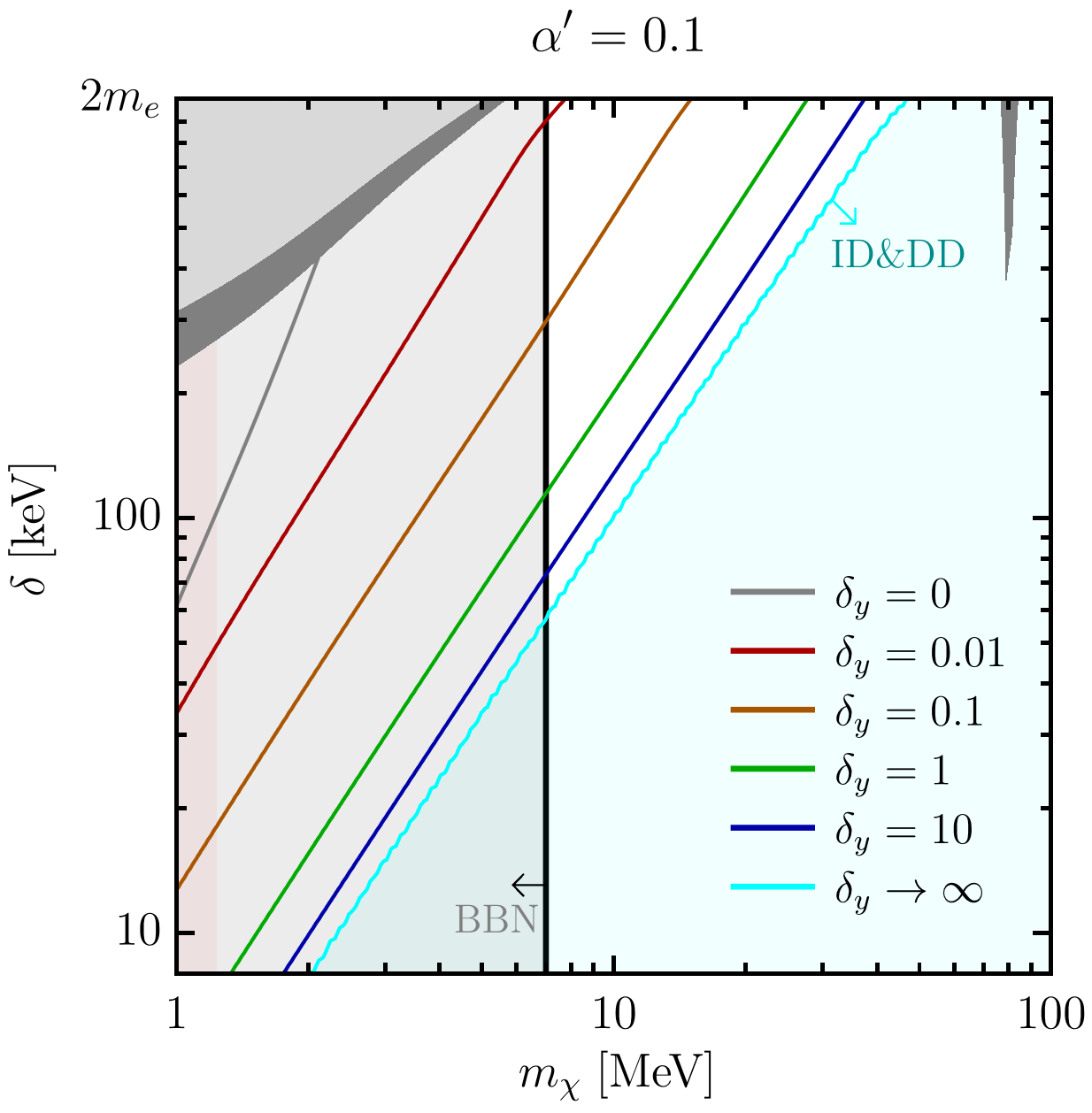}\vspace{0.25cm}\hspace{0.95cm}
    \includegraphics[width=\sepf\columnwidth]{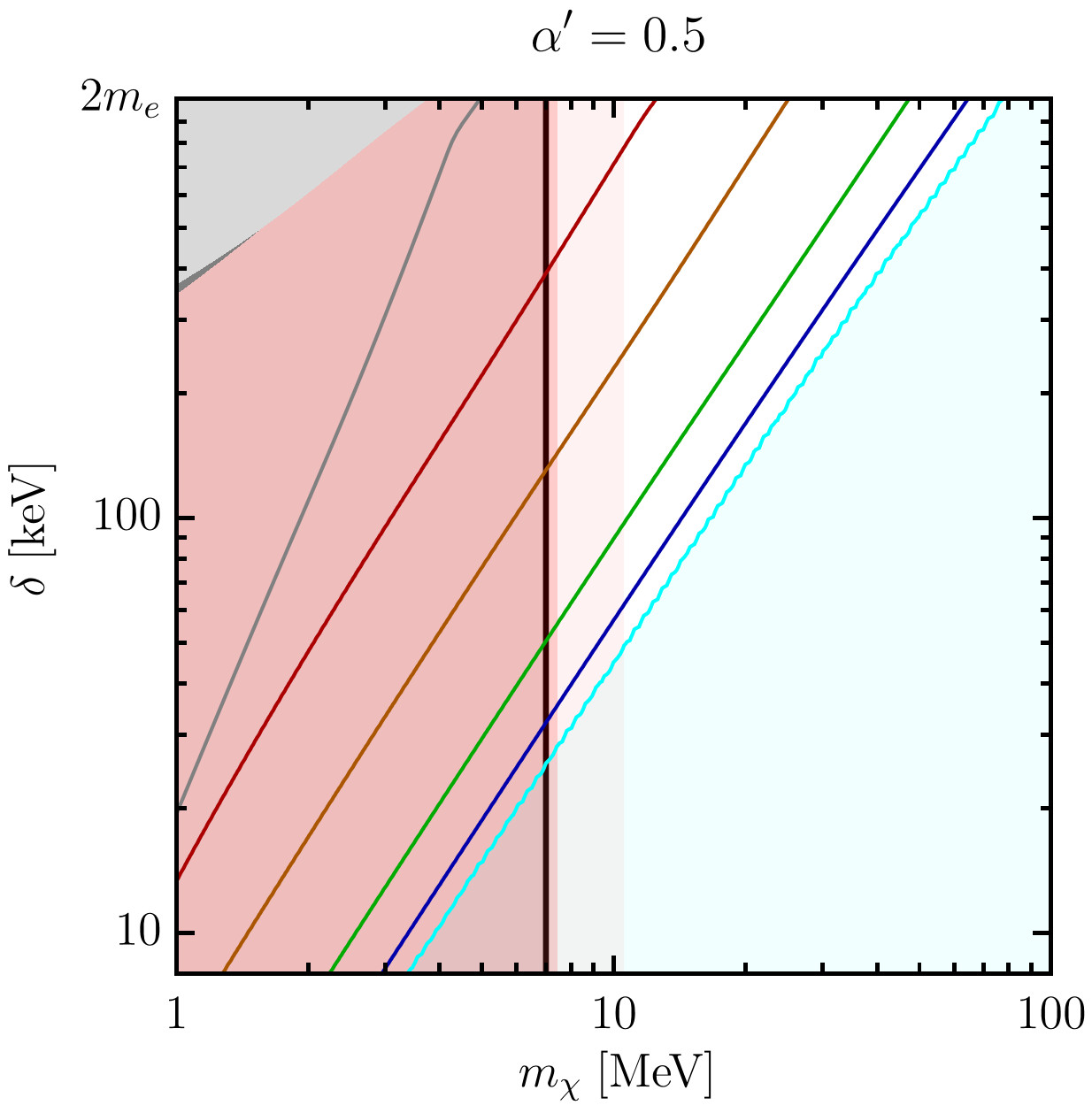}
    \includegraphics[width=\sepf\columnwidth]{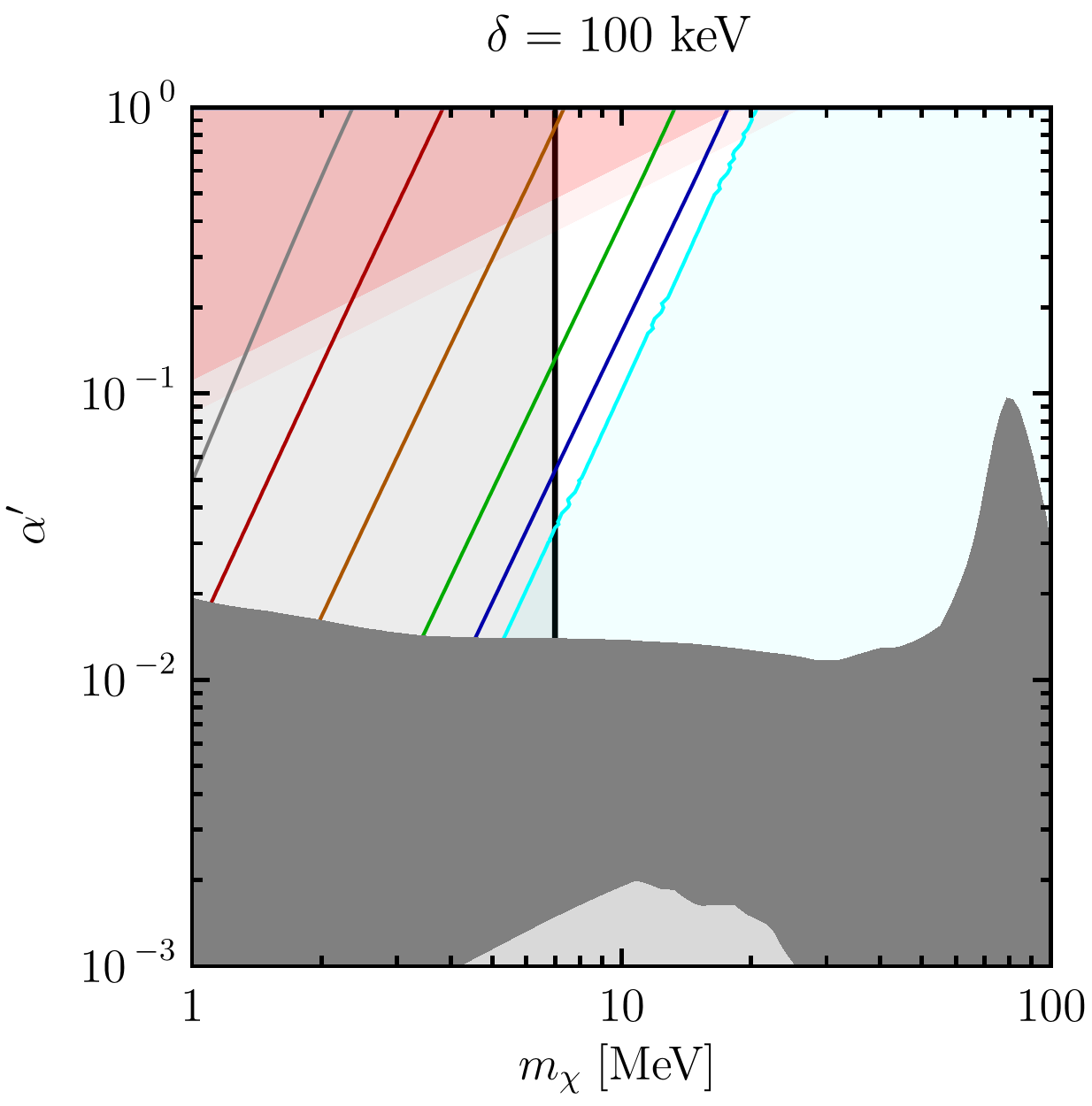} \hspace{0.95cm}
    \includegraphics[width=\sepf\columnwidth]{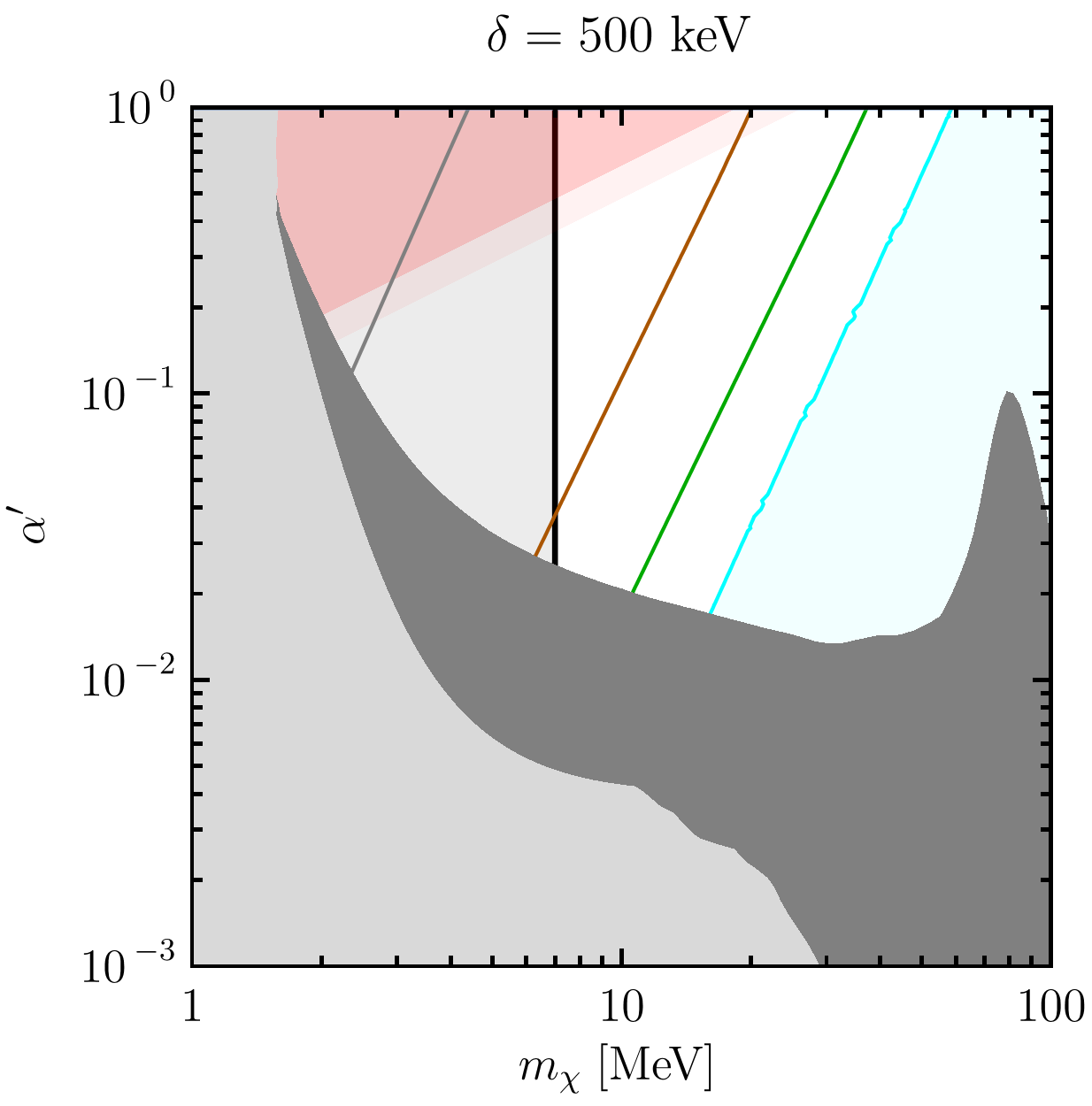}
    \caption{ 
     Top (bottom) panels present contours of $f_0=10^{-11}$ for different values of the parity-violating parameter $\delta_y$  in the plane of mass splitting $\delta$ (dark fine-structure constant $\alpha'$) and DM mass $m_\chi$ for fixed values of $\alpha'$ ($\delta$)---on the left $\alpha'=0.1$ ($\delta=100$~keV) and on the right $\alpha'=0.5$ ($\delta=500$~keV)---adopting the benchmark choice of $m_{A'}=3m_\chi$. Note that current constraints from indirect and direct searches are absent for parity-violating scenarios with initial fractions below $f_0<10^{-11}$.  }\label{fig:parity}
\end{figure}

 In order to understand the impact of non-maximal parity violation,  we show in \cref{fig:parity} the initial excited-state fraction contours corresponding to $f_0=10^{-11}$ for different values of the parity-violating parameter $\delta_y$.  As in \cref{fig:exclusionIDMvsNIDMmassxdelta,fig:exclusionIDMvsNIDMmassxalpha}, the regions above these contours and right to the BBN limit line are free from DD and ID constraints. This highlights that even small parity violation effects, $\delta_y \sim10^{-2}$, can already open up new parameter space. Note that, as mentioned before, limits from DM self-interactions and colliders, as well as those from BBN, present negligible variations due to parity violation in the parameter space under consideration; thus, we simply report the bounds for the parity-conserving scenario.

\begin{figure}[p]
    \centering \includegraphics[width=0.775\linewidth]{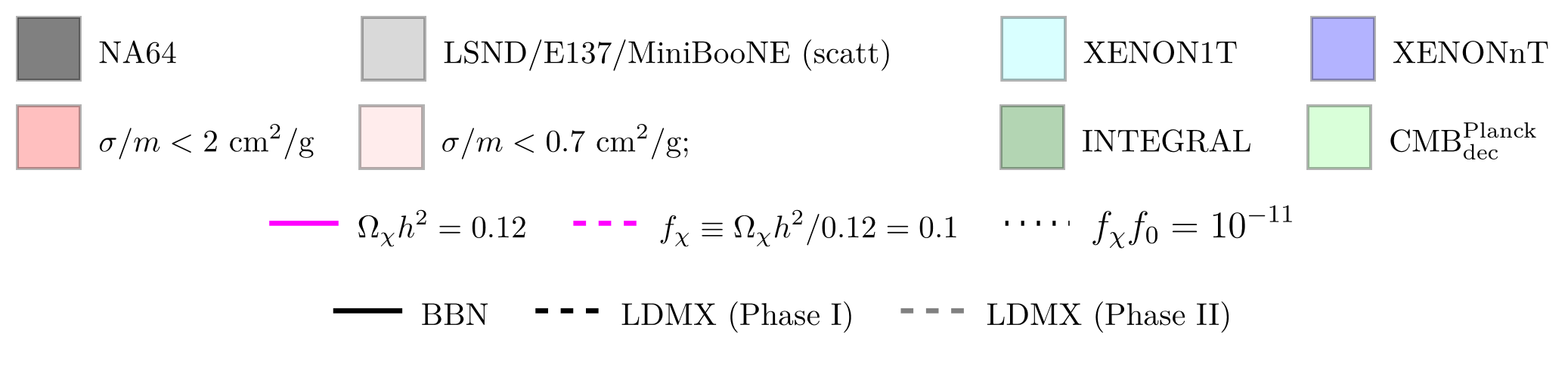}\vspace{0.1cm}\\
    \includegraphics[width=0.45\linewidth]{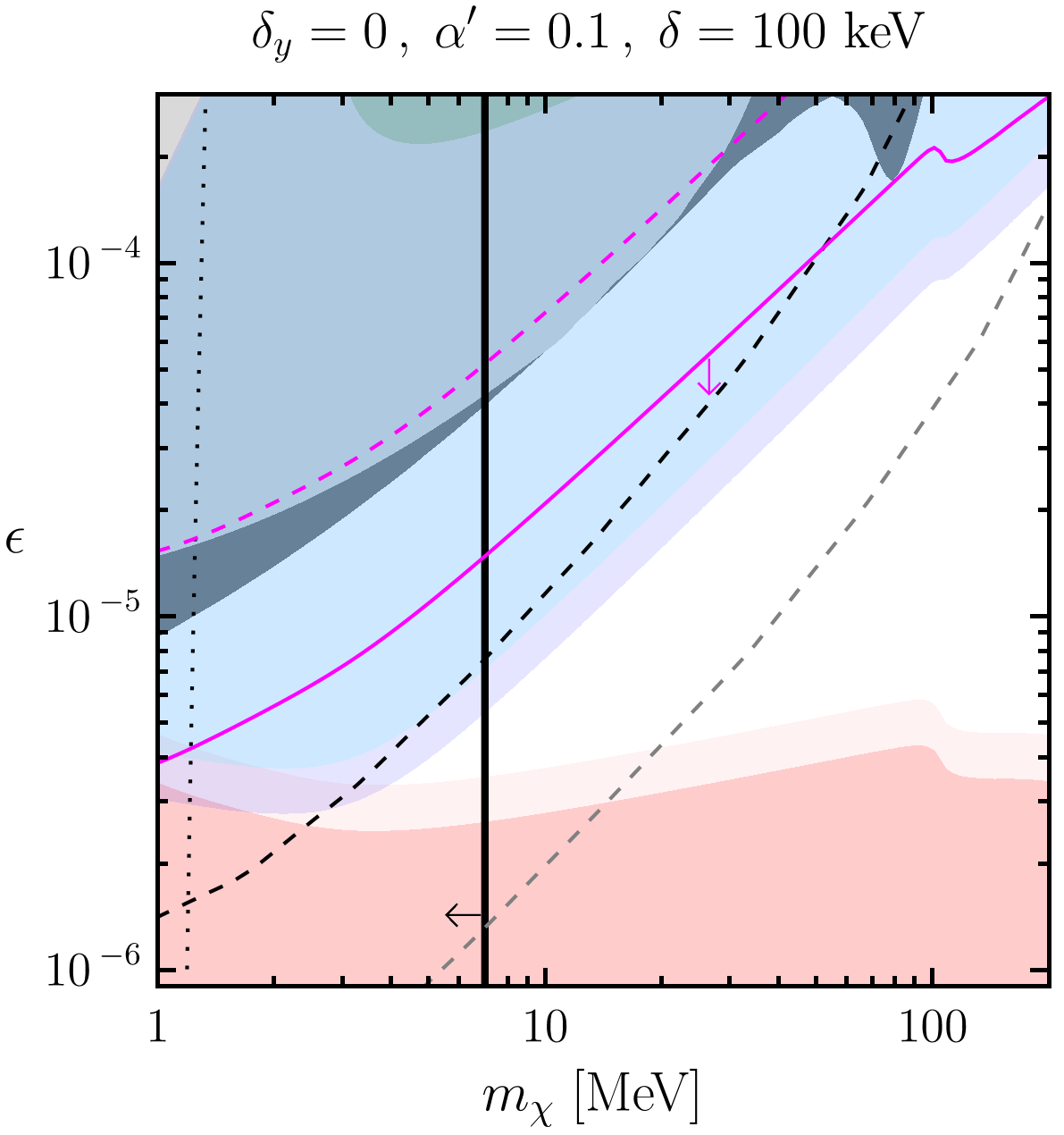}\vspace{0.25cm}\hspace{0.95cm}
    \includegraphics[width=0.45\linewidth]{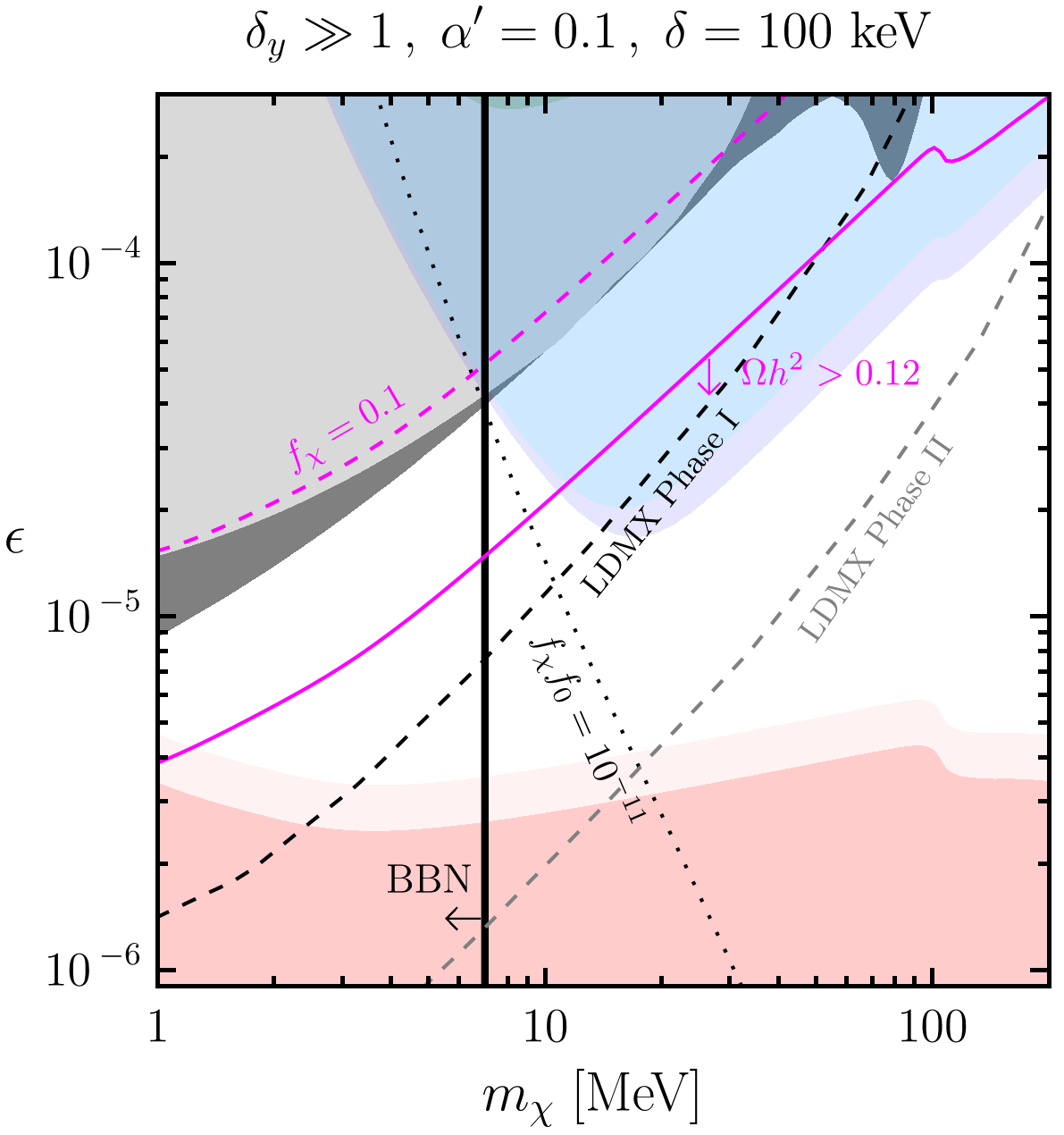}
    \includegraphics[width=0.45\linewidth]{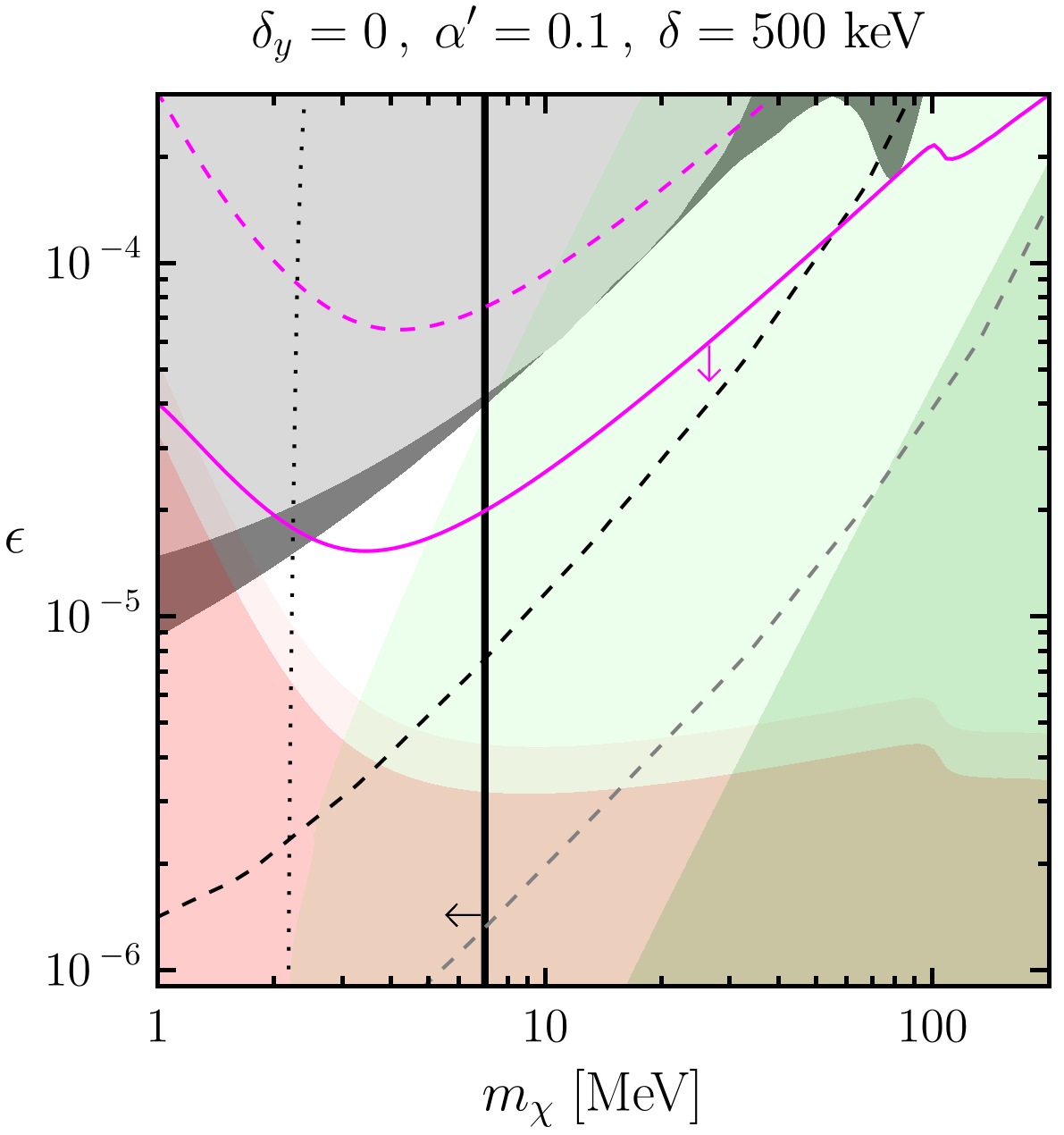}\hspace{0.95cm}
    \includegraphics[width=0.45\linewidth]{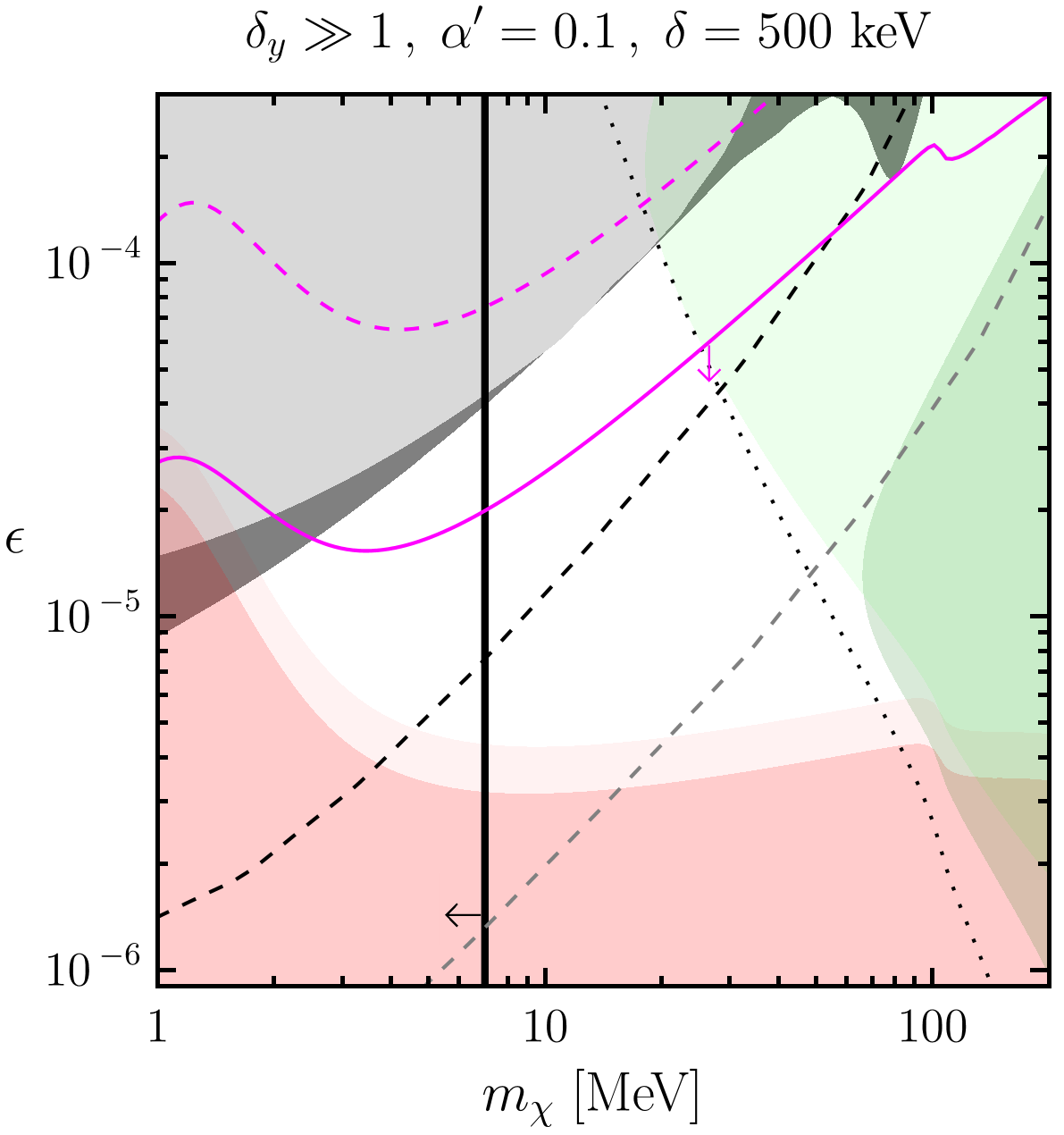}
    \caption{Same constraints as in \cref{fig:exclusionIDMvsNIDMmassxdelta}, shown in the $(m_\chi, \epsilon)$ plane for fixed mass splittings 
        $\delta = 100~\text{keV}$ (top) and $\delta = 500~\text{keV}$ (bottom), assuming $\alpha' = 0.1$. 
        At each point, the dark fermion abundance $f_\chi \equiv \Omega / \Omega_{\rm obs}$ is computed using \texttt{MicrOMEGAs}. 
        Left (right) panels correspond to the parity-conserving (maximally parity-violating) scenarios. 
        The legend and color scheme are identical to those in \cref{fig:exclusionIDMvsNIDMmassxdelta}, except that the projected reaches of the ``Phase~I'' and ``Phase~II'' LDMX runs~\cite{LDMX:2018cma} are shown by the dashed black and gray lines, respectively, and dotted contours now indicate an excited-state fraction of the observed DM abundance $f_\chi f_0 = 10^{-11}$. 
        The magenta solid line marks the thermal target, while dashed magenta lines correspond to the case where dark fermions constitute $10\%$ of the total DM density. 
    }
    \label{fig:exclusionIDMvsNIDMmassxEPSalpha01}
\end{figure}

\begin{figure}[t]
    \centering \includegraphics[width=0.775\linewidth]{plots/toplegend_figEPS.png}\vspace{0.1cm}\\
    \centering
    \includegraphics[width=0.45\linewidth]{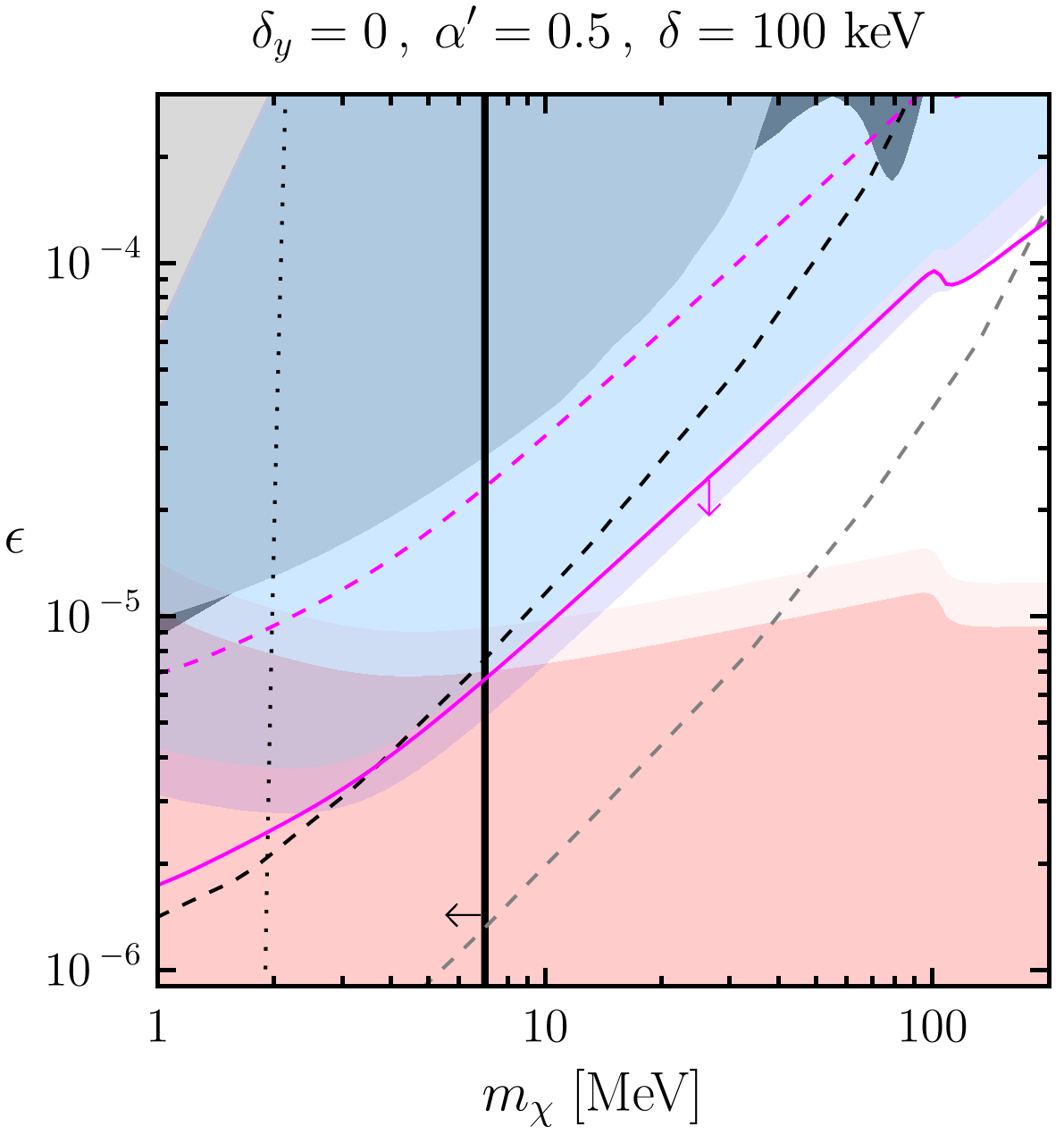}\vspace{0.25cm}\hspace{0.95cm}
    \includegraphics[width=0.45\linewidth]{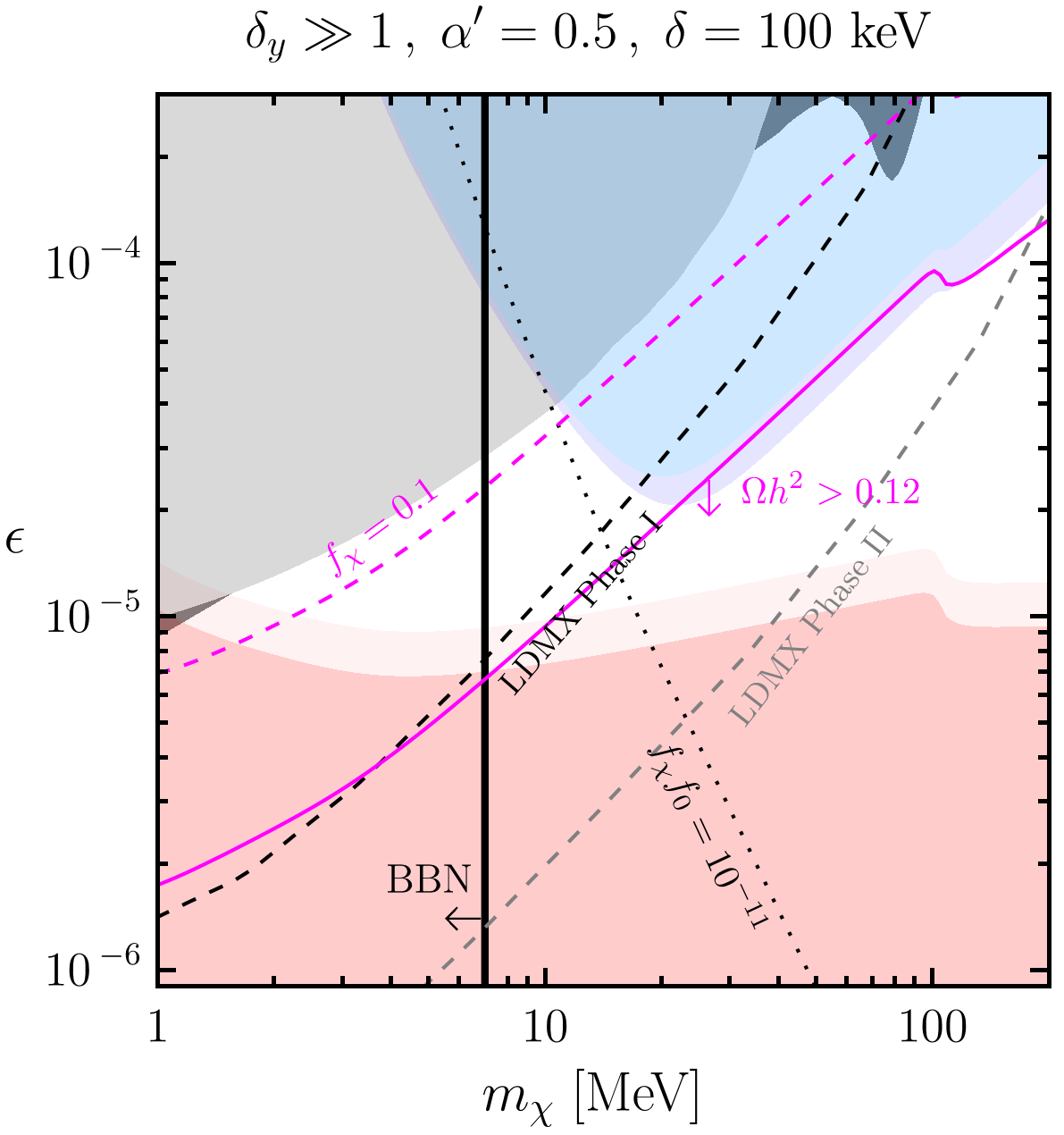}
    \includegraphics[width=0.45\linewidth]{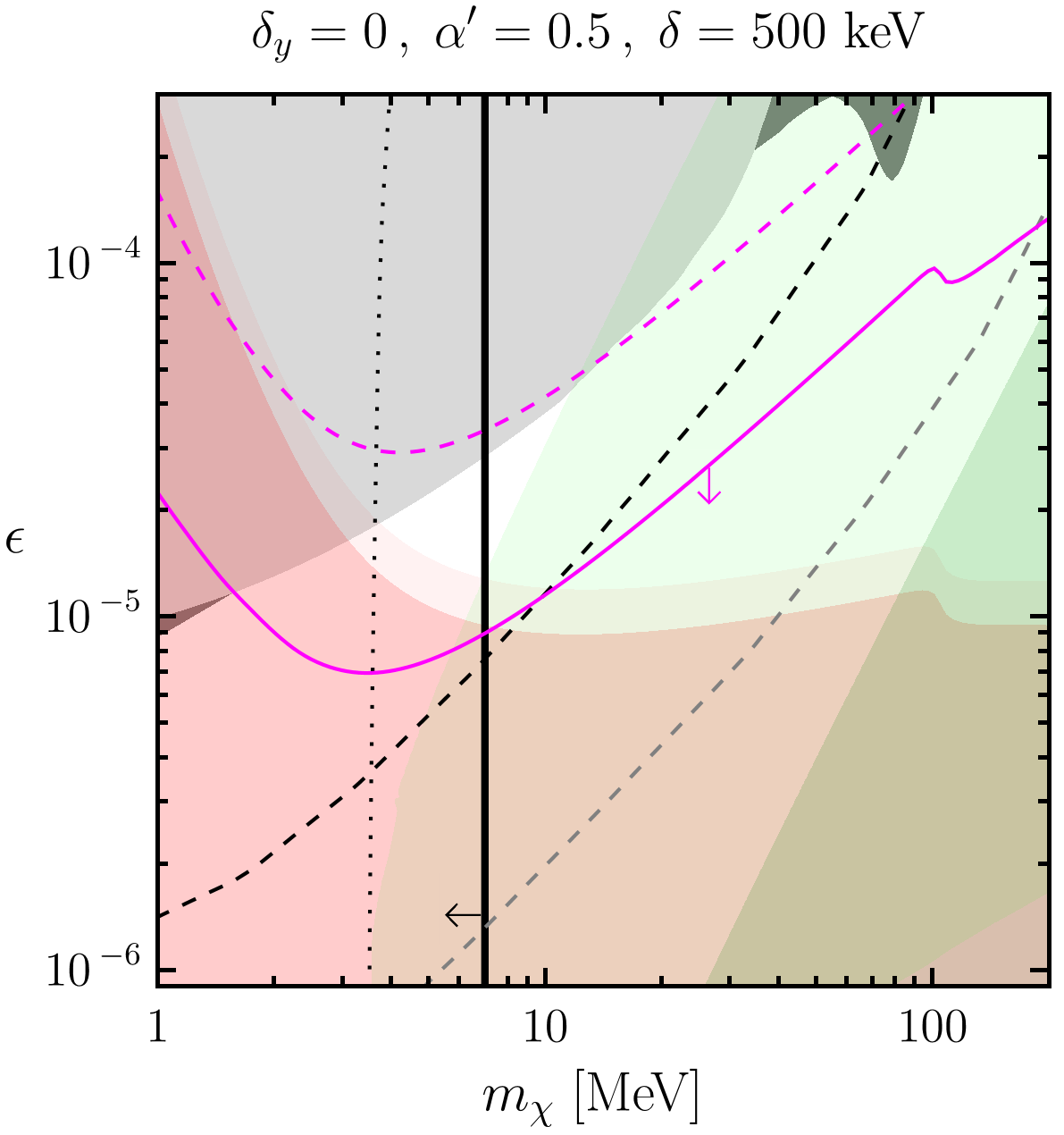}\hspace{0.95cm}
    \includegraphics[width=0.45\linewidth]{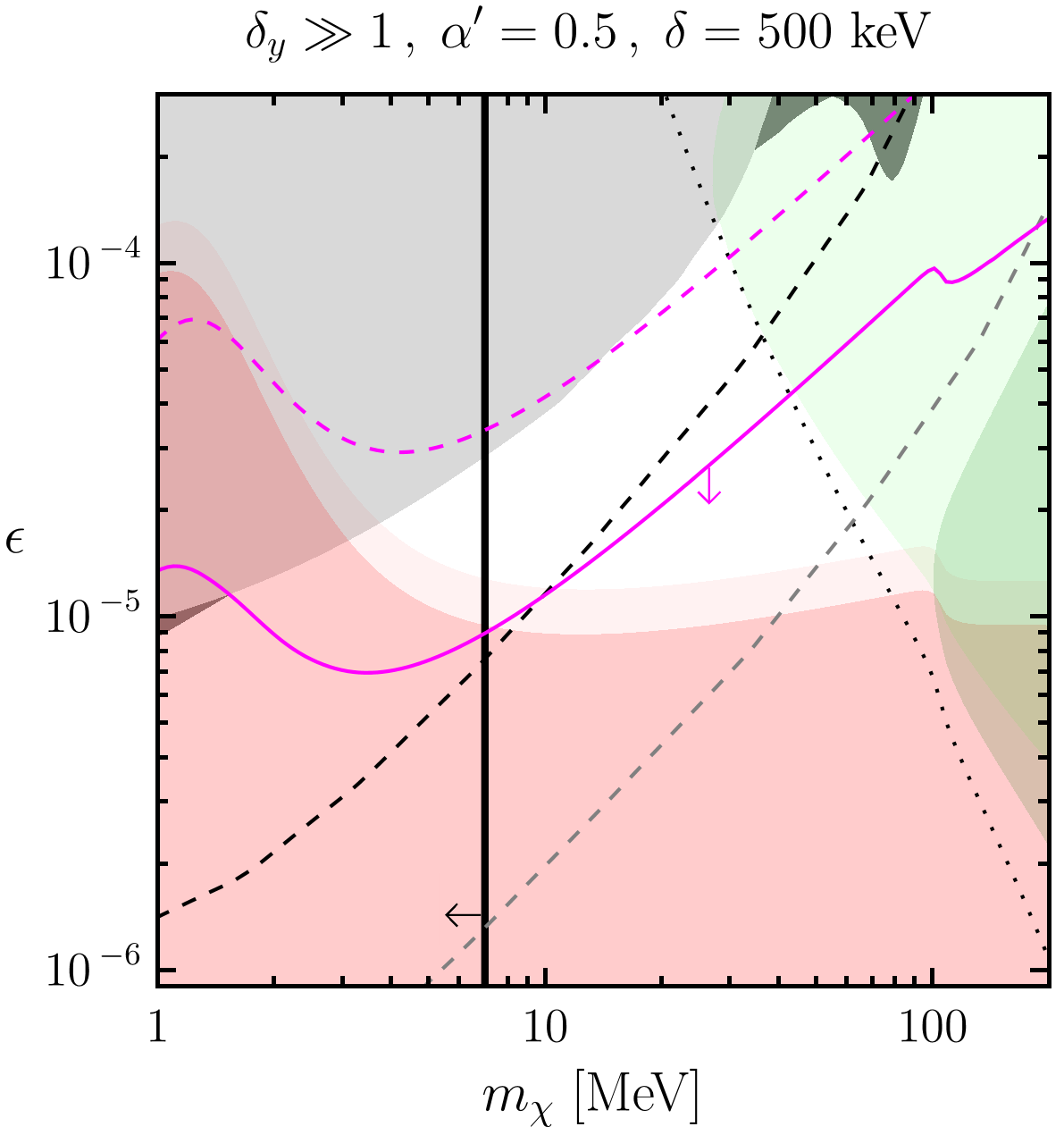}
    \caption{Same as \cref{fig:exclusionIDMvsNIDMmassxEPSalpha01}, but for $\alpha' = 0.5$.}
    \label{fig:exclusionIDMvsNIDMmassxEPSalpha05}
\end{figure}

 Finally, in \cref{fig:exclusionIDMvsNIDMmassxEPSalpha01,fig:exclusionIDMvsNIDMmassxEPSalpha05} we present our bounds in the conventional plane of kinetic mixing $\epsilon$ versus DM mass. Here, for each point we compute the total abundance of both dark fermions $\chi$ and $\chi^\ast$, which we denote by $\Omega=\Omega_\chi+\Omega_{\chi^*}$. This is used to define the DM energy density fraction $f_\chi$:
\begin{equation}
f_\chi\equiv\dfrac{\Omega}{\Omega_{\rm obs}}\;.
\end{equation} In this way, we consider the possibility that the total abundance of $\chi$ and $\chi^*$ does not satisfy the Planck bound, relevant for the situation where our candidates are a sub-component of the total DM. 
In both figures, we present the lines with $f_\chi=1$ ($0.1$) in solid (dashed) magenta. Note that, in standard cosmology, the regions below the $f_\chi=1$ line are excluded, due to overproduction of DM.\footnote{For examples of non-standard cosmology controlling DM overabundance via entropy injection, see Refs.~\cite{Acharya:2008bk,Drees:2017iod,Arias:2019uol,Bernal:2022wck,Haque:2023yra,Silva-Malpartida:2023yks,Silva-Malpartida:2024emu,Bernal:2025szh}. } This exclusion is expected to hold down to values of the kinetic mixing $\epsilon\sim 10^{-11}$, where DM does not reach thermal equilibrium with the SM~\cite{Heeba:2023bik}.

In this context, one needs to recast the previous bounds on our parameter space. Firstly, since the fraction of DM corresponding to the excited state $\chi^\ast$ is $f_{\chi}f_0$, all constraints depending on $f_0$ can be recalculated by replacing $f_0 \to f_{\chi}f_0$. Secondly, self-interaction bounds are reinterpreted by imposing the upper limit given in  \cref{eq:SIDM_bound} on $f_\chi^2\,\sigma/m$.  Finally, collider constraints do not depend on either $f_\chi$ or $f_0$.    

Results for $\alpha'=0.1$ are shown in \cref{fig:exclusionIDMvsNIDMmassxEPSalpha01}. The figure takes $\delta=100$~keV (500~keV) on the upper (lower) rows. The parity-conserving case ($\delta_y=0$) is shown on the left column, and the parity-violating scenario ($\delta_y\gg1$) on the right one. Once again, we find the aforementioned opening of the parameter space when parity is violated. Indeed, for $\delta=100$ (500)~keV case, the parity-conserving scenario is completely excluded by DD (ID and BBN) limits, while the parity-violating case remains viable for masses $m_\chi \lesssim 10$ (30)~MeV. Additionally, in \cref{fig:exclusionIDMvsNIDMmassxEPSalpha05}, we show the corresponding results  for $\alpha'=0.5$. Conclusions are similar, with the exception that  a small triangular region with $f_\chi <1$ is also allowed for the parity-conserving scenario.

The figures also show the expected sensitivities for LDMX, in both ``Phase I" and ``Phase II". The former will probe the entire unconstrained space of our model for $\alpha'=0.1$ while only the latter can fully explore the case of $\alpha'=0.5$---adopting the benchmark $m_{A'}=3m_\chi$.

Finally, in both figures, for parity violation, we find that $f_\chi f_0 = 10^{-11}$ is still a good indicator of where ID loses sensitivity, and also for DD provided that $f_\chi\gtrsim0.1$.  Furthermore, note that we find a lower limit on fraction of the sub-component DM of about $f_\chi\gtrsim0.1$ (irrespective of parity considerations) giving that  larger values of $\epsilon$ are required in such cases and collider constraints  remain unchanged. 

\section{Conclusions} \label{sec:conc}

Sub-GeV dark matter (DM) has attracted significant attention due to the relative weakness of DD and collider constraints, as well as the numerous experiments proposed to probe this mass range. Inelastic sub-GeV DM scenarios can further evade the stringent ID bounds that exclude standard $s$-wave annihilating candidates, making them particularly attractive. Nevertheless, the phenomenologically viable parameter space remains highly constrained, especially for long-lived excited states (exothermic DM). In this work, we have shown that parity violation, which generates (suppressed) elastic interactions, plays a crucial role in reopening the parameter space of the model. Importantly, the framework developed here is largely model-independent and can be readily applied to a broad class of exothermic sub-GeV DM models.

A key ingredient in assessing the phenomenology of these models is the population of long-lived excited states, which naturally arises for $\delta < 2 m_e$. We have computed the thermal history of the ground and excited states with significantly improved accuracy by solving the integrated Boltzmann equation for their relative abundance, fully accounting for up- and down-scattering with nucleons and electrons, as well as dark-sector conversion processes, and by computing in detail the dark-sector temperature evolution. Crucially, the (suppressed) new interactions generated through parity violation can maintain $\chi^\ast \leftrightarrow \chi$ conversions in equilibrium down to $T_\chi \lesssim \delta$, reducing the number of excited states and thereby relaxing direct- and indirect-detection constraints.

Using these results, we have revisited the full set of phenomenological limits, incorporating the most recent experimental data. The updated parameter space contains substantial regions that remain viable, demonstrating that parity violation can reopen territory previously regarded as excluded. In particular, even parity-violating effects as small as percent-level (e.g., $m_R \sim 1.01\, m_L$) restore the viability of significant portions of parameter space.  For $m_A'=3 m_\chi$ and $\alpha' =0.1$, we have obtained that the allowed parity-violating region of parameter space corresponds to roughly $m_\chi \in [7,50]$ MeV and $\delta \in [50,1000]$ keV, with $\epsilon \in [10^{-5},10^{-4}]$. For larger (smaller) $\alpha'$, limits from self-interactions and DD demand somewhat larger (smaller) DM masses, corresponding to similar values of the kinetic mixing parameter. Moreover, the larger the parity-violating interactions, the larger (smaller) the allowed DM mass (mass splitting). 

Overall, our findings establish parity-violating inelastic DM as a compelling and testable framework for sub-GeV DM, motivating further experimental scrutiny. Future searches with enhanced sensitivity to low-mass scattering and de-excitation signatures, such as LDMX Phase I and, especially, Phase II, will be particularly powerful for probing the allowed parameter space identified in this work.

\vspace{0.5cm}
\section*{Acknowledgements}

We are grateful to Stefan Clementz and Nicholas Leerdam for participating in the early stages of a related project, and Nicholas Leerdam for also sharing with us his numerical code. The authors thank Nicolás Bernal for discussions related to the dark sector temperature and Sandra Robles for discussions related to Ref.~\cite{Bell:2018pkk}.  GG is supported by the Doctoral School ``Karlsruhe School of Elementary and Astroparticle Physics: Science and Technology (KSETA)” through the GSSP program of the German Academic Exchange Service (DAAD). JHG is supported by the \enquote{Consolidación Investigadora Grant CNS2022-135592}, funded also by \enquote{European Union NextGenerationEU/PRTR} and by the \enquote{Generalitat Valenciana} through the GenT Excellence Program (CIESGT2024-007). JS and JJP acknowledge funding by the {\it Dirección de Gestión de la Investigación} at PUCP, through grant DFI-PUCP-PI1144. This work is partially supported by the Spanish \emph{Agencia Estatal de Investigación} MICINN/AEI (10.13039/501100011033) grant PID2023-148162NB-C21 and the \emph{Severo Ochoa} project MCIU/AEI CEX2023-001292-S. This project has received funding from the European Union’s Horizon Europe research and innovation programme under the Marie Skłodowska-Curie Staff Exchange grant agreement No 101086085 – ASYMMETRY.

\vspace{0.5cm}

\appendix
 
\section{Numerical integration} 
\label{app:int}

The Boltzmann equation is stiff at early times when the abundance is close to its equilibrium value, which makes standard explicit solvers either unstable or prohibitively inefficient due to the very small steps required by the rapidly varying source terms. To robustly handle this stiffness, we adopt an implicit scheme following DarkSUSY \cite{Bringmann:2018lay} and, more recently, DRAKE \cite{Binder:2021bmg}. We evolve the independent variable $x$ geometrically, $\log x_{i+1}=\log x_i+h$, where $h$ is the adaptive step size. An adaptive controller adjusts $h$ based on local error estimates and the convergence of the implicit solver.

Our integration strategy is based on an embedded pair of implicit methods: a trapezoidal update and an implicit Euler update. First, given a function $y(x)$, we expand $y_i\equiv y(x_i)$ using a trapezoidal discretization of the evolution between $x_i$ and $x_{i+1}$, yielding
\begin{align}
    y_{i+1}^{T} = y_i + \frac{1}{2}\,h\,(y'_i+y'_{i+1}) + \mathcal{O}(h^2)\,,
\end{align}
while the corresponding implicit Euler update is
\begin{align}
  y_{i+1}^{E} = y_i + h\,y'_{i+1}\,.
\end{align}
The difference between these two estimates provides a built-in measure of the truncation error. Expanding the trapezoidal rule, one finds
\begin{align}
  \frac{1}{2}h^2 \lvert y'' \rvert 
  \;\approx\; \frac{1}{2}h\,\lvert y'_{i+1}-y'_i\rvert
  \;=\; \lvert y_{i+1}^{T}-y_{i+1}^{E}\rvert\,,
\end{align}
so that \(\lvert y_{i+1}^{T}-y_{i+1}^{E}\rvert\) directly estimates the second–order term in the local expansion. A step is accepted when this local error indicator is small compared to the solution,
\begin{align}
  \lvert y_{i+1}^{T}-y_{i+1}^{E}\rvert 
  \;<\; \varepsilon\,\lvert y_{i+1}^{T}\rvert\,,
\end{align}
and throughout this work we take \(\varepsilon = 10^{-2}\). The same error estimate is then used by an adaptive controller to increase or decrease \(h\) for the next step. In our application the evolved quantity is the relative abundance
\begin{align}
  f_i \equiv f(x_i)\,,
\end{align}
so we now specialize the above construction to $y=f$. The Boltzmann equation \eqref{f boltzmann eq} can be recast as
\begin{align}
  \frac{df}{dx} \;=\; \lambda(x)\,\big(A\,f^2 - B\,f + C\big)\,,
  \label{eq:BEnew}
\end{align}
where the effective interaction rate $\lambda(x)$ collects the effects of the Hubble parameter and the entropy evolution of the plasma, and is given by 
\begin{align}
\lambda(x) &= \frac{1}{xH} \left ( 1+ \frac{m_{\chi}}{3xg_s} \frac{\mathrm{d} g_s}{\mathrm{d} T} \right )  \,.
\end{align}
and the coefficients $A$, $B$ and $C$ encode the relevant number changing and conversion processes,
\begin{align}
A &=
\begin{aligned}[t]
 &\Gamma_{\chi^*\chi\to\chi\chi}
 + \Gamma_{\chi\chi\to\chi^*\chi}
 - \Gamma_{\chi^*\chi^*\to\chi^*\chi} \\
 &\hspace{3em}
 - \Gamma_{\chi^*\chi\to\chi^*\chi^*}
 - \Gamma_{\chi^*\chi^*\to\chi\chi}
 + \Gamma_{\chi\chi\to\chi^*\chi^*}
\end{aligned}\,\\[0.6em]
B &=
\begin{aligned}[t]
 &\Gamma_{\chi^* e^{\pm}\to\chi e^{\pm}}
 + \Gamma_{\chi e^{\pm}\to\chi^* e^{\pm}}
 + \Gamma_{\chi^*\chi\to\chi\chi} \\
 &\hspace{3em}
 + 2\,\Gamma_{\chi\chi\to\chi^*\chi}
 - \Gamma_{\chi^*\chi\to\chi^*\chi^*}
 + 2\,\Gamma_{\chi\chi\to\chi^*\chi^*}
\end{aligned}\,\\[0.6em]
C &=
\Gamma_{\chi e^{\pm}\to\chi^* e^{\pm}}
+ \Gamma_{\chi\chi\to\chi^*\chi}
+ \Gamma_{\chi\chi\to\chi^*\chi^*}\,.
\end{align}
Applying the discretization discussed above to \cref{eq:BEnew} gives closed form updates for the trapezoidal and implicit–Euler estimates. Solving explicitly for the next step we obtain
\begin{align}
f_{i+1}^{T} &=
\frac{2\,c_T}{\,2+u_{i+1}\, B_{i+1}
+\sqrt{\big(2+u_{i+1}\, B_{i+1}\big)^2-4\,u_{i+1}\,A_{i+1}\,c_T}\,}\,,\\
f_{i+1}^{E} &=
\frac{2\,c_E}{\,1+u_{i+1}\, B_{i+1}
+\sqrt{\big(1+u_{i+1}\, B_{i+1}\big)^2-4\,u_{i+1}\,A_{i+1}\,c_E}\,}\,,
\end{align}
where \(u_i \equiv h\,\lambda_i\) and
\begin{align}
c_T &= 2f_i + u_i\big(A_i f_i^2 - B_i f_i + C_i\big) + u_{i+1}C_{i+1}\,,\\
c_E &= f_i + u_{i+1}C_{i+1}\,.
\end{align}

\begin{figure}
  \centering
\includegraphics[width=.8\textwidth]{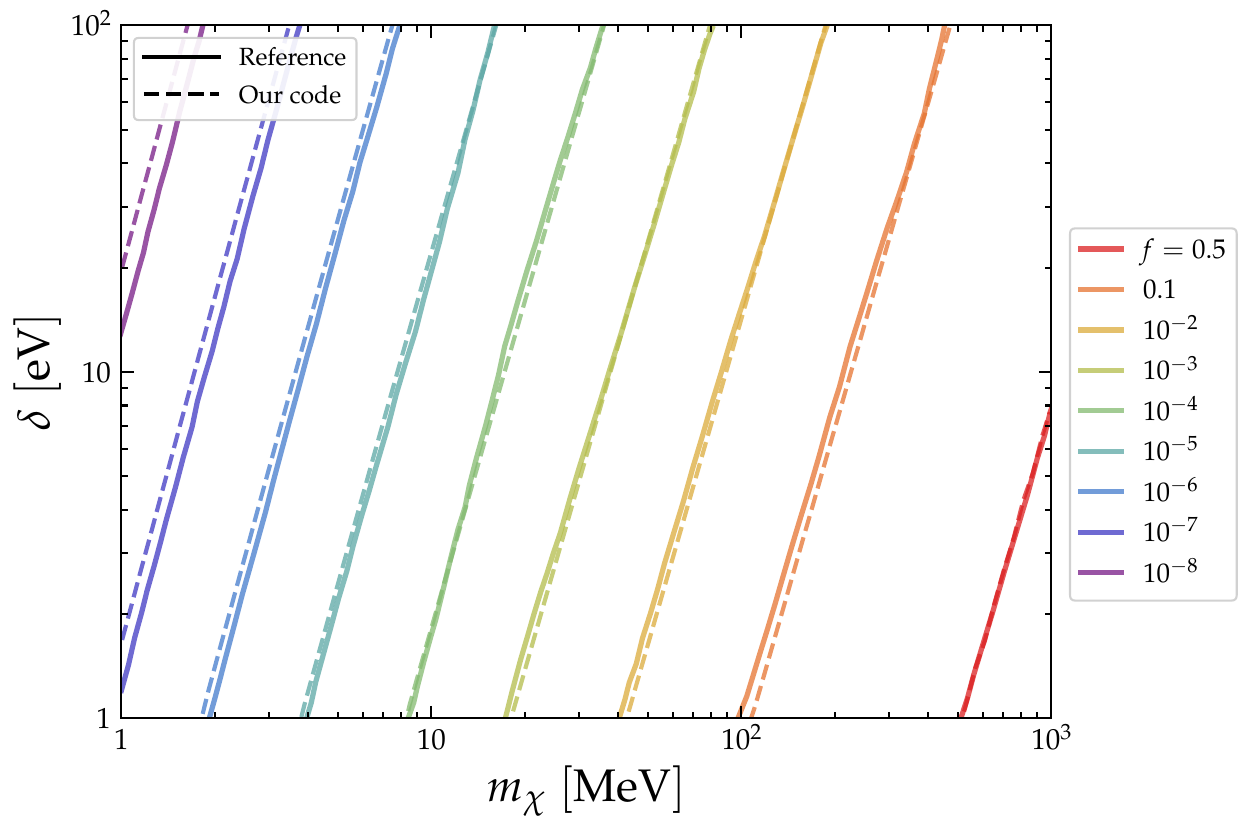}
  \caption{Comparison of predictions for the relative abundance $f$, contrasting our implicit discretization (``Our Code'') with digitized results from Ref.~\cite{Berlin:2023qco} (``Reference''). Shown is the purely inelastic case, $\delta_y=0$. The benchmark parameters are the same as in the referenced study, namely $m_A = 3 m_\chi$, $\alpha' = 0.5$, and a value of $\epsilon$ chosen to reproduce $\Omega_{\rm obs} h^2$.}
  \label{fig:test}
\end{figure}
We select the physical branch of the square root to ensure $f_{i+1}>0$. This implicit embedded pair provides both stiffness control and an internal error estimate for step adaptation. We implemented this scheme in our code and validated it against the benchmark of Ref.~\cite{Berlin:2023qco}. The comparison is shown in \cref{fig:test}. In the plot, the curve labeled ``Reference'' corresponds to data digitized from the published results of Ref.~\cite{Berlin:2023qco}, whereas the curve labeled ``Our Code'' is produced by our \texttt{C++} implementation of the implicit trapezoidal scheme with embedded backward–Euler error control described above. We use exactly the parameter choices reported in Ref.~\cite{Berlin:2023qco}. The two curves display consistent trends, indicating that our implementation captures the qualitative behavior reported in the reference.

\section{Dark sector temperature}  
\label{app:temperatureDS}

In this appendix we derive the evolution of the DS temperature relative to the SM bath temperature, under the assumption of \textit{entropy conservation}. We assume an \textit{instantaneous kinetic decoupling} between the two sectors and \textit{full thermal equilibrium} within the DS (i.e. both chemical and kinetic equilibrium).  This allows us to define two independent entropy densities, each associated with its own temperature.

We begin by recalling some general thermodynamic relations relevant to the early Universe. The contribution of a non-relativistic particle species $i$ to the total entropy density is given by~\cite{Pereira:2024vdk}
\begin{equation}
    s_i =  \sum_i \left(\frac{5}{2} +\frac{m_i-\mu_i}{T_i} \right)n_i \,,
\end{equation}
where $m_i$ is the particle mass, $\mu_i$ its chemical potential, $T_i$ its temperature, and $n_i = e^{\mu_i/T_i} \,n_i^{(\rm eq)}$ its number density, with
\begin{equation}
    n_{i}^{(\rm eq)} = g_i \left( \frac{m_i T_i}{2\pi} \right)^{3/2} e^{-m_i/T_i} \, ,
\end{equation}
and $g_i$ the number of internal degrees of freedom. The chemical potential can then be expressed as
\begin{equation}
    \mu_i = T_i \left( \ln{n_i} - \ln{n_i^{(\rm eq)}}\right) ,
\end{equation}
which quantifies the deviation from the  $\mu_i=0$ case---the typical situation for DM in chemical equilibrium with the SM.

We further define the yield as
\begin{equation}
\label{eq:YieldGeneralMu}
    Y_i = \frac{n_i}{s_{\rm SM}} = \frac{45}{2\pi^2} \frac{g_i}{g_{\ast s}}
    \left( \frac{m_i }{2\pi} \frac{T_i}{T_{\rm SM}^2}\right)^{3/2} 
    e^{-(m_i -\mu_i)/T_i} \, ,
\end{equation}
where $g_{\ast s}$ is the effective number of SM relativistic degrees of freedom for the entropy.  
In terms of the yield, the chemical potential becomes
\begin{equation} \label{eq:chemicalPotentialWithY}
    \mu_i = T_i \left( \ln{Y_i} - \ln{Y_i^{(\rm eq)}}\right) \, ,
\end{equation}
where $Y_i^{(\rm eq)} = Y_i(\mu_i=0)$. The logarithm of the equilibrium yield reads
\begin{equation} \label{eq:LNofYeq}
    \ln{Y_i^{(\rm eq)}} = 
    \ln{\left( \frac{45}{2\pi^2} \frac{g_i}{g_{\ast s}}\right)}
    + \frac{3}{2}\ln{\left( \frac{m_i }{2\pi}\right)}
    + \frac{3}{2}\ln{\left( \frac{T_i}{T_{\rm SM}^2}\right)}   
    - \frac{m_i}{T_i}
    \equiv C_i + \frac{3}{2}\ln{\left( \frac{T_i}{T_{\rm SM}^2}\right)} - \frac{m_i}{T_i} \,,
\end{equation}
where all temperature-independent terms have been absorbed into the constant $C_i$.

In the following, we briefly discuss the validity of our main assumptions. We then derive the temperature evolution for a single-component DM candidate and subsequently generalize the discussion to inelastic DM and to a multi-component DS.

\subsection{Kinetic equilibrium assumption}

Before proceeding with the derivations, it is useful to comment on the validity of our assumptions.
First, entropy conservation is a standard approximation in early-Universe thermodynamics and is widely adopted in the literature (see e.g.~Refs.~\cite{Pereira:2024vdk,Aoki:2022tek}).

The assumption of instantaneous kinetic decoupling from the SM plasma is, of course, an idealization of a more gradual process. However, since the precise behavior of the transition around the kinetic decoupling temperature $T_{\rm kin}$ is expected to have a negligible impact on the conversion freeze-out at $T_{f^\ast}$ (as long as $T_{f^\ast} \ll T_{\rm kin}$), this simplification is well justified for the parameter space relevant to this work.

Finally, the assumption of kinetic equilibrium within the DS down to $T_{f^*}$ requires more care.
In contrast to WIMP scenarios, where scattering off relativistic SM particles efficiently maintains equilibrium, here the non-relativistic DS particles must scatter among themselves to redistribute momentum.
Such processes are typically less efficient, with only a mild enhancement compared to DS number-changing processes due to the larger abundance of the lightest dark species, roughly $\sim e^{\delta_i/T}$,  where $\delta_i = m_i - m_\chi$ is the mass splitting between the species $\chi_i$ and the lightest state $\chi$, and $T$ is the DS temperature. 

Indeed, because a relevant amount of momentum exchange is required (rather than simply counting the total number of collisions, as in chemical equilibrium), the momentum-transfer rate $\gamma(T)$—e.g.\ from processes such as $\chi_i \chi \to \chi\, \chi_i$—is generally smaller than the chemical equilibrium rate $\Gamma(T)$—typically from reactions like $\chi_i \chi_i \to \chi \chi$.  
Their ratio scales approximately as~\cite{Feng:2010zp}
\begin{equation}
    \frac{\gamma}{\Gamma} \sim e^{\delta_i/T}\frac{T}{m_i} \, .
\end{equation} 
Therefore, results at very low temperatures $T \ll m_i$ should be interpreted with some caution.  
This is particularly relevant for small mass splittings $\delta_i \ll m_i$, since the partial reheating (due to the conversion of $\delta$ into kinetic energy) becomes significant only below $T \sim \delta_i$.  
Note also that $\Gamma(T)$ rapidly decreases for $T \lesssim \delta_i$ due to the Boltzmann suppression of the heavier states, implying that their chemical decoupling typically occur shortly thereafter. Thus, in this regime, one expects a competition between the factors $e^{\delta_i/T}$ and $T/m_i$, making a general treatment nontrivial.  A detailed study of this interplay will be explored in future work.

 After kinetic decoupling, the momenta of the heavier states simply redshift with the expansion. If the light species remain in internal kinetic equilibrium---so that a single DS temperature can still be defined---the heavier states continue to transfer energy to them, still leading to a partial reheating. If not, the light states retain a similarly redshifted momentum distribution, potentially modified by small additional kicks from conversions or from the decays of the heavier states. In either case, a detailed treatment of kinetic decoupling within the DS lies beyond the scope of this work. 

Finally, in our numerical results, we find that the resulting temperature variations have only a minor impact on the excited-state fraction $f$. For this reason, we justify no further analysis on the DS kinetic decoupling in this work, though such effects could become relevant in models where temperature changes are more pronounced.

\subsection{Single-component dark matter }

In the following, we consider a dark bath consisting of a single DM species with mass $m$.\footnote{Since only one species is present, we omit the index $i$ for all DS quantities.}  
After kinetic decoupling, the DS no longer exchanges entropy with the SM, and the ratio of their entropy densities is a constant $c$,\footnote{Throughout this appendix, $c$, $c'$, $c''$, and $c'''$ denote constants independent of temperature; primes are used to distinguish successive algebraic redefinitions.}
\begin{equation}
      \frac{s}{s_{\rm SM}} = \left(\frac{5}{2} + \frac{m - \mu}{T}\right) Y =c\, .
\end{equation}
As the DM number density is conserved after decoupling from the SM, the yield $Y$ remains constant,\footnote{Note that for cannibal DM~\cite{Pappadopulo:2016pkp}, $Y$ may still evolve even after decoupling through number-changing interactions such as $3 \leftrightarrow 2$ processes. These cannibal processes convert rest mass into kinetic energy and thereby delay the cooling of the DS~\cite{Berlin:2016gtr}; in that case, $\mu=0$ throughout the cannibal phase.} and we can write
\begin{equation}
    \frac{m - \mu}{T} = c' \, .
\end{equation} 
Using  \cref{eq:chemicalPotentialWithY}, we obtain
\begin{equation}
    \frac{m}{T} - \left(\ln{Y} - \ln{Y^{(\rm eq)}}\right) = c'
    \quad \Rightarrow \quad 
    \frac{m}{T} + \ln{Y^{(\rm eq)}} = c'' \, .
\end{equation}
Substituting \cref{eq:LNofYeq} then yields
\begin{equation}
     C + \frac{3}{2}\ln{\left( \frac{T}{T_{\rm SM}^2}\right)} = c'' 
     \quad \Rightarrow \quad
     \frac{T}{T_{\rm SM}^2} = c''' \, .
\end{equation}
Since the two sectors shared a common temperature at kinetic decoupling, $T = T_{\rm SM} = T_{\rm kin}$, we have $c''' = 1/T_{\rm kin}$. Therefore, the DS temperature evolves as
\begin{equation}
    T = \frac{T_{\rm SM}^2}{T_{\rm kin}} \, ,
\end{equation}
which is the expected redshift of non-relativistic DM momenta.

\subsection{Inelastic dark matter}

We now extend this reasoning to a two-component inelastic DM system with a mass splitting $\delta$.  
Assuming both components remain in thermal equilibrium among themselves with a common chemical potential $\mu_\chi = \mu_{\chi^\ast} \equiv \mu$,\footnote{This is ensured by processes such as $\chi^\ast\chi^\ast \leftrightarrow \chi\chi$ and $\chi^\ast\chi \leftrightarrow \chi\chi$, which are efficient in the model considered in this work for temperatures right below $T_{\rm kin}$. } and that the DS is completely decoupled from the SM after $T_{\rm kin}$, the ratio of entropies becomes
\begin{equation}
    \frac{s}{s_{\rm SM}} = 
    \left(\frac{5}{2} + \frac{m_\chi - \mu}{T}\right) Y_\chi
    + \left(\frac{5}{2} + \frac{m_\chi + \delta - \mu}{T}\right) Y_{\chi^\ast} = c \, ,
\end{equation} where we have neglected subdominant DS contributions (e.g.\ from the heavy dark photon, which is Boltzmann suppressed and  also decays rapidly). 

Introducing the excited-state fraction $f$ defined in \cref{eq:FratioDEF}, we can rewrite the ratio as
\begin{equation}
    \left[\left(\frac{5}{2} + \frac{m_\chi - \mu}{T}\right)(1-f)
    + \left(\frac{5}{2} + \frac{m_\chi + \delta - \mu}{T}\right)f \right] Y_{\rm DM} = c \, ,
\end{equation}
where $Y_{\rm DM} = Y_\chi + Y_{\chi^\ast}$ is the total DM yield.  
For a non-cannibal DS, $Y_{\rm DM}$ is constant, leading to
\begin{equation}
   \frac{m_\chi - \mu + f \delta}{T} = c' \, .
\end{equation}
Using \cref{eq:chemicalPotentialWithY}, we then find 
\begin{equation}
    \frac{m_\chi + f \delta}{T} - 
    \ln{\!\left(\frac{(1-f)Y_{\rm DM}}{Y_\chi^{(\rm eq)}}\right)} = c' 
    \quad \Rightarrow \quad 
    \frac{m_\chi + f \delta}{T} + \ln{Y_\chi^{(\rm eq)}} - \ln{(1-f)} = c'' \, .
\end{equation} 
Substituting \cref{eq:LNofYeq}, we obtain
\begin{equation}\label{eq:temperatureCaseIDM}
    f \frac{\delta}{T} + C_\chi + \frac{3}{2}\ln{\!\left( \frac{T}{T_{\rm SM}^2}\right)} - \ln{(1-f)} = c''
    \quad \Rightarrow \quad
    \left(\frac{e^{f\delta/T}}{1-f}\right)^{2/3} \frac{T}{T_{\rm SM}^2} = c''' \, .
\end{equation}

Given thermal equilibrium (justified since we aim to compute the temperature $T_{f^\ast}$ at which conversions freeze out), 
the excited-state fraction can be written as~\cite{DallaValleGarcia:2024zva}
\begin{equation}\label{eq:fAsChemicalEquilibrium}
    f = \frac{\xi}{1+\xi} \, , \qquad 
    \xi = \left(1+\frac{\delta}{m_\chi}\right)^{3/2} e^{-\delta/T} \, ,
\end{equation} where we have used \cref{eq:YieldGeneralMu}.
With this expression and the initial condition $T=T_{\rm SM}$ at $T_{\rm kin}$, \cref{eq:temperatureCaseIDM} can be solved to obtain $T(T_{\rm SM},T_{\rm kin})$. 
Although we do not find a closed-form solution to $T(T_{\rm SM},T_{\rm kin})$, important insights can still be gained by studying two limiting regimes.  

At high temperatures, right after kinetic decoupling ($\delta\ll T\lesssim T_{\rm kin}$), the excited and ground states   remain nearly equally populated $f \simeq 1/2$, see \cref{eq:fAsChemicalEquilibrium}. Imposing the boundary condition $T = T_{\rm SM} = T_{\rm kin}$ then fixes
\begin{equation}\label{eq:cprime3value}
    c''' \simeq \frac{2^{2/3}}{T_{\rm kin}} \, ,
\end{equation}
leading to $T = T_{\rm SM}^2 / T_{\rm kin}$, identical to the single-component case. 
This behavior is expected, since for $T \gg \delta$ the energy injection from mass conversions between states is negligible. 

At low temperatures ($T \ll \delta$), the excited-state abundance becomes exponentially suppressed ($f \simeq 0$), and combining \cref{eq:temperatureCaseIDM} with \cref{eq:cprime3value}  yields
\begin{equation}
\label{eq:Scaling.1.6}
    T \simeq 2^{2/3} \frac{T_{\rm SM}^2}{T_{\rm kin}} \approx 1.6 \frac{T_{\rm SM}^2}{T_{\rm kin}} \, ,
\end{equation}
corresponding to a $\sim 60\%$ temperature increase, originating from the conversion of $\chi^\ast$ mass energy into kinetic energy of the dark bath. 
The scaling $T \propto T_{\rm SM}^2$ persists, as expected from momentum redshift, since the $\chi^\ast$ population has effectively disappeared.

Note that if for some initial temperature $T_I$ (not necessarily identified with $T_{\rm kin}$) the  DS was already decoupled  from the SM bath and the excited-state fraction  were fixed at $f_I=\mathcal{O}(1)$ by an external mechanism (for instance, through fast decays of a third particle), the DS  temperature would instead strongly increase with the mass splitting $\delta$.  This is due to the exponential increase of $c'''$ at $T_I$ which later, for $f\simeq0$, leads to an  increased temperature $T\propto e^{2f_I\delta/3T_I}T_{\rm SM}^2$.
This is not the case in the present work where, under chemical equilibrium at $T_I = T_{\rm kin}$, the excited-state population follows the Boltzmann suppression given in \cref{eq:fAsChemicalEquilibrium}. 

\begin{figure}
  \centering
  \includegraphics[width=.7\textwidth]{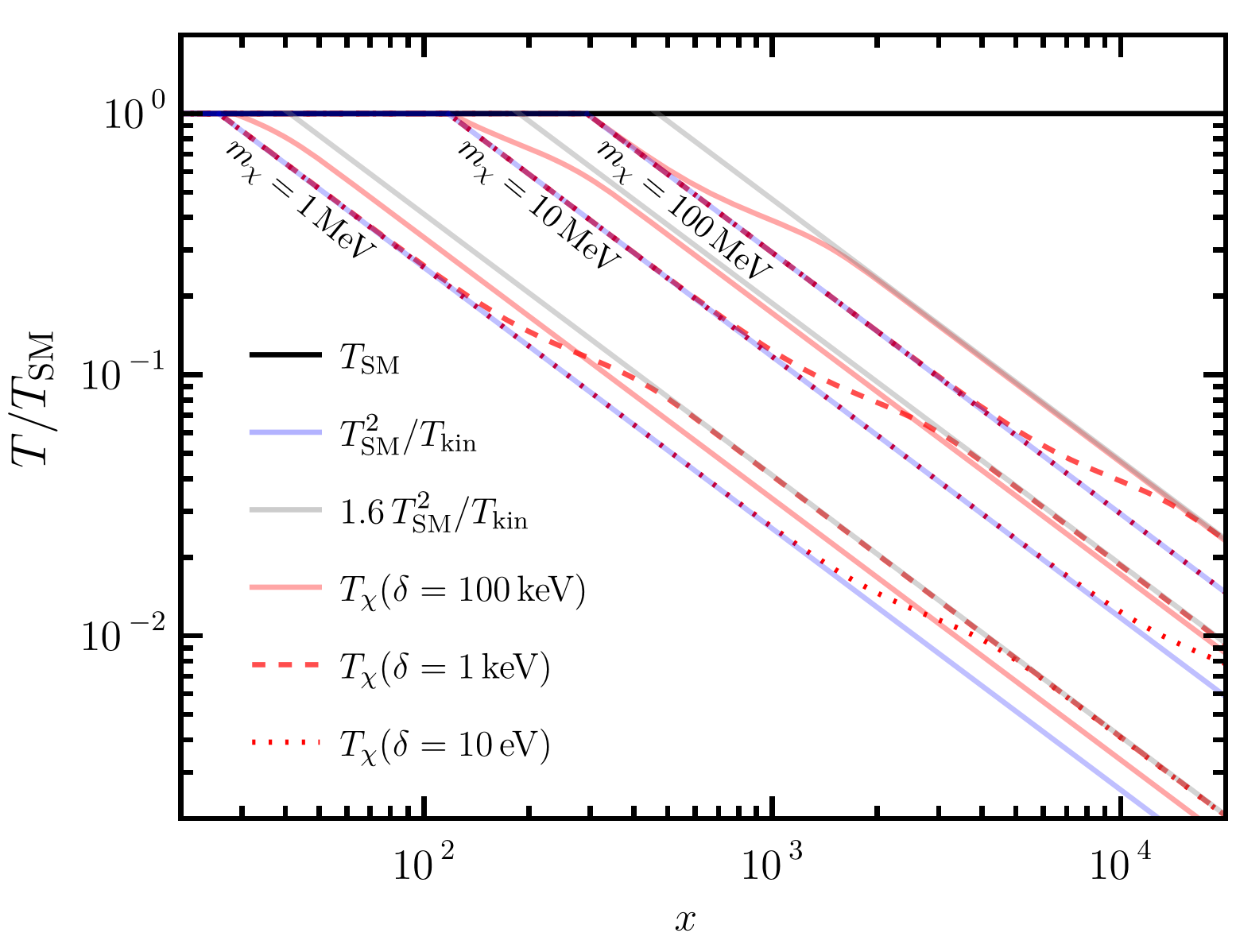}
  \caption{Evolution of the DS temperature $T_\chi$ as a function of $x=m_\chi/T_{\rm SM}$ computed with our prescription for several values of $\delta$ and $m_\chi$. The results are compared with the usual approximation $T_\chi = T_{\rm SM}^{2}/T_{\rm kin}$ found in the literature, as well as with the limiting case $T_\chi = 1.6\,T_{\rm SM}^{2}/T_{\rm kin}$ discussed in the main text.  }
  \label{fig:TxEvol}
\end{figure}

Finally, to capture the full behavior, we solve \cref{eq:temperatureCaseIDM} numerically with the initial condition $T = T_{\rm SM}$ at $T_{\rm kin}$. In \cref{fig:TxEvol}, we show in red the DS temperature evolution  (normalized to the SM one)  for several representative cases with $m_\chi = 1,\ 10,\ 100~\mathrm{MeV}$  (left, middle, right) and $\delta = 10~\mathrm{eV},\ 1~\mathrm{keV},\ 100~\mathrm{keV}$  (dotted, dashed, solid), compared with the analytic approximations $T = T_{\rm SM}^2 / T_{\rm kin}$  (blue) and  $T = 1.6 \,T_{\rm SM}^2 / T_{\rm kin}$  (gray).  In all cases we neglect a possible kinetic decoupling between DM states. For the red curves, $T_{\rm kin}$ is determined by the scattering cross section of $\chi^\ast e^- \leftrightarrow \chi\, e^-$, which is fixed by requiring the correct relic abundance to be reproduced via visible freeze-out for $m_{A^{\prime}} = 3 m_\chi$. For the blue and gray lines, we set $T_{\rm kin}$ equal to the corresponding values for the red $\delta=10$~eV curves.

We first focus on the case with relatively small mass splitting, i.e.\ the dotted and dashed red curves, where we find a perfect match with our expectations at high and low temperatures. We see that, the smaller $\delta$, the lower the temperature where deviations from the $T^2_{\rm SM}$ scaling appear. The reason for this is that, for $T \gg \delta$, the injected kinetic energy is negligible compared to the kinetic energy of the DM particles ($\sim T$). 

If we now turn to a larger mass splitting (solid red curves), we find that for $m_{\chi}=1$ and $10~\text{MeV}$ the kinetic decoupling temperature satisfies $T_{\rm kin} \lesssim \delta$. In this regime, the excited-state abundance is already Boltzmann suppressed at kinetic decoupling ($f \ll 1$), which modifies the coefficient $c'''$ in \cref{eq:cprime3value} and consequently alters the factor $1.6$ in \cref{eq:Scaling.1.6}. 
Furthermore, the suppression of the heavier state requires a larger kinetic mixing $\epsilon$ to reproduce the observed relic abundance, which in turn delays kinetic decoupling, leading to a smaller $T_{\rm kin}$. 
Therefore, for these specific cases, the red curves are not expected to coincide with the gray ones—as is indeed observed.

\subsection{$(N+1)$-component dark matter}

We finally generalize our study to a DS composed of $N$ heavier flavors $\chi_i$ in addition to a ground state $\chi$, with mass splittings $\delta_i = m_i - m_\chi$.  
Assuming that efficient conversions and scatterings maintain a common temperature and chemical potential $\mu$ (and that cannibalism processes are negligible),  conservation of entropy  implies  
\begin{equation}
     \left(\frac{e^{f_i \delta_i / T}}{1-f}\right)^{2/3} \frac{T}{T_{\rm SM}^2} = c \, ,
\end{equation}
where $f_i = n_i / n_{\rm DM}$ is the fractional abundance of the heavier state~$i$ and 
\begin{equation}
    f = \sum_{i=1}^{N} f_i
\end{equation}
denotes the total excited-state fraction. 
For compactness, we adopt the Einstein summation convention for repeated indices, 
i.e.\ $f_i \delta_i \equiv \sum_{i=1}^{N} f_i \delta_i$.

In thermal equilibrium (where each $f_i$ is given by \cref{eq:fAsChemicalEquilibrium} with $\delta \to \delta_i$), the constant $c$ can be expressed as
\begin{equation}
    c \simeq \frac{k^{2/3}}{T_{\rm kin}} \, ,
\end{equation}
where $k \leq N$ denotes the number of light states satisfying $T_{\rm kin} \gg \delta_i$. We neglect the smaller contribution from the $N-k$ heavier states with $\delta_i \gtrsim T_{\rm kin}$ since their abundances are strongly Boltzmann suppressed.  
At low temperatures $T \ll \delta_i$, the DS temperature can therefore reach
\begin{equation}
    \frac{T}{T_z} \sim k^{2/3} \, ,
\end{equation}
where $T_z = T_{\rm SM}^{2}/T_{\rm kin}$ is the standard redshifted single-component temperature. For $k=2$, we recover the $60\,\%$ increase, see \cref{eq:Scaling.1.6}), of the two-state scenario studied in this work.  This shows that mass-splitting–induced reheating can become particularly significant for a DS containing a large number of light states ($k \gg 1$).

\section{Constraints and sensitivities from invisible $Z$ boson decays} \label{app:Zinv}

Even though in this work we focus entirely on DM--SM interactions mediated by dark photons, our model also allows DM to couple directly to $Z$ bosons. In the following, we present the corresponding interaction terms, and calculate the partial widths for invisible $Z\to\chi\chi,\ \chi\chi^{*},\, \chi^{*}\chi^{*}$ decays.

The interaction terms for $\chi$ and $\chi^*$ stem from the original covariant derivative, which can be written as:
\begin{eqnarray}
e'\, A_{D\mu}\left(\bar\chi_L\gamma^\mu\chi_L-\bar\chi^c_R\gamma^\mu\chi^c_R\right) &=& 
-\frac{e'}{2}\cos2\theta\,\bar{\chi}\gamma^\mu\gamma^5\chi\,A_{D\mu} +\frac{e'}{2}\cos2\theta\,\bar{\chi}^*\gamma^\mu\gamma^5\chi^*\,A_{D\mu} \nonumber \\
 && +i\,\frac{e'}{2}\sin2\theta\,\bar{\chi}\gamma^\mu\,\chi^* A_{D\mu}
 -i\,\frac{e'}{2}\sin2\theta\,\bar{\chi}^*\gamma^\mu\,\chi\,A_{D\mu}~,
\end{eqnarray}
where $A_D^\mu$ is the dark photon on the interaction basis and the angle $\theta$ is defined in \cref{eq:cos2theta}. Note that the $\chi^{(*)}$ fermions on the right-hand side are Majorana particles.

Writing $A_D^\mu$ on its mass basis, a coupling with the $Z$ boson arises due to the kinetic mixing $\epsilon$. To a very good approximation, we find the following interaction terms between DM and the $Z$:
\begin{multline}
\mathcal L \supset \frac{e'}{2}\frac{m_{\hat Z}^2}{m_{\hat Z}^2-m_{\hat A}^2}\tan\theta_w\,\epsilon\left[-\cos2\theta\,\bar\chi\gamma^\mu\gamma^5\chi
-\cos2\theta\,\bar\chi^*\gamma^\mu\gamma^5\chi^* \right. \\
\left.+i\sin2\theta\,\bar\chi\gamma^\mu\chi^*
-i\sin2\theta\,\bar\chi^*\gamma^\mu\chi\right]Z_\mu~,
\end{multline}
where $m_{\hat Z}=\sqrt{g^2+g^{\prime2}}\,v_H/2$ and $m_{\hat A}= 2\,e'v_\Phi$, with $v_H$ and $v_\Phi$ being the vacuum expectation values of the SM and dark Higgs bosons, respectively. With these couplings, and considering $m_{\chi^{(*)}},\,m_{A'}\ll m_Z$, one can obtain:
\begin{eqnarray}
\Gamma(Z\to \chi\chi)=\Gamma(Z\to \chi^*\chi^*)&=&\frac{e^{\prime\,2}}{24\pi}\cos^22\theta\,\tan^2\theta_w\,\epsilon^2\,m_Z \\
\Gamma(Z\to \chi\chi^*)&=&\frac{e^{\prime\,2}}{12\pi}\sin^22\theta\,\tan^2\theta_w\,\epsilon^2\,m_Z
\end{eqnarray}
This leads to a contribution to the width for $Z$ invisible decay
\begin{equation}
\Delta\Gamma(Z\to {\rm inv})=(e^{\prime\,2}/12\pi)\tan^2\theta_w\,\epsilon^2 \,m_Z\,,
\end{equation}
which is not sensitive to parity violation.

In order to place limits from the $Z\to{\rm invisible}$ measurement, we use Table~D.1 of Ref.~\cite{ALEPH:2005ab}, which shows the most constraining bounds reported by PDG (see also Refs.~\cite{L3:1998uub,CMS:2022ett,ATLAS:2023ynf}). Here we have $\Delta\Gamma(Z\to {\rm inv})<2.0~$MeV, at 95\% C.L., in other words:
\begin{equation}
 \epsilon<\sqrt{\frac{6.0\times10^{-3}}{\alpha_D\,\tan^2\theta_w\,m_Z[\rm GeV]}}
\end{equation}
Taking $m_Z=91$~GeV, $\tan^2\theta_w=0.3$ and $\alpha_D=0.5$, we find $\epsilon<0.021$, which is much larger than the maximum value of $\epsilon$ used in our work.

Regarding prospects, the FCC-ee CDR~\cite{FCC:2018byv} expresses the $Z$ invisible width by the number of neutrinos $N_\nu$. This was measured by LEP to be $N_\nu=2.984\pm0.008$~\cite{ALEPH:2005ab}. The report in Section~3.2.3 of Ref.~\cite{FCC:2018byv} considers the possibility of improving precision to a factor $0.0004$. In the following, we will provide a rough estimate of what this means.

The number of neutrinos at LEP is derived from the comparison between the SM expectation of the ratio $\Gamma_{\nu\bar\nu}/\Gamma_{\ell\ell}$ (the partial width for decays into a single neutrino species over the corresponding width for one generation of massless charged leptons) and its measured value $R^0_{\rm inv}$:
\begin{equation}
 R^0_{\rm inv}=N_\nu\left(\frac{\Gamma_{\nu\bar\nu}}{\Gamma_{\ell\ell}}\right)_{\rm SM}
\end{equation}
If we now assume an extra contribution to the $Z$ invisible width, we will have:
\begin{equation}
 R^0_{\rm inv}=\left(\frac{3\Gamma_{\nu\bar\nu}+\Delta\Gamma}{\Gamma_{\ell\ell}}\right)
 =\frac{3\Gamma_{\nu\bar\nu}+\Delta\Gamma}{\Gamma_{\nu\bar\nu}}\left(\frac{\Gamma_{\nu\bar\nu}}{\Gamma_{\ell\ell}}\right)_{\rm SM}
\end{equation}
Thus, if we write $N_\nu=3+\Delta\Gamma/\Gamma_{\nu\bar\nu}$, we can expect a bound $\Delta\Gamma/\Gamma_{\nu\bar\nu}\lesssim2\times0.0004$.

The SM prediction for the partial width into one neutrino pair is:
\begin{equation}
 \Gamma_{\nu\bar\nu}=\frac{G_F\, m_Z^3}{12\sqrt2\pi}=0.166~{\rm GeV}\,,
\end{equation}
so this means:
\begin{equation}
 \Delta\Gamma\lesssim 0.133~{\rm MeV}\,.
\end{equation}
With this rough analysis, we see that the constraint on the invisible width could improve by one order of magnitude, which would imply $\epsilon\lesssim5\times10^{-3}$.

Although the numerical values obtained here are not directly relevant for the parameter space analyzed in this work, the resulting limits are entirely independent of parity-violating effects and therefore apply straightforwardly to the large–mass-splitting and heavier DS scenarios studied in Ref.~\cite{Garcia:2024uwf}. In particular, the thermal target around $m_\chi \simeq 5~\mathrm{GeV}$ remains inaccessible to upcoming searches at $B$-factories and to DD experiments, as discussed in detail in Ref.~\cite{Garcia:2024uwf}. By contrast, our analysis of invisible $Z$ decays has the potential to close this gap: the expected FCC-ee sensitivity to the invisible width would allow a full exploration of the thermal target for not-so-inelastic DM in the regime of sizable mass splittings, $\delta/m_\chi \gtrsim 0.1$, which continues to evade all previously studied probes.

\bibliographystyle{JHEP}

\bibliography{biblio}

@article{LDMX:2018cma,
    author = "{\r{A}}kesson, Torsten and others",
    collaboration = "LDMX",
    title = "{Light Dark Matter eXperiment (LDMX)}",
    eprint = "1808.05219",
    archivePrefix = "arXiv",
    primaryClass = "hep-ex",
    reportNumber = "FERMILAB-PUB-18-324-A, SLAC-PUB-17303",
    month = "8",
    year = "2018"
}

@phdthesis{Leerdam:2024jxn,
    author = "Leerdam, Nicholas Dean",
    title = "{Dark Matter in Beyond the Standard Model Physics}",
    school = "Adelaide U.",
    year = "2024"
}

@article{Garcia:2024uwf,
    author = "Garcia, Giovani Dalla Valle and Kahlhoefer, Felix and Ovchynnikov, Maksym and Schwetz, Thomas",
    title = "{Not-so-inelastic Dark Matter}",
    eprint = "2405.08081",
    archivePrefix = "arXiv",
    primaryClass = "hep-ph",
    reportNumber = "P3H-24-028, TTP24-011",
    doi = "10.1007/JHEP02(2025)127",
    journal = "JHEP",
    volume = "02",
    pages = "127",
    year = "2025"
}

@article{Bloch:2020uzh,
	Archiveprefix = {arXiv},
	Author = {Bloch, Itay M. and Caputo, Andrea and Essig, Rouven and Redigolo, Diego and Sholapurkar, Mukul and Volansky, Tomer},
	Eprint = {2006.14521},
	Month = {6},
	Primaryclass = {hep-ph},
	Title = {{Exploring New Physics with O(keV) Electron Recoils in Direct Detection Experiments}},
	Year = {2020}}

@article{Bramante:2020zos,
	Archiveprefix = {arXiv},
	Author = {Bramante, Joseph and Song, Ningqiang},
	Eprint = {2006.14089},
	Month = {6},
	Primaryclass = {hep-ph},
	Title = {{Electric But Not Eclectic: Thermal Relic Dark Matter for the XENON1T Excess}},
	Year = {2020}}

@article{Harigaya:2020ckz,
	Archiveprefix = {arXiv},
	Author = {Harigaya, Keisuke and Nakai, Yuichiro and Suzuki, Motoo},
	Eprint = {2006.11938},
	Month = {6},
	Primaryclass = {hep-ph},
	Title = {{Inelastic Dark Matter Electron Scattering and the XENON1T Excess}},
	Year = {2020}}

@article{Aprile:2020tmw,
	Archiveprefix = {arXiv},
	Author = {Aprile, E. and others},
	Collaboration = {XENON},
	Eprint = {2006.09721},
	Month = {6},
	Primaryclass = {hep-ex},
	Title = {{Observation of Excess Electronic Recoil Events in XENON1T}},
	Year = {2020}}

@article{Batell:2009vb,
	Archiveprefix = {arXiv},
	Author = {Batell, Brian and Pospelov, Maxim and Ritz, Adam},
	Doi = {10.1103/PhysRevD.79.115019},
	Eprint = {0903.3396},
	Journal = {Phys. Rev. D},
	Pages = {115019},
	Primaryclass = {hep-ph},
	Title = {{Direct Detection of Multi-component Secluded WIMPs}},
	Volume = {79},
	Year = {2009}}

@article{Blennow:2015hzp,
	Archiveprefix = {arXiv},
	Author = {Blennow, Mattias and Clementz, Stefan and Herrero-Garcia, Juan},
	Doi = {10.1088/1475-7516/2016/04/004},
	Eprint = {1512.03317},
	Journal = {JCAP},
	Number = {04},
	Pages = {004},
	Primaryclass = {hep-ph},
	Slaccitation = {%%CITATION = ARXIV:1512.03317;%%},
	Title = {{Pinning down inelastic dark matter in the Sun and in direct detection}},
	Volume = {1604},
	Year = {2016}}

@article{Chang:2010en,
	Archiveprefix = {arXiv},
	Author = {Chang, Spencer and Weiner, Neal and Yavin, Itay},
	Doi = {10.1103/PhysRevD.82.125011},
	Eprint = {1007.4200},
	Journal = {Phys. Rev.},
	Pages = {125011},
	Primaryclass = {hep-ph},
	Slaccitation = {%%CITATION = ARXIV:1007.4200;%%},
	Title = {{Magnetic Inelastic Dark Matter}},
	Volume = {D82},
	Year = {2010}}

@article{Cui:2009xq,
	Archiveprefix = {arXiv},
	Author = {Cui, Yanou and Morrissey, David E. and Poland, David and Randall, Lisa},
	Doi = {10.1088/1126-6708/2009/05/076},
	Eprint = {0901.0557},
	Journal = {JHEP},
	Pages = {076},
	Primaryclass = {hep-ph},
	Slaccitation = {%%CITATION = ARXIV:0901.0557;%%},
	Title = {{Candidates for Inelastic Dark Matter}},
	Volume = {05},
	Year = {2009}}

@article{Hall:1997ah,
	Archiveprefix = {arXiv},
	Author = {Hall, Lawrence J. and Moroi, Takeo and Murayama, Hitoshi},
	Doi = {10.1016/S0370-2693(98)00196-8},
	Eprint = {hep-ph/9712515},
	Journal = {Phys. Lett.},
	Pages = {305-312},
	Primaryclass = {hep-ph},
	Reportnumber = {LBNL-41199, LBL-41199, UCB-PTH-97-69},
	Slaccitation = {%%CITATION = HEP-PH/9712515;%%},
	Title = {{Sneutrino cold dark matter with lepton number violation}},
	Volume = {B424},
	Year = {1998}}

@article{TuckerSmith:2001hy,
	Archiveprefix = {arXiv},
	Author = {Tucker-Smith, David and Weiner, Neal},
	Doi = {10.1103/PhysRevD.64.043502},
	Eprint = {hep-ph/0101138},
	Journal = {Phys. Rev.},
	Pages = {043502},
	Primaryclass = {hep-ph},
	Reportnumber = {UCB-PTH-00-43, LBNL-47234, UW-PT-00-17},
	Slaccitation = {%%CITATION = HEP-PH/0101138;%%},
	Title = {{Inelastic dark matter}},
	Volume = {D64},
	Year = {2001}}

@article{Lee_2021,
title={Exothermic dark matter for XENON1T excess},
volume={2021},
ISSN={1029-8479},
url={http://dx.doi.org/10.1007/JHEP01(2021)019},
DOI={10.1007/jhep01(2021)019},
number={1},
journal={Journal of High Energy Physics},
publisher={Springer Science and Business Media LLC},
author={Lee, Hyun Min},
year={2021},
month={Jan}
}

@article{Bernabei:2010mq,
    author = "Bernabei, R. and others",
    collaboration = "DAMA, LIBRA",
    title = "{New results from DAMA/LIBRA}",
    eprint = "1002.1028",
    archivePrefix = "arXiv",
    primaryClass = "astro-ph.GA",
    doi = "10.1140/epjc/s10052-010-1303-9",
    journal = "Eur. Phys. J. C",
    volume = "67",
    pages = "39--49",
    year = "2010"
}

@article{TuckerSmith:2004jv,
    author = "Tucker-Smith, David and Weiner, Neal",
    title = "{The Status of inelastic dark matter}",
    eprint = "hep-ph/0402065",
    archivePrefix = "arXiv",
    reportNumber = "UW-PT-04-01",
    doi = "10.1103/PhysRevD.72.063509",
    journal = "Phys. Rev. D",
    volume = "72",
    pages = "063509",
    year = "2005"
}

@article{Bozorgnia:2013hsa,
    author = "Bozorgnia, Nassim and Herrero-Garcia, Juan and Schwetz, Thomas and Zupan, Jure",
    title = "{Halo-independent methods for inelastic dark matter scattering}",
    eprint = "1305.3575",
    archivePrefix = "arXiv",
    primaryClass = "hep-ph",
    doi = "10.1088/1475-7516/2013/07/049",
    journal = "JCAP",
    volume = "07",
    pages = "049",
    year = "2013"
}

@article{Graham:2010ca,
    author = "Graham, Peter W. and Harnik, Roni and Rajendran, Surjeet and Saraswat, Prashant",
    title = "{Exothermic Dark Matter}",
    eprint = "1004.0937",
    archivePrefix = "arXiv",
    primaryClass = "hep-ph",
    reportNumber = "FERMILAB-PUB-10-062-T, MIT-CTP-4140, SU-ITP-10-15",
    doi = "10.1103/PhysRevD.82.063512",
    journal = "Phys. Rev. D",
    volume = "82",
    pages = "063512",
    year = "2010"
}

@article{Frandsen:2014ima,
    author = "Frandsen, Mads T. and Shoemaker, Ian M.",
    title = "{Up-shot of inelastic down-scattering at CDMS-Si}",
    eprint = "1401.0624",
    archivePrefix = "arXiv",
    primaryClass = "hep-ph",
    reportNumber = "CP3-ORIGINS-2013-053, DIAS-2013-53",
    doi = "10.1103/PhysRevD.89.051701",
    journal = "Phys. Rev. D",
    volume = "89",
    number = "5",
    pages = "051701",
    year = "2014"
}

@article{Chen:2014tka,
    author = "Chen, Nan and Wang, Qing and Zhao, Wei and Lin, Shin-Ted and Yue, Qian and Li, Jin",
    title = "{Exothermic isospin-violating dark matter after SuperCDMS and CDEX}",
    eprint = "1404.6043",
    archivePrefix = "arXiv",
    primaryClass = "hep-ph",
    doi = "10.1016/j.physletb.2015.02.043",
    journal = "Phys. Lett. B",
    volume = "743",
    pages = "205--212",
    year = "2015"
}

@article{Binder:2021bmg,
    author = "Binder, Tobias and Bringmann, Torsten and Gustafsson, Michael and Hryczuk, Andrzej",
    title = "{DRAKE: Dark matter Relic Abundance beyond Kinetic Equilibrium}",
    eprint = "2103.01944",
    archivePrefix = "arXiv",
    primaryClass = "hep-ph",
    doi = "10.1140/epjc/s10052-021-09357-5",
    journal = "Eur. Phys. J. C",
    volume = "81",
    pages = "577",
    year = "2021"
}

@article{Bringmann:2018lay,
    author = {Bringmann, Torsten and Edsj\"o, Joakim and Gondolo, Paolo and Ullio, Piero and Bergstr\"om, Lars},
    title = "{DarkSUSY 6 : An Advanced Tool to Compute Dark Matter Properties Numerically}",
    eprint = "1802.03399",
    archivePrefix = "arXiv",
    primaryClass = "hep-ph",
    doi = "10.1088/1475-7516/2018/07/033",
    journal = "JCAP",
    volume = "07",
    pages = "033",
    year = "2018"
}

@article{Planck:2018vyg,
    author = "Aghanim, N. and others",
    collaboration = "Planck",
    title = "{Planck 2018 results. VI. Cosmological parameters}",
    eprint = "1807.06209",
    archivePrefix = "arXiv",
    primaryClass = "astro-ph.CO",
    doi = "10.1051/0004-6361/201833910",
    journal = "Astron. Astrophys.",
    volume = "641",
    pages = "A6",
    year = "2020",
    note = "[Erratum: Astron.Astrophys. 652, C4 (2021)]"
}

@article{Arkani-Hamed:2008hhe,
    author = "Arkani-Hamed, Nima and Finkbeiner, Douglas P. and Slatyer, Tracy R. and Weiner, Neal",
    title = "{A Theory of Dark Matter}",
    eprint = "0810.0713",
    archivePrefix = "arXiv",
    primaryClass = "hep-ph",
    doi = "10.1103/PhysRevD.79.015014",
    journal = "Phys. Rev. D",
    volume = "79",
    pages = "015014",
    year = "2009"
}

@article{Chen:2009dm,
    author = "Chen, Fang and Cline, James M. and Frey, Andrew R.",
    title = "{A New twist on excited dark matter: Implications for INTEGRAL, PAMELA/ATIC/PPB-BETS, DAMA}",
    eprint = "0901.4327",
    archivePrefix = "arXiv",
    primaryClass = "hep-ph",
    doi = "10.1103/PhysRevD.79.063530",
    journal = "Phys. Rev. D",
    volume = "79",
    pages = "063530",
    year = "2009"
}

@article{Finkbeiner:2009mi,
    author = "Finkbeiner, Douglas P. and Slatyer, Tracy R. and Weiner, Neal and Yavin, Itay",
    title = "{PAMELA, DAMA, INTEGRAL and Signatures of Metastable Excited WIMPs}",
    eprint = "0903.1037",
    archivePrefix = "arXiv",
    primaryClass = "hep-ph",
    doi = "10.1088/1475-7516/2009/09/037",
    journal = "JCAP",
    volume = "09",
    pages = "037",
    year = "2009"
}

@article{Blennow:2018xwu,
    author = "Blennow, Mattias and Clementz, Stefan and Herrero-Garcia, Juan",
    title = "{The distribution of inelastic dark matter in the Sun}",
    eprint = "1802.06880",
    archivePrefix = "arXiv",
    primaryClass = "hep-ph",
    reportNumber = "IFT-UAM-CSIC-18-019, ADP-18-4-T1052, IFT-UAM/CSIC-18-019, ADP-18-4/T1052",
    doi = "10.1140/epjc/s10052-018-5863-4",
    journal = "Eur. Phys. J. C",
    volume = "78",
    number = "5",
    pages = "386",
    year = "2018",
    note = "[Erratum: Eur.Phys.J.C 79, 407 (2019)]"
}

@article{Blennow:2016gde,
    author = "Blennow, Mattias and Clementz, Stefan and Herrero-Garcia, Juan",
    title = "{Self-interacting inelastic dark matter: A viable solution to the small scale structure problems}",
    eprint = "1612.06681",
    archivePrefix = "arXiv",
    primaryClass = "hep-ph",
    reportNumber = "ADP-16-48-T1004",
    doi = "10.1088/1475-7516/2017/03/048",
    journal = "JCAP",
    volume = "03",
    pages = "048",
    year = "2017"
}

@article{Sabti:2019mhn,
    author = "Sabti, Nashwan and Alvey, James and Escudero, Miguel and Fairbairn, Malcolm and Blas, Diego",
    title = "{Refined Bounds on MeV-scale Thermal Dark Sectors from BBN and the CMB}",
    eprint = "1910.01649",
    archivePrefix = "arXiv",
    primaryClass = "hep-ph",
    reportNumber = "KCL-2019-75",
    doi = "10.1088/1475-7516/2020/01/004",
    journal = "JCAP",
    volume = "01",
    pages = "004",
    year = "2020"
}

@article{Berlin:2023qco,
    author = "Berlin, Asher and Krnjaic, Gordan and Pinetti, Elena",
    title = "{Reviving MeV-GeV indirect detection with inelastic dark matter}",
    eprint = "2311.00032",
    archivePrefix = "arXiv",
    primaryClass = "hep-ph",
    reportNumber = "FERMILAB-PUB-21-457-T",
    doi = "10.1103/PhysRevD.110.035015",
    journal = "Phys. Rev. D",
    volume = "110",
    number = "3",
    pages = "035015",
    year = "2024"
}

@article{Foguel:2024lca,
    author = "Foguel, Ana Luisa and Reimitz, Peter and Funchal, Renata Zukanovich",
    title = "{Unlocking the inelastic Dark Matter window with vector mediators}",
    eprint = "2410.00881",
    archivePrefix = "arXiv",
    primaryClass = "hep-ph",
    doi = "10.1007/JHEP05(2025)001",
    journal = "JHEP",
    volume = "05",
    pages = "001",
    year = "2025"
}

@article{Belanger:2020gnr,
    author = "Belanger, Genevieve and Mjallal, Ali and Pukhov, Alexander",
    title = "{Recasting direct detection limits within micrOMEGAs and implication for non-standard Dark Matter scenarios}",
    eprint = "2003.08621",
    archivePrefix = "arXiv",
    primaryClass = "hep-ph",
    doi = "10.1140/epjc/s10052-021-09012-z",
    journal = "Eur. Phys. J. C",
    volume = "81",
    number = "3",
    pages = "239",
    year = "2021"
}

@article{Izaguirre:2015zva,
    author = "Izaguirre, Eder and Krnjaic, Gordan and Shuve, Brian",
    title = "{Discovering Inelastic Thermal-Relic Dark Matter at Colliders}",
    eprint = "1508.03050",
    archivePrefix = "arXiv",
    primaryClass = "hep-ph",
    doi = "10.1103/PhysRevD.93.063523",
    journal = "Phys. Rev. D",
    volume = "93",
    number = "6",
    pages = "063523",
    year = "2016"
}

@article{Beacham:2019nyx,
    author = "Beacham, J. and others",
    title = "{Physics Beyond Colliders at CERN: Beyond the Standard Model Working Group Report}",
    eprint = "1901.09966",
    archivePrefix = "arXiv",
    primaryClass = "hep-ex",
    reportNumber = "CERN-PBC-REPORT-2018-007",
    doi = "10.1088/1361-6471/ab4cd2",
    journal = "J. Phys. G",
    volume = "47",
    number = "1",
    pages = "010501",
    year = "2020"
}

@article{Baryakhtar:2020rwy,
    author = "Baryakhtar, Masha and Berlin, Asher and Liu, Hongwan and Weiner, Neal",
    title = "{Electromagnetic signals of inelastic dark matter scattering}",
    eprint = "2006.13918",
    archivePrefix = "arXiv",
    primaryClass = "hep-ph",
    doi = "10.1007/JHEP06(2022)047",
    journal = "JHEP",
    volume = "06",
    pages = "047",
    year = "2022"
}

@article{CarrilloGonzalez:2021lxm,
    author = "Carrillo Gonz{\'a}lez, Mariana and Toro, Natalia",
    title = "{Cosmology and signals of light pseudo-Dirac dark matter}",
    eprint = "2108.13422",
    archivePrefix = "arXiv",
    primaryClass = "hep-ph",
    reportNumber = "Imperial/TP/2021/MC/03",
    doi = "10.1007/JHEP04(2022)060",
    journal = "JHEP",
    volume = "04",
    pages = "060",
    year = "2022"
}

@article{NA64:2023wbi,
    author = "Andreev, Yu. M. and others",
    collaboration = "NA64",
    title = "{Search for Light Dark Matter with NA64 at CERN}",
    eprint = "2307.02404",
    archivePrefix = "arXiv",
    primaryClass = "hep-ex",
    reportNumber = "CERN-EP-2023-130",
    doi = "10.1103/PhysRevLett.131.161801",
    journal = "Phys. Rev. Lett.",
    volume = "131",
    number = "16",
    pages = "161801",
    year = "2023"
}

@article{NA64:2025ddk,
    author = "Andreev, Yu. M. and others",
    collaboration = "NA64",
    title = "{Searching for Light Dark Matter and Dark Sectors with the NA64 experiment at the CERN SPS}",
    eprint = "2505.14291",
    archivePrefix = "arXiv",
    primaryClass = "hep-ex",
    month = "5",
    year = "2025"
}

@article{Berlin:2018pwi,
    author = "Berlin, Asher and Gori, Stefania and Schuster, Philip and Toro, Natalia",
    title = "{Dark Sectors at the Fermilab SeaQuest Experiment}",
    eprint = "1804.00661",
    archivePrefix = "arXiv",
    primaryClass = "hep-ph",
    reportNumber = "SLAC-PUB-17238",
    doi = "10.1103/PhysRevD.98.035011",
    journal = "Phys. Rev. D",
    volume = "98",
    number = "3",
    pages = "035011",
    year = "2018"
}

@article{LSND:2001akn,
    author = "Auerbach, L. B. and others",
    collaboration = "LSND",
    title = "{Measurement of electron - neutrino - electron elastic scattering}",
    eprint = "hep-ex/0101039",
    archivePrefix = "arXiv",
    doi = "10.1103/PhysRevD.63.112001",
    journal = "Phys. Rev. D",
    volume = "63",
    pages = "112001",
    year = "2001"
}

@article{deNiverville:2011it,
    author = "deNiverville, Patrick and Pospelov, Maxim and Ritz, Adam",
    title = "{Observing a light dark matter beam with neutrino experiments}",
    eprint = "1107.4580",
    archivePrefix = "arXiv",
    primaryClass = "hep-ph",
    doi = "10.1103/PhysRevD.84.075020",
    journal = "Phys. Rev. D",
    volume = "84",
    pages = "075020",
    year = "2011"
}

@article{Bjorken:1988as,
    author = "Bjorken, J. D. and Ecklund, S. and Nelson, W. R. and Abashian, A. and Church, C. and Lu, B. and Mo, L. W. and Nunamaker, T. A. and Rassmann, P.",
    title = "{Search for Neutral Metastable Penetrating Particles Produced in the SLAC Beam Dump}",
    reportNumber = "FERMILAB-PUB-88-044, PRINT-88-0352 (FERMILAB)",
    doi = "10.1103/PhysRevD.38.3375",
    journal = "Phys. Rev. D",
    volume = "38",
    pages = "3375",
    year = "1988"
}

@article{Batell:2014mga,
    author = "Batell, Brian and Essig, Rouven and Surujon, Ze'ev",
    title = "{Strong Constraints on Sub-GeV Dark Sectors from SLAC Beam Dump E137}",
    eprint = "1406.2698",
    archivePrefix = "arXiv",
    primaryClass = "hep-ph",
    doi = "10.1103/PhysRevLett.113.171802",
    journal = "Phys. Rev. Lett.",
    volume = "113",
    number = "17",
    pages = "171802",
    year = "2014"
}

@article{MiniBooNE:2017nqe,
    author = "Aguilar-Arevalo, A. A. and others",
    collaboration = "MiniBooNE",
    title = "{Dark Matter Search in a Proton Beam Dump with MiniBooNE}",
    eprint = "1702.02688",
    archivePrefix = "arXiv",
    primaryClass = "hep-ex",
    reportNumber = "FERMILAB-PUB-17-059-AD-E-ND",
    doi = "10.1103/PhysRevLett.118.221803",
    journal = "Phys. Rev. Lett.",
    volume = "118",
    number = "22",
    pages = "221803",
    year = "2017"
}

@article{Duerr:2019dmv,
    author = "Duerr, Michael and Ferber, Torben and Hearty, Christopher and Kahlhoefer, Felix and Schmidt-Hoberg, Kai and Tunney, Patrick",
    title = "{Invisible and displaced dark matter signatures at Belle II}",
    eprint = "1911.03176",
    archivePrefix = "arXiv",
    primaryClass = "hep-ph",
    reportNumber = "DESY-19-141, OUTP-19-10P, TTK-19-46",
    doi = "10.1007/JHEP02(2020)039",
    journal = "JHEP",
    volume = "02",
    pages = "039",
    year = "2020"
}

@article{Yang:2022zlh,
    author = "Yang, Kwei-Chou",
    title = "{Freeze-out forbidden dark matter in the hidden sector in the mass range from sub-GeV to TeV}",
    eprint = "2209.10827",
    archivePrefix = "arXiv",
    primaryClass = "hep-ph",
    reportNumber = "CYCU-HEP-22-08",
    doi = "10.1007/JHEP11(2022)083",
    journal = "JHEP",
    volume = "11",
    pages = "083",
    year = "2022"
}

@article{Bernal:2017mqb,
    author = "Bernal, Nicol{\'a}s and Chu, Xiaoyong and Pradler, Josef",
    title = "{Simply split strongly interacting massive particles}",
    eprint = "1702.04906",
    archivePrefix = "arXiv",
    primaryClass = "hep-ph",
    doi = "10.1103/PhysRevD.95.115023",
    journal = "Phys. Rev. D",
    volume = "95",
    number = "11",
    pages = "115023",
    year = "2017"
}

@article{McDermott:2017qcg,
    author = "McDermott, Samuel D. and Patel, Hiren H. and Ramani, Harikrishnan",
    title = "{Dark Photon Decay Beyond The Euler-Heisenberg Limit}",
    eprint = "1705.00619",
    archivePrefix = "arXiv",
    primaryClass = "hep-ph",
    reportNumber = "ACFI-T17-08, YITP-SB-17-14",
    doi = "10.1103/PhysRevD.97.073005",
    journal = "Phys. Rev. D",
    volume = "97",
    number = "7",
    pages = "073005",
    year = "2018"
}

@article{Duerr:2016tmh,
    author = "Duerr, Michael and Kahlhoefer, Felix and Schmidt-Hoberg, Kai and Schwetz, Thomas and Vogl, Stefan",
    title = "{How to save the WIMP: global analysis of a dark matter model with two s-channel mediators}",
    eprint = "1606.07609",
    archivePrefix = "arXiv",
    primaryClass = "hep-ph",
    reportNumber = "DESY-16-113",
    doi = "10.1007/JHEP09(2016)042",
    journal = "JHEP",
    volume = "09",
    pages = "042",
    year = "2016"
}

@article{Emken:2021vmf,
    author = "Emken, Timon and Frerick, Jonas and Heeba, Saniya and Kahlhoefer, Felix",
    title = "{Electron recoils from terrestrial upscattering of inelastic dark matter}",
    eprint = "2112.06930",
    archivePrefix = "arXiv",
    primaryClass = "hep-ph",
    reportNumber = "TTK-21-53, DESY-21-216",
    doi = "10.1103/PhysRevD.105.055023",
    journal = "Phys. Rev. D",
    volume = "105",
    number = "5",
    pages = "055023",
    year = "2022"
}

@article{Slatyer:2015jla,
    author = "Slatyer, Tracy R.",
    title = "{Indirect dark matter signatures in the cosmic dark ages. I. Generalizing the bound on s-wave dark matter annihilation from Planck results}",
    eprint = "1506.03811",
    archivePrefix = "arXiv",
    primaryClass = "hep-ph",
    reportNumber = "MIT-CTP-4682",
    doi = "10.1103/PhysRevD.93.023527",
    journal = "Phys. Rev. D",
    volume = "93",
    number = "2",
    pages = "023527",
    year = "2016"
}

@article{Cirelli:2024ssz,
    author = "Cirelli, Marco and Strumia, Alessandro and Zupan, Jure",
    title = "{Dark Matter}",
    eprint = "2406.01705",
    archivePrefix = "arXiv",
    primaryClass = "hep-ph",
    month = "6",
    year = "2024"
}

@article{DelaTorreLuque:2023olp,
    author = "De la Torre Luque, Pedro and Balaji, Shyam and Koechler, Jordan",
    title = "{Importance of Cosmic-Ray Propagation on Sub-GeV Dark Matter Constraints}",
    eprint = "2311.04979",
    archivePrefix = "arXiv",
    primaryClass = "hep-ph",
    doi = "10.3847/1538-4357/ad41e0",
    journal = "Astrophys. J.",
    volume = "968",
    number = "1",
    pages = "46",
    year = "2024"
}

@article{Mitridate:2018iag,
    author = "Mitridate, Andrea and Podo, Alessandro",
    title = "{Bounds on Dark Matter decay from 21 cm line}",
    eprint = "1803.11169",
    archivePrefix = "arXiv",
    primaryClass = "hep-ph",
    doi = "10.1088/1475-7516/2018/05/069",
    journal = "JCAP",
    volume = "05",
    pages = "069",
    year = "2018"
}

@article{Slatyer:2016qyl,
    author = "Slatyer, Tracy R. and Wu, Chih-Liang",
    title = "{General Constraints on Dark Matter Decay from the Cosmic Microwave Background}",
    eprint = "1610.06933",
    archivePrefix = "arXiv",
    primaryClass = "astro-ph.CO",
    reportNumber = "MIT-CTP-4842",
    doi = "10.1103/PhysRevD.95.023010",
    journal = "Phys. Rev. D",
    volume = "95",
    number = "2",
    pages = "023010",
    year = "2017"
}

@article{Planck:2015fie,
    author = "Ade, P. A. R. and others",
    collaboration = "Planck",
    title = "{Planck 2015 results. XIII. Cosmological parameters}",
    eprint = "1502.01589",
    archivePrefix = "arXiv",
    primaryClass = "astro-ph.CO",
    doi = "10.1051/0004-6361/201525830",
    journal = "Astron. Astrophys.",
    volume = "594",
    pages = "A13",
    year = "2016"
}

@article{Wadekar:2021qae,
    author = "Wadekar, Digvijay and Wang, Zihui",
    title = "{Strong constraints on decay and annihilation of dark matter from heating of gas-rich dwarf galaxies}",
    eprint = "2111.08025",
    archivePrefix = "arXiv",
    primaryClass = "hep-ph",
    doi = "10.1103/PhysRevD.106.075007",
    journal = "Phys. Rev. D",
    volume = "106",
    number = "7",
    pages = "075007",
    year = "2022"
}

@article{Essig:2013goa,
    author = "Essig, Rouven and Kuflik, Eric and McDermott, Samuel D. and Volansky, Tomer and Zurek, Kathryn M.",
    title = "{Constraining Light Dark Matter with Diffuse X-Ray and Gamma-Ray Observations}",
    eprint = "1309.4091",
    archivePrefix = "arXiv",
    primaryClass = "hep-ph",
    reportNumber = "YITP-SB-29-13, FERMILAB-PUB-13-377-A-T, MCTP-13-27",
    doi = "10.1007/JHEP11(2013)193",
    journal = "JHEP",
    volume = "11",
    pages = "193",
    year = "2013"
}

@article{Balazs:2022tjl,
    author = "Bal{\'a}zs, Csaba and others",
    title = "{Cosmological constraints on decaying axion-like particles: a global analysis}",
    eprint = "2205.13549",
    archivePrefix = "arXiv",
    primaryClass = "astro-ph.CO",
    reportNumber = "gambit-physics-2022, KCL-PH-TH/2022-23, TTP22-034",
    doi = "10.1088/1475-7516/2022/12/027",
    journal = "JCAP",
    volume = "12",
    pages = "027",
    year = "2022"
}

@article{Fixsen:1996nj,
    author = "Fixsen, D. J. and Cheng, E. S. and Gales, J. M. and Mather, John C. and Shafer, R. A. and Wright, E. L.",
    title = "{The Cosmic Microwave Background spectrum from the full COBE FIRAS data set}",
    eprint = "astro-ph/9605054",
    archivePrefix = "arXiv",
    doi = "10.1086/178173",
    journal = "Astrophys. J.",
    volume = "473",
    pages = "576",
    year = "1996"
}

@article{Acharya:2019owx,
    author = "Acharya, Sandeep Kumar and Khatri, Rishi",
    title = "{New CMB spectral distortion constraints on decaying dark matter with full evolution of electromagnetic cascades before recombination}",
    eprint = "1903.04503",
    archivePrefix = "arXiv",
    primaryClass = "astro-ph.CO",
    doi = "10.1103/PhysRevD.99.123510",
    journal = "Phys. Rev. D",
    volume = "99",
    number = "12",
    pages = "123510",
    year = "2019"
}

@article{Linden:2024fby,
    author = "Linden, Tim and Nguyen, Thong T. Q. and Tait, Tim M. P.",
    title = "{X-ray constraints on dark photon tridents}",
    eprint = "2406.19445",
    archivePrefix = "arXiv",
    primaryClass = "hep-ph",
    reportNumber = "UCI-HEP-TR-2024-10",
    doi = "10.1103/37gn-x3y1",
    journal = "Phys. Rev. D",
    volume = "112",
    number = "2",
    pages = "023026",
    year = "2025"
}

@article{Bolliet:2020ofj,
    author = "Bolliet, Boris and Chluba, Jens and Battye, Richard",
    title = "{Spectral distortion constraints on photon injection from low-mass decaying particles}",
    eprint = "2012.07292",
    archivePrefix = "arXiv",
    primaryClass = "astro-ph.CO",
    doi = "10.1093/mnras/stab1997",
    journal = "Mon. Not. Roy. Astron. Soc.",
    volume = "507",
    number = "3",
    pages = "3148--3178",
    year = "2021"
}

@article{Bell:2018zra,
    author = "Bell, Nicole F. and Busoni, Giorgio and Sanderson, Isaac W.",
    title = "{Loop Effects in Direct Detection}",
    eprint = "1803.01574",
    archivePrefix = "arXiv",
    primaryClass = "hep-ph",
    doi = "10.1088/1475-7516/2018/08/017",
    journal = "JCAP",
    volume = "08",
    pages = "017",
    year = "2018",
    note = "[Erratum: JCAP 01, E01 (2019)]"
}

@article{Hook:2010tw,
    author = "Hook, Anson and Izaguirre, Eder and Wacker, Jay G.",
    title = "{Model Independent Bounds on Kinetic Mixing}",
    eprint = "1006.0973",
    archivePrefix = "arXiv",
    primaryClass = "hep-ph",
    reportNumber = "SLAC-PUB-14131",
    doi = "10.1155/2011/859762",
    journal = "Adv. High Energy Phys.",
    volume = "2011",
    pages = "859762",
    year = "2011"
}

@article{Berlin:2018bsc,
    author = "Berlin, Asher and Blinov, Nikita and Krnjaic, Gordan and Schuster, Philip and Toro, Natalia",
    title = "{Dark Matter, Millicharges, Axion and Scalar Particles, Gauge Bosons, and Other New Physics with LDMX}",
    eprint = "1807.01730",
    archivePrefix = "arXiv",
    primaryClass = "hep-ph",
    reportNumber = "FERMILAB-PUB-18-310-A, SLAC-PUB-17297",
    doi = "10.1103/PhysRevD.99.075001",
    journal = "Phys. Rev. D",
    volume = "99",
    number = "7",
    pages = "075001",
    year = "2019"
}

@article{Randall:2008ppe,
    author = "Randall, Scott W. and Markevitch, Maxim and Clowe, Douglas and Gonzalez, Anthony H. and Bradac, Marusa",
    title = "{Constraints on the Self-Interaction Cross-Section of Dark Matter from Numerical Simulations of the Merging Galaxy Cluster 1E 0657-56}",
    eprint = "0704.0261",
    archivePrefix = "arXiv",
    primaryClass = "astro-ph",
    doi = "10.1086/587859",
    journal = "Astrophys. J.",
    volume = "679",
    pages = "1173--1180",
    year = "2008"
}

@article{Finkbeiner:2011dx,
    author = "Finkbeiner, Douglas P. and Galli, Silvia and Lin, Tongyan and Slatyer, Tracy R.",
    title = "{Searching for Dark Matter in the CMB: A Compact Parameterization of Energy Injection from New Physics}",
    eprint = "1109.6322",
    archivePrefix = "arXiv",
    primaryClass = "astro-ph.CO",
    doi = "10.1103/PhysRevD.85.043522",
    journal = "Phys. Rev. D",
    volume = "85",
    pages = "043522",
    year = "2012"
}

@article{ParticleDataGroup:2024cfk,
    author = "Navas, S. and others",
    collaboration = "Particle Data Group",
    title = "{Review of particle physics}",
    doi = "10.1103/PhysRevD.110.030001",
    journal = "Phys. Rev. D",
    volume = "110",
    number = "3",
    pages = "030001",
    year = "2024"
}

@article{Fitzpatrick:2021cij,
    author = "Fitzpatrick, Patrick J. and Liu, Hongwan and Slatyer, Tracy R. and Tsai, Yu-Dai",
    title = "{New thermal relic targets for inelastic vector-portal dark matter}",
    eprint = "2105.05255",
    archivePrefix = "arXiv",
    primaryClass = "hep-ph",
    reportNumber = "MIT-CTP/5298, FERMILAB-PUB-21-189-AE-T",
    doi = "10.1103/PhysRevD.106.083507",
    journal = "Phys. Rev. D",
    volume = "106",
    number = "8",
    pages = "083507",
    year = "2022"
}

@article{Li:2023vpv,
    author = "Li, Shao-Ping and Xu, Xun-Jie",
    title = "{Production rates of dark photons and Z' in the Sun and stellar cooling bounds}",
    eprint = "2304.12907",
    archivePrefix = "arXiv",
    primaryClass = "hep-ph",
    doi = "10.1088/1475-7516/2023/09/009",
    journal = "JCAP",
    volume = "09",
    pages = "009",
    year = "2023"
}

@article{Chang:2018rso,
    author = "Chang, Jae Hyeok and Essig, Rouven and McDermott, Samuel D.",
    title = "{Supernova 1987A Constraints on Sub-GeV Dark Sectors, Millicharged Particles, the QCD Axion, and an Axion-like Particle}",
    eprint = "1803.00993",
    archivePrefix = "arXiv",
    primaryClass = "hep-ph",
    reportNumber = "YITP-SB-18-01, FERMILAB-PUB-17-432-T",
    doi = "10.1007/JHEP09(2018)051",
    journal = "JHEP",
    volume = "09",
    pages = "051",
    year = "2018"
}

@article{Kogut:2011xw,
    author = "Kogut, A. and others",
    title = "{The Primordial Inflation Explorer (PIXIE): A Nulling Polarimeter for Cosmic Microwave Background Observations}",
    eprint = "1105.2044",
    archivePrefix = "arXiv",
    primaryClass = "astro-ph.CO",
    doi = "10.1088/1475-7516/2011/07/025",
    journal = "JCAP",
    volume = "07",
    pages = "025",
    year = "2011"
}

@article{Berlin:2014tja,
    author = "Berlin, Asher and Hooper, Dan and McDermott, Samuel D.",
    title = "{Simplified Dark Matter Models for the Galactic Center Gamma-Ray Excess}",
    eprint = "1404.0022",
    archivePrefix = "arXiv",
    primaryClass = "hep-ph",
    reportNumber = "FERMILAB-PUB-14-060-A, MCTP-14-07",
    doi = "10.1103/PhysRevD.89.115022",
    journal = "Phys. Rev. D",
    volume = "89",
    number = "11",
    pages = "115022",
    year = "2014"
}

@article{Smith-Orlik:2023kyl,
    author = "Smith-Orlik, Adam and others",
    title = "{The impact of the Large Magellanic Cloud on dark matter direct detection signals}",
    eprint = "2302.04281",
    archivePrefix = "arXiv",
    primaryClass = "astro-ph.GA",
    doi = "10.1088/1475-7516/2023/10/070",
    journal = "JCAP",
    volume = "10",
    pages = "070",
    year = "2023"
}

@article{Duerr:2020muu,
    author = "Duerr, Michael and Ferber, Torben and Garcia-Cely, Camilo and Hearty, Christopher and Schmidt-Hoberg, Kai",
    title = "{Long-lived Dark Higgs and Inelastic Dark Matter at Belle II}",
    eprint = "2012.08595",
    archivePrefix = "arXiv",
    primaryClass = "hep-ph",
    reportNumber = "DESY-20-226",
    doi = "10.1007/JHEP04(2021)146",
    journal = "JHEP",
    volume = "04",
    pages = "146",
    year = "2021"
}

@article{Herrera:2023fpq,
    author = "Herrera, Gonzalo and Ibarra, Alejandro and Shirai, Satoshi",
    title = "{Enhanced prospects for direct detection of inelastic dark matter from a non-galactic diffuse component}",
    eprint = "2301.00870",
    archivePrefix = "arXiv",
    primaryClass = "hep-ph",
    reportNumber = "TUM-HEP 1447/22, IPMU22-0070",
    doi = "10.1088/1475-7516/2023/04/026",
    journal = "JCAP",
    volume = "04",
    pages = "026",
    year = "2023"
}

@article{Bell:2021xff,
    author = "Bell, Nicole F. and Dent, James B. and Dutta, Bhaskar and Ghosh, Sumit and Kumar, Jason and Newstead, Jayden L. and Shoemaker, Ian M.",
    title = "{Cosmic-ray upscattered inelastic dark matter}",
    eprint = "2108.00583",
    archivePrefix = "arXiv",
    primaryClass = "hep-ph",
    reportNumber = "MI-HET-758",
    doi = "10.1103/PhysRevD.104.076020",
    journal = "Phys. Rev. D",
    volume = "104",
    pages = "076020",
    year = "2021"
}

@article{Guha:2024mjr,
    author = "Guha, Atanu and Park, Jong-Chul",
    title = "{Constraints on cosmic-ray boosted dark matter with realistic cross section}",
    eprint = "2401.07750",
    archivePrefix = "arXiv",
    primaryClass = "hep-ph",
    reportNumber = "Journal of Cosmology and Astroparticle Physics, Volume 2024, July
  2024",
    doi = "10.1088/1475-7516/2024/07/074",
    journal = "JCAP",
    volume = "07",
    pages = "074",
    year = "2024"
}

@article{Bhowmick:2022zkj,
    author = "Bhowmick, Supritha and Ghosh, Diptimoy and Sachdeva, Divya",
    title = "{Blazar boosted dark matter {\textemdash} direct detection constraints on {\ensuremath{\sigma}}e{\ensuremath{\chi}}: role of energy dependent cross sections}",
    eprint = "2301.00209",
    archivePrefix = "arXiv",
    primaryClass = "hep-ph",
    doi = "10.1088/1475-7516/2023/07/039",
    journal = "JCAP",
    volume = "07",
    pages = "039",
    year = "2023"
}

@article{CDEX:2024qzq,
    author = "Xu, R. and others",
    collaboration = "CDEX",
    title = "{Constraints on the Blazar-Boosted Dark Matter from the CDEX-10 Experiment}",
    eprint = "2403.20276",
    archivePrefix = "arXiv",
    primaryClass = "hep-ex",
    month = "3",
    year = "2024"
}

@article{Jeesun:2025gzt,
    author = "Jeesun, Sk",
    title = "{Blazar boosted ALP and vector portal dark matter confronting light mediator searches}",
    eprint = "2501.11569",
    archivePrefix = "arXiv",
    primaryClass = "hep-ph",
    doi = "10.1103/PhysRevD.111.103022",
    journal = "Phys. Rev. D",
    volume = "111",
    number = "10",
    pages = "103022",
    year = "2025"
}

@article{Gondolo:2012vh,
    author = "Gondolo, Paolo and Hisano, Junji and Kadota, Kenji",
    title = "{The Effect of quark interactions on dark matter kinetic decoupling and the mass of the smallest dark halos}",
    eprint = "1205.1914",
    archivePrefix = "arXiv",
    primaryClass = "hep-ph",
    doi = "10.1103/PhysRevD.86.083523",
    journal = "Phys. Rev. D",
    volume = "86",
    pages = "083523",
    year = "2012"
}

@article{Bertoni:2014mva,
    author = "Bertoni, Bridget and Ipek, Seyda and McKeen, David and Nelson, Ann E.",
    title = "{Constraints and consequences of reducing small scale structure via large dark matter-neutrino interactions}",
    eprint = "1412.3113",
    archivePrefix = "arXiv",
    primaryClass = "hep-ph",
    doi = "10.1007/JHEP04(2015)170",
    journal = "JHEP",
    volume = "04",
    pages = "170",
    year = "2015"
}

@article{Pereira:2024vdk,
    author = "Pereira, David Silva and Ferraz, Jo{\~a}o and Lobo, Francisco S. N. and Mimoso, Jos{\'e} Pedro",
    title = "{Thermodynamics of the Primordial Universe}",
    eprint = "2411.03018",
    archivePrefix = "arXiv",
    primaryClass = "gr-qc",
    doi = "10.3390/e26110947",
    journal = "Entropy",
    volume = "26",
    number = "11",
    pages = "947",
    year = "2024"
}

@article{Berlin:2016gtr,
    author = "Berlin, Asher and Hooper, Dan and Krnjaic, Gordan",
    title = "{Thermal Dark Matter From A Highly Decoupled Sector}",
    eprint = "1609.02555",
    archivePrefix = "arXiv",
    primaryClass = "hep-ph",
    reportNumber = "FERMILAB-PUB-16-318-A",
    doi = "10.1103/PhysRevD.94.095019",
    journal = "Phys. Rev. D",
    volume = "94",
    number = "9",
    pages = "095019",
    year = "2016"
}

@article{Pappadopulo:2016pkp,
    author = "Pappadopulo, Duccio and Ruderman, Joshua T. and Trevisan, Gabriele",
    title = "{Dark matter freeze-out in a nonrelativistic sector}",
    eprint = "1602.04219",
    archivePrefix = "arXiv",
    primaryClass = "hep-ph",
    doi = "10.1103/PhysRevD.94.035005",
    journal = "Phys. Rev. D",
    volume = "94",
    number = "3",
    pages = "035005",
    year = "2016"
}

@article{DallaValleGarcia:2024zva,
    author = "Dalla Valle Garcia, Giovani",
    title = "{A minimalistic model for inelastic dark matter}",
    eprint = "2411.02147",
    archivePrefix = "arXiv",
    primaryClass = "hep-ph",
    doi = "10.1016/j.physletb.2025.139320",
    journal = "Phys. Lett. B",
    volume = "862",
    pages = "139320",
    year = "2025"
}

@article{Feng:2010zp,
    author = "Feng, Jonathan L. and Kaplinghat, Manoj and Yu, Hai-Bo",
    title = "{Sommerfeld Enhancements for Thermal Relic Dark Matter}",
    eprint = "1005.4678",
    archivePrefix = "arXiv",
    primaryClass = "hep-ph",
    reportNumber = "UCI-TR-2010-06",
    doi = "10.1103/PhysRevD.82.083525",
    journal = "Phys. Rev. D",
    volume = "82",
    pages = "083525",
    year = "2010"
}

@article{Bell:2018pkk,
    author = "Bell, Nicole F. and Busoni, Giorgio and Robles, Sandra",
    title = "{Heating up Neutron Stars with Inelastic Dark Matter}",
    eprint = "1807.02840",
    archivePrefix = "arXiv",
    primaryClass = "hep-ph",
    doi = "10.1088/1475-7516/2018/09/018",
    journal = "JCAP",
    volume = "09",
    pages = "018",
    year = "2018"
}

@article{Gustafson:2024aom,
    author = "Gustafson, R. Andrew and Herrera, Gonzalo and Mukhopadhyay, Mainak and Murase, Kohta and Shoemaker, Ian M.",
    title = "{Cosmic-ray cooling in active galactic nuclei as a new probe of inelastic dark matter}",
    eprint = "2408.08947",
    archivePrefix = "arXiv",
    primaryClass = "hep-ph",
    doi = "10.1103/5m57-vgt2",
    journal = "Phys. Rev. D",
    volume = "111",
    number = "12",
    pages = "L121303",
    year = "2025"
}

@article{Gustafson:2025dff,
    author = "Gustafson, R. Andrew and Herrera, Gonzalo and Mukhopadhyay, Mainak and Murase, Kohta and Shoemaker, Ian M.",
    title = "{Cosmic-ray boosted inelastic dark matter from neutrino-emitting active galactic nuclei}",
    eprint = "2508.20984",
    archivePrefix = "arXiv",
    primaryClass = "hep-ph",
    reportNumber = "FERMILAB-PUB-25-0620-T",
    month = "8",
    year = "2025"
}

@article{Mishra:2025juk,
    author = "Mishra, Arvind Kumar and Liu, Ning and Lu, Chih-Ting",
    title = "{Probing gauged U(1) sub-GeV dark matter via cosmic ray cooling in active galactic nuclei}",
    eprint = "2504.03409",
    archivePrefix = "arXiv",
    primaryClass = "hep-ph",
    doi = "10.1016/j.dark.2025.102050",
    journal = "Phys. Dark Univ.",
    volume = "49",
    pages = "102050",
    year = "2025"
}

@article{Robertson:2016xjh,
    author = "Robertson, Andrew and Massey, Richard and Eke, Vincent",
    title = "{What does the Bullet Cluster tell us about self-interacting dark matter?}",
    eprint = "1605.04307",
    archivePrefix = "arXiv",
    primaryClass = "astro-ph.CO",
    doi = "10.1093/mnras/stw2670",
    journal = "Mon. Not. Roy. Astron. Soc.",
    volume = "465",
    number = "1",
    pages = "569--587",
    year = "2017"
}

@article{Wittman:2017gxn,
    author = "Wittman, David and Golovich, Nathan and Dawson, William A.",
    title = "{The Mismeasure of Mergers: Revised Limits on Self-interacting Dark Matter in Merging Galaxy Clusters}",
    eprint = "1701.05877",
    archivePrefix = "arXiv",
    primaryClass = "astro-ph.CO",
    doi = "10.3847/1538-4357/aaee77",
    journal = "Astrophys. J.",
    volume = "869",
    number = "2",
    pages = "104",
    year = "2018"
}

@article{XENON:2020rca,
    author = "Aprile, E. and others",
    collaboration = "XENON",
    title = "{Excess electronic recoil events in XENON1T}",
    eprint = "2006.09721",
    archivePrefix = "arXiv",
    primaryClass = "hep-ex",
    doi = "10.1103/PhysRevD.102.072004",
    journal = "Phys. Rev. D",
    volume = "102",
    number = "7",
    pages = "072004",
    year = "2020"
}

@article{XENON:2022ltv,
    author = "Aprile, E. and others",
    collaboration = "XENON",
    title = "{Search for New Physics in Electronic Recoil Data from XENONnT}",
    eprint = "2207.11330",
    archivePrefix = "arXiv",
    primaryClass = "hep-ex",
    doi = "10.1103/PhysRevLett.129.161805",
    journal = "Phys. Rev. Lett.",
    volume = "129",
    number = "16",
    pages = "161805",
    year = "2022"
}

@article{PandaX:2024cic,
    author = "Zeng, Xinning and others",
    collaboration = "PandaX",
    title = "{Exploring New Physics with PandaX-4T Low Energy Electronic Recoil Data}",
    eprint = "2408.07641",
    archivePrefix = "arXiv",
    primaryClass = "hep-ex",
    doi = "10.1103/PhysRevLett.134.041001",
    journal = "Phys. Rev. Lett.",
    volume = "134",
    number = "4",
    pages = "041001",
    year = "2025"
}

@article{An:2014twa,
    author = "An, Haipeng and Pospelov, Maxim and Pradler, Josef and Ritz, Adam",
    title = "{Direct Detection Constraints on Dark Photon Dark Matter}",
    eprint = "1412.8378",
    archivePrefix = "arXiv",
    primaryClass = "hep-ph",
    reportNumber = "CALT-TH-2014-173",
    doi = "10.1016/j.physletb.2015.06.018",
    journal = "Phys. Lett. B",
    volume = "747",
    pages = "331--338",
    year = "2015"
}

@misc{NIST_XCOM_2010,
  author       = {M. J. Berger and J. H. Hubbell and S. M. Seltzer and J. Chang and J. S. Coursey and D. S. Zucker and K. Olsen},
  title        = {{XCOM: Photon Cross Sections Database}},
  howpublished = {\url{https://www.nist.gov/pml/xcom-photon-cross-sections-database}},
  year         = {2010},
  note         = {NIST Standard Reference Database 8 (XGAM). DOI: 10.18434/T48G6X},
  organization = {National Institute of Standards and Technology},
  address      = {Gaithersburg, MD, USA},
}

@article{Bouchet:2008rp,
    author = "Bouchet, L. and Jourdain, E. and Roques, J. P. and Strong, A. and Diehl, R. and Lebrun, F. and Terrier, R.",
    title = "{INTEGRAL SPI All-Sky View in Soft Gamma Rays: Study of Point Source and Galactic Diffuse Emissions}",
    eprint = "0801.2086",
    archivePrefix = "arXiv",
    primaryClass = "astro-ph",
    doi = "10.1086/529489",
    journal = "Astrophys. J.",
    volume = "679",
    pages = "1315",
    year = "2008"
}

@article{Aoki:2022tek,
    author = "Aoki, Sinya and Kawana, Kiyoharu",
    title = "{Entropy and its conservation in expanding Universe}",
    eprint = "2210.03323",
    archivePrefix = "arXiv",
    primaryClass = "hep-th",
    reportNumber = "YITP-22-115",
    doi = "10.1142/S0217751X23500720",
    journal = "Int. J. Mod. Phys. A",
    volume = "38",
    number = "14",
    pages = "2350072",
    year = "2023"
}

@misc{Na64SPSC,
  author = "Celentano, A. and Sieber, H.",
  title = "\href{https://indico.cern.ch/event/1303571/contributions/5482379/attachments/2702152/4702863/NA64-SPSC-2023-10.pdf}{Status and plans of the NA64e and NA64mu Experiments}",
  year = {2023}
}

@article{ALEPH:2005ab,
    author = "Schael, S. and others",
    collaboration = "ALEPH, DELPHI, L3, OPAL, SLD, LEP Electroweak Working Group, SLD Electroweak Group, SLD Heavy Flavour Group",
    title = "{Precision electroweak measurements on the $Z$ resonance}",
    eprint = "hep-ex/0509008",
    archivePrefix = "arXiv",
    reportNumber = "SLAC-R-774",
    doi = "10.1016/j.physrep.2005.12.006",
    journal = "Phys. Rept.",
    volume = "427",
    pages = "257--454",
    year = "2006"
}

@article{L3:1998uub,
    author = "Acciarri, M. and others",
    collaboration = "L3",
    title = "{Determination of the number of light neutrino species from single photon production at LEP}",
    reportNumber = "CERN-EP-98-025, CERN-EP-98-25",
    doi = "10.1016/S0370-2693(98)00519-X",
    journal = "Phys. Lett. B",
    volume = "431",
    pages = "199--208",
    year = "1998"
}

@article{CMS:2022ett,
    author = "Tumasyan, Armen and others",
    collaboration = "CMS",
    title = "{Precision measurement of the Z boson invisible width in pp collisions at s=13 TeV}",
    eprint = "2206.07110",
    archivePrefix = "arXiv",
    primaryClass = "hep-ex",
    reportNumber = "CMS-SMP-18-014, CERN-EP-2022-088",
    doi = "10.1016/j.physletb.2022.137563",
    journal = "Phys. Lett. B",
    volume = "842",
    pages = "137563",
    year = "2023"
}

@article{ATLAS:2023ynf,
    author = "Aad, Georges and others",
    collaboration = "ATLAS",
    title = "{Measurement of the Z boson invisible width at s=13 TeV with the ATLAS detector}",
    eprint = "2312.02789",
    archivePrefix = "arXiv",
    primaryClass = "hep-ex",
    reportNumber = "CERN-EP-2023-232",
    doi = "10.1016/j.physletb.2024.138705",
    journal = "Phys. Lett. B",
    volume = "854",
    pages = "138705",
    year = "2024"
}

@article{FCC:2018byv,
    author = "Abada, A. and others",
    collaboration = "FCC",
    title = "{FCC Physics Opportunities}: {Future Circular Collider Conceptual Design Report Volume 1}",
    reportNumber = "CERN-ACC-2018-0056",
    doi = "10.1140/epjc/s10052-019-6904-3",
    journal = "Eur. Phys. J. C",
    volume = "79",
    number = "6",
    pages = "474",
    year = "2019"
}

@article{Brahma:2023psr,
    author = "Brahma, Nirmalya and Heeba, Saniya and Schutz, Katelin",
    title = "{Resonant pseudo-Dirac dark matter as a sub-GeV thermal target}",
    eprint = "2308.01960",
    archivePrefix = "arXiv",
    primaryClass = "hep-ph",
    doi = "10.1103/PhysRevD.109.035006",
    journal = "Phys. Rev. D",
    volume = "109",
    number = "3",
    pages = "035006",
    year = "2024"
}

@article{Heeba:2023bik,
    author = "Heeba, Saniya and Lin, Tongyan and Schutz, Katelin",
    title = "{Inelastic freeze-in}",
    eprint = "2304.06072",
    archivePrefix = "arXiv",
    primaryClass = "hep-ph",
    doi = "10.1103/PhysRevD.108.095016",
    journal = "Phys. Rev. D",
    volume = "108",
    number = "9",
    pages = "095016",
    year = "2023"
}

@article{Silva-Malpartida:2024emu,
    author = "Silva-Malpartida, Javier and Bernal, Nicol{\'a}s and Jones-P{\'e}rez, Joel and Lineros, Roberto A.",
    title = "{From WIMPs to FIMPs: impact of early matter domination}",
    eprint = "2408.08950",
    archivePrefix = "arXiv",
    primaryClass = "hep-ph",
    doi = "10.1088/1475-7516/2025/03/003",
    journal = "JCAP",
    volume = "03",
    pages = "003",
    year = "2025"
}

@article{Silva-Malpartida:2023yks,
    author = "Silva-Malpartida, Javier and Bernal, Nicol{\'a}s and Jones-P{\'e}rez, Joel and Lineros, Roberto A.",
    title = "{From WIMPs to FIMPs with low~reheating~temperatures}",
    eprint = "2306.14943",
    archivePrefix = "arXiv",
    primaryClass = "hep-ph",
    doi = "10.1088/1475-7516/2023/09/015",
    journal = "JCAP",
    volume = "09",
    pages = "015",
    year = "2023"
}

@article{Bernal:2025szh,
    author = "Bernal, Nicol{\'a}s and Neto, Jacinto P. and Silva-Malpartida, Javier and Queiroz, Farinaldo S.",
    title = "{Enabling thermal dark matter within the vanilla L{\ensuremath{\mu}}-L{\ensuremath{\tau}} model}",
    eprint = "2507.02048",
    archivePrefix = "arXiv",
    primaryClass = "hep-ph",
    doi = "10.1103/8g85-c8sh",
    journal = "Phys. Rev. D",
    volume = "112",
    number = "7",
    pages = "075042",
    year = "2025"
}

@article{Haque:2023yra,
    author = "Haque, MD Riajul and Maity, Debaprasad and Mondal, Rajesh",
    title = "{WIMPs, FIMPs, and Inflaton phenomenology via reheating, CMB and {\ensuremath{\Delta}}N$_{eff}$}",
    eprint = "2301.01641",
    archivePrefix = "arXiv",
    primaryClass = "hep-ph",
    doi = "10.1007/JHEP09(2023)012",
    journal = "JHEP",
    volume = "09",
    pages = "012",
    year = "2023"
}

@article{Bernal:2022wck,
    author = "Bernal, Nicol{\'a}s and Xu, Yong",
    title = "{WIMPs during reheating}",
    eprint = "2209.07546",
    archivePrefix = "arXiv",
    primaryClass = "hep-ph",
    reportNumber = "PI/UAN-2022-722FT",
    doi = "10.1088/1475-7516/2022/12/017",
    journal = "JCAP",
    volume = "12",
    pages = "017",
    year = "2022"
}

@article{Acharya:2008bk,
    author = "Acharya, Bobby Samir and Kumar, Piyush and Bobkov, Konstantin and Kane, Gordon and Shao, Jing and Watson, Scott",
    title = "{Non-thermal Dark Matter and the Moduli Problem in String Frameworks}",
    eprint = "0804.0863",
    archivePrefix = "arXiv",
    primaryClass = "hep-ph",
    reportNumber = "UCB-PTH-08-06, MCTP-08-10",
    doi = "10.1088/1126-6708/2008/06/064",
    journal = "JHEP",
    volume = "06",
    pages = "064",
    year = "2008"
}

@article{Drees:2017iod,
    author = "Drees, Manuel and Hajkarim, Fazlollah",
    title = "{Dark Matter Production in an Early Matter Dominated Era}",
    eprint = "1711.05007",
    archivePrefix = "arXiv",
    primaryClass = "hep-ph",
    doi = "10.1088/1475-7516/2018/02/057",
    journal = "JCAP",
    volume = "02",
    pages = "057",
    year = "2018"
}

@article{Arias:2019uol,
    author = "Arias, Paola and Bernal, Nicol{\'a}s and Herrera, Alan and Maldonado, Carlos",
    title = "{Reconstructing Non-standard Cosmologies with Dark Matter}",
    eprint = "1906.04183",
    archivePrefix = "arXiv",
    primaryClass = "hep-ph",
    reportNumber = "PI/UAN-2019-649FT",
    doi = "10.1088/1475-7516/2019/10/047",
    journal = "JCAP",
    volume = "10",
    pages = "047",
    year = "2019"
}

@article{Totani:2025fxx,
    author = "Totani, Tomonori",
    title = "{20 GeV halo-like excess of the Galactic diffuse emission and implications for dark matter annihilation}",
    eprint = "2507.07209",
    archivePrefix = "arXiv",
    primaryClass = "astro-ph.HE",
    doi = "10.1088/1475-7516/2025/11/080",
    journal = "JCAP",
    volume = "11",
    pages = "080",
    year = "2025"
}

@article{Darme:2018jmx,
    author = "Darm{\'e}, Luc and Rao, Soumya and Roszkowski, Leszek",
    title = "{Signatures of dark Higgs boson in light fermionic dark matter scenarios}",
    eprint = "1807.10314",
    archivePrefix = "arXiv",
    primaryClass = "hep-ph",
    doi = "10.1007/JHEP12(2018)014",
    journal = "JHEP",
    volume = "12",
    pages = "014",
    year = "2018"
}

@article{DeSimone:2010tf,
    author = "De Simone, Andrea and Sanz, Veronica and Sato, Hiromitsu Phil",
    title = "{Pseudo-Dirac Dark Matter Leaves a Trace}",
    eprint = "1004.1567",
    archivePrefix = "arXiv",
    primaryClass = "hep-ph",
    reportNumber = "MIT-CTP-4142",
    doi = "10.1103/PhysRevLett.105.121802",
    journal = "Phys. Rev. Lett.",
    volume = "105",
    pages = "121802",
    year = "2010"
}

\end{document}